\documentclass{aa}
\usepackage[varg]{txfonts}
\usepackage[dvipsnames]{xcolor}
\usepackage{fancyhdr}
\usepackage[english]{babel}
\usepackage{graphicx}
\usepackage{amssymb}
\usepackage{geometry}
\usepackage{amsmath}
\usepackage{footnote}
\usepackage{natbib}
\usepackage{pdflscape}
\usepackage{longtable}
\bibpunct{(}{)}{;}{a}{}{,}
\usepackage[pdftex,colorlinks]{hyperref}
\begin{document}
\title{SUPER IV. CO($J$=3$-$2) properties of active galactic nucleus hosts at cosmic noon revealed by ALMA
}
\titlerunning{SUPER IV. CO($J$=3$-$2) properties of AGN hosts at Cosmic Noon}

\author{C.~Circosta\inst{\ref*{UCL}}\thanks{\email{c.circosta@ucl.ac.uk}}
 \and V.~Mainieri\inst{\ref*{ESO}}
 \and I.~Lamperti\inst{\ref*{ESO},\ref*{UCL}}
 \and P.~Padovani\inst{\ref*{ESO}}
 \and M.~Bischetti\inst{\ref*{OAtrieste}}
 \and C.~M.~Harrison\inst{\ref*{newcastle}}
 \and D.~Kakkad\inst{\ref*{ESOchile}}
 \and A.~Zanella\inst{\ref*{pd}}
 \and G.~Vietri\inst{\ref*{milano}}
 \and G.~Lanzuisi\inst{\ref*{DIFA},\ref*{OABO}}
 \and M.~Salvato\inst{\ref*{MPE}}
 \and M.~Brusa\inst{\ref*{DIFA},\ref*{OABO}}
 \and S.~Carniani\inst{\ref*{pisa}}
 \and C.~Cicone\inst{\ref*{oslo}}
 \and G.~Cresci\inst{\ref*{OAarcetri}}
 \and C.~Feruglio\inst{\ref*{OAtrieste}}
 \and B.~Husemann\inst{\ref*{MPIA}}
 \and F.~Mannucci\inst{\ref*{OAarcetri}}
 \and A.~Marconi\inst{\ref*{uniFI},\ref*{OAarcetri}}
 \and M.~Perna\inst{\ref*{cab},\ref*{OAarcetri}}
 \and E.~Piconcelli\inst{\ref*{OAroma}}
 \and A.~Puglisi\inst{\ref*{durham}}
 \and A.~Saintonge\inst{\ref*{UCL}}
 \and M.~Schramm\inst{\ref*{japan}}
 \and C.~Vignali\inst{\ref*{DIFA},\ref*{OABO}}
 \and L.~Zappacosta\inst{\ref*{OAroma}}
  }
 
\institute{Department of Physics \& Astronomy, University College London, Gower Street, London WC1E 6BT, UK\label{UCL}
 \and European Southern Observatory, Karl-Schwarzschild-Str. 2, 85748 Garching bei M\"{u}nchen, Germany\label{ESO}
 \and INAF - Osservatorio Astronomico di Trieste, via G.B. Tiepolo 11, 34143 Trieste, Italy\label{OAtrieste}
 \and School of Mathematics, Statistics and Physics, Newcastle University, NE1 7RU, UK\label{newcastle}
 \and European Southern Observatory, Alonso de Cordova 3107, Casilla 19, Santiago 19001, Chile\label{ESOchile}
  \and INAF - Osservatorio Astronomico di Padova, Vicolo dell’Osservatorio 5, I-35122, Padova, Italy\label{pd}
 \and INAF IASF-Milano, Via Alfonso Corti 12, 20133 Milano\label{milano}
 \and Dipartimento di Fisica e Astronomia dell’Universit\`a degli Studi di Bologna, via P. Gobetti 93/2, 40129 Bologna, Italy\label{DIFA}
 \and INAF/OAS, Osservatorio di Astrofisica e Scienza dello Spazio di Bologna, via P. Gobetti 93/3, 40129 Bologna, Italy\label{OABO}
 \and MPE, Giessenbach-Str. 1, 85748 Garching bei M\"{u}nchen, Germany\label{MPE}
 \and Scuola Normale Superiore, Piazza dei Cavalieri 7, I-56126 Pisa, Italy\label{pisa}
 \and Institute of Theoretical Astrophysics, University of Oslo, P.O. Box 1029, Blindern, 0315 Oslo, Norway\label{oslo}
 \and INAF - Osservatorio Astrofisico di Arcetri, Largo E. Fermi 5, 50125 Firenze, Italy\label{OAarcetri}
 \and Max-Planck-Institut f\"{u}r Astronomie, K\"{o}nigstuhl 17, 69117 Heidelberg, Germany\label{MPIA}
 \and Dipartimento di Fisica e Astronomia, Università di Firenze, Via G. Sansone 1, 50019 Sesto Fiorentino (Firenze), Italy\label{uniFI}
 \and Centro de Astrobiología (CAB, CSIC–INTA), Departamento de Astrofísica, Cra. de Ajalvir Km. 4, 28850 – Torrejón de Ardoz, Madrid, Spain\label{cab}
 \and INAF - Osservatorio Astronomico di Roma, Via Frascati 33, 00078 Monte Porzio Catone (Roma), Italy\label{OAroma}
 \and Centre for Extragalactic Astronomy, Department of Physics, Durham University, South Road, Durham, DH1 3LE, UK\label{durham}
 \and Graduate school of Science and Engineering, Saitama Univ. 255 Shimo-Okubo, Sakura-ku, Saitama City, Saitama 338-8570, Japan\label{japan}
}

\abstract{Feedback from active galactic nuclei (AGN) is thought to be key in shaping the life cycle of their host galaxies by regulating star-formation activity. Therefore, to understand the impact of AGN on star formation, it is essential to
trace the molecular gas out of which stars form. 
In this paper we present the first systematic study of the CO properties of AGN hosts at $z\approx2$ for a sample of 27 X-ray selected AGN spanning two orders of magnitude in AGN bolometric luminosity ($\log L_{\textnormal{bol}}/\textnormal{erg s}^{-1}=44.7-46.9$) by using ALMA Band 3 observations of the CO(3-2) transition ($\sim$1$''$ angular resolution). To search for evidence of AGN feedback on the CO properties of the host galaxies, 
we compared our AGN with a sample of inactive (i.e., non-AGN) galaxies from the PHIBSS survey with similar redshift, stellar masses, and star-formation rates (SFRs). We used the same CO transition as a consistent proxy for the gas mass for the two samples in order to avoid systematics involved when assuming conversion factors (e.g., excitation corrections and $\alpha_\textnormal{CO}$). 
By adopting a Bayesian approach to take upper limits into account, we analyzed CO luminosities as a function of stellar masses and SFRs, as well as the ratio $L'_{\textnormal{CO(3-2)}}/M_{\ast}$ (a proxy for the gas fraction). The two samples show statistically consistent trends in the $L'_{\textnormal{CO(3-2)}}$-$L_{\textnormal{FIR}}$ and $L'_{\textnormal{CO(3-2)}}$-$M_{\ast}$ planes. However, there are indications that AGN feature lower CO(3-2) luminosities (0.4$-$0.7 dex) than inactive galaxies at the 2$-$3$\sigma$ level when we focus on the subset of parameters where the results are better constrained (i.e., $L_{\textnormal{FIR}}\approx10^{12.2}$ $L_{\odot}$ and $M_{\ast}>10^{11}$ M$_{\odot}$) and on the distribution of the mean $\log(L'_{\textnormal{CO(3-2)}}/M_{\ast})$. Therefore, even by conservatively assuming the same excitation factor $r_{31}$, we would find lower molecular gas masses in AGN, and assuming higher $r_{31}$ would exacerbate this difference. We interpret our result as a hint of the potential effect of AGN activity (such as radiation and outflows), which may be able to heat, excite, dissociate, and/or deplete the gas reservoir of the host galaxies. Better SFR measurements and deeper CO observations for AGN as well as larger and more uniformly selected samples of both AGN and inactive galaxies are required to confirm whether there is a true difference between the two populations.
}
\keywords{galaxies: active -- galaxies: evolution -- galaxies:ISM -- quasars: general -- surveys -- submillimeter:ISM} 
\maketitle
\section{Introduction}

The active phases supermassive black holes (SMBHs) 
go through while accreting material, when they are visible as active galactic nuclei (AGN), are thought to play a key role in galaxy evolution. 
During these phases, the central engine releases a huge amount of energy, which is injected into the surrounding interstellar medium (ISM). This energy, if coupled efficiently, could remove, heat, and/or dissociate the molecular gas, the fuel out of which stars form, as well as heat up the halo and suppress further gas accretion. Such a process, referred to as AGN feedback, may regulate the growth of the host galaxy
\citep{fabian12,kormendy13,king15,harrison17}. 
Theoretically, AGN feedback is invoked to reproduce the observed properties of the galaxy population, for example the lack of very massive galaxies in the most massive galaxy haloes \citep{somerville08}, the bimodal color distribution of galaxies \citep{strateva01}, and the correlations between SMBH mass and host galaxy properties \citep{magorrian98,ferrarese00}. 
Although AGN feedback is a necessary ingredient in models of galaxy evolution, proving its role observationally remains a challenge. In particular, powerful outflows have been observed in many AGN host galaxies \citep[e.g.,][]{rupke13,cicone14,genzel14,harrison14,carniani15,cresci15,kakkad16,nesvadba17a,brusa18,davies20,veilleux20}
, but their long-term impact on the global star-forming activity of the hosts is still an open issue \citep[e.g.,][]{cresci18,gallagher19,scholtz20}.

To understand the link between AGN and star formation, several authors have investigated the star-formation rates (SFRs) of AGN host galaxies as a function of AGN luminosity \citep{lutz10,harrison12,page12,rosario12,rovilos12,delvecchio15,stanley15,harris16}. Many of the apparently discrepant results in these works can be explained by selection effects (e.g., controlling for stellar mass and redshift evolution; e.g., \citealt{stanley17}). However, while the observational results are broadly consistent with simple feedback models \citep[e.g.,][]{harrison17}, they currently provide limited diagnostic power on the specifics of how AGN feedback works, at least in the absense of a wide range of theoretical predictions \citep[e.g.,][]{scholtz18,schulze19}.

Observations of cold molecular gas are more promising since any activity from the AGN (radiation, outflows, or jets) will have an impact on the molecular gas reservoir first, and then on the SFR, which previous studies have focused on. The molecular gas provides an instantaneous measure of the raw fuel from which stars form and can be used as a more direct tracer of potential feedback effects. 
Over the last decade, large observational efforts have been devoted to map the molecular gas reservoir of galaxies \citep[e.g.,][]{daddi10gas,garcia12,bauermeister13,bothwell13,tacconi13,genzel15,silverman15,decarli16,cicone17,saintonge17,tacconi18,fluetsch19,freundlich19}. These studies are largely based on observations of carbon monoxide (CO) rotational emission lines, used as a tracer of cold molecular hydrogen $H_{2}$ (the ground-state rotational transition in particular traces the total molecular gas best), but there are also many studies based on dust emission (e.g., \citealt{tacconi18} and references therein). The main targets of such CO campaigns have primarily been inactive\footnote{Throughout the paper, we refer to AGN and inactive (i.e., non-AGN) galaxies as those showing and lacking, respectively, an active nucleus at their center, regardless of their SFR.} galaxies that mostly lie on the main sequence of star-forming galaxies \citep[e.g.,][]{noeske07,schreiber15}, where the majority of the cosmic star-formation activity occurs. 
The fundamental relation between SFR and the molecular gas content of galaxies, the Schmidt-Kennicutt relation, provides precious information about how efficiently galaxies turn their gas into stars \citep[][]{schmidt59,kennicutt89}. This star-formation law is usually presented in terms of surface densities, therefore requiring spatially resolved measurements of galaxies. However, in high-redshift studies, an integrated form of this relation with global measurements of the SFR and molecular mass is normally used \citep[e.g.,][]{carilli13,sargent14}. 

A similar observational effort is needed to systematically characterize the cold gas phase of AGN host galaxies 
to understand whether this is different from inactive galaxies and quantify any potential effect of AGN feedback on the host galaxy ISM. 
Local studies agree in reporting no clear evidence for AGN to affect the ISM component of the host, by tracing the molecular gas phase \citep[e.g.,][]{saintonge12,husemann17CO,saintonge17,rosario18,jarvis20,shangguan20}, the atomic one \citep[e.g.,][]{ellison19}, dust mass and dust absorption as a proxy of the gas mass \citep[e.g.,][]{shangguan19,yesuf20}. AGN appear to follow the same star-formation law as inactive galaxies. Interestingly, studies at redshift $z>1$ present contrasting results. Since the redshift $1<z<3$, the so-called cosmic noon, corresponds to the peak of accretion activity of SMBHs \citep{madau14,aird15}, when the energy injected into the host galaxy may be maximized, this cosmic epoch is a crucial laboratory to look for AGN feedback effects. In particular, some studies find reduced molecular gas fractions (i.e., the molecular gas mass per unit stellar mass, $f_{\textnormal{gas}}=M_{\textnormal{mol}}/M_{\ast}$) and depletion timescales (i.e., the timescale needed to consume the available molecular gas content given the current SFR, $t_{\textnormal{dep}}=M_{\textnormal{mol}}/\textnormal{SFR}$) of AGN compared with the parent population of inactive galaxies \citep[][but see also \citealt{kirkpatrick19}; \citealt{herreracamus19}; \citealt{spingola20}]{carilli13,brusa15,carniani17,kakkad17,fiore17,perna18,talia18,loiacono19}. This has been interpreted as evidence for highly efficient gas consumption possibly related to AGN feedback affecting the gas reservoir of the host galaxies. For a few targets, fast ionized/molecular outflows were also detected \citep[e.g.,][]{brusa15,carniani17,brusa18}. Complementary observations tracing the possible presence of outflows are therefore needed to generally confirm such conclusions \citep[e.g.,][]{vayner17,brusa18,herreracamus19,loiacono19}. 

Nevertheless, these studies can be affected by assumptions and limitations. 
CO measurements at $z>1$ are usually performed using various high-J transitions, and excitation corrections are needed to estimate the luminosity of the ground-state transition. The CO spectral line energy distribution (SLED) of a given target is rarely known \citep[e.g.,][]{weiss07,carilli13,mashian15} and therefore the excitation correction is typically highly uncertain. 
In addition, calculating gas masses from CO luminosities ($L'_{\textnormal{CO}}$) requires the assumption of a conversion factor $\alpha_{\textnormal{CO}}$, that depends on the conditions of the ISM and typically ranges between 0.8 and 4 M$_{\odot}$/(K km s$^{-1}$ pc$^2$) for solar-metallicity galaxies \citep{carilli13,accurso17}.
When dealing with SFRs of AGN hosts, an additional complication is the difficulty to properly account for the AGN contribution to the far-infrared (FIR) luminosity. As recently shown by \citet{kirkpatrick19}, different methods to estimate the AGN contribution can lead to completely different results and place the AGN population on the same star-formation law of inactive galaxies. 
Finally, AGN samples at high redshift are usually small and likely biased toward brighter objects \citep[e.g.,][]{brusa18} or are heterogeneous (e.g., different selection criteria, CO lines observed) when assembled from literature data \citep[e.g.,][]{fiore17,perna18,kirkpatrick19,bischetti20}. 

In this paper we 
present the first systematic and uniform analysis of the CO(3-2) emission of 27 X-ray selected AGN at $z\approx2$ to infer whether their activity affects the CO properties of the host galaxy. The targeted transition is the lowest-J transition accessible with the Atacama Large Millimeter/submillimeter Array (ALMA) at $z\approx2$. In this work, we controlled for many of the biases mentioned above by performing a uniform selection and multiwavelength characterization of a fairly representative sample of X-ray AGN. The sample includes less extreme sources, which cover a wider range in AGN bolometric luminosity and uniform distribution on the main sequence. We compared the CO emission properties of our AGN with those of inactive galaxies. The comparison sample was constructed as carefully as possible, taking into account stellar masses, SFRs, redshift and observed CO transition in order to minimize systematic differences.
In particular, by considering targets observed in the same CO transition, we avoid assumptions on conversion factors, which could bias our conclusions.

The paper is structured as follows: In Sec.~\ref{s:sample} we describe the sample selection criteria and the multiwavelength properties of our AGN; the sample of inactive galaxies used for comparison is presented in Sec.~\ref{s:lit_sample}; 
in Sec.~\ref{s:alma_data} we outline the ALMA observations and data analysis; we present and discuss our results in Sec.~\ref{s:results} and \ref{s:discussion}, respectively. Our conclusions are given in Sec.~\ref{s:Conclusions}. 
In this work we adopted a \textit{WMAP9} cosmology \citep{hinshaw13}, $H_{0} = 69.3$ km s$^{-1}$ Mpc$^{-1}$, $\Omega_{\textnormal{M}} = 0.287$ and $\Omega_{\Lambda} = 0.713$. 

\section{The AGN sample}\label{s:sample}

\subsection{\label{s:selection}Target selection}

As a tracer of AGN activity we used X-ray emission, which is very efficient thanks to the low contamination from the host galaxy. We combined catalogs from a deep and small-area as well as a shallow and wide-area X-ray survey, in order to include in our sample both high- and low-luminosity AGN and cover a wide range in AGN bolometric luminosity $L_{\textnormal{bol}}$ \citep{brandt15,circosta18}. The targets were drawn from the following surveys, by adopting as a threshold an absorption-corrected 2$-$10 keV X-ray luminosity $L_{\textnormal{X}} \geq 10^{42}$ erg s$^{-1}$. (i) The \textit{COSMOS-Legacy} survey \citep{civano16,marchesi16}, a 4.6 Ms \textit{Chandra} observation of the COSMOS field, with a deep exposure over an area of about 2.2 deg$^{2}$ at a limiting depth of $8.9 \times 10^{-16}$ erg cm$^{-2}$ s$^{-1}$ in the $0.5-10$ keV band. (ii) The wide-area \textit{XMM-Newton} XXL survey North \citep{pierre16}, a $\sim$25 deg$^{2}$ field surveyed for about 3 Ms by \textit{XMM-Newton}, with a sensitivity in the full $0.5-10$ keV band of $2 \times 10^{-15}$ erg cm$^{-2}$ s$^{-1}$. The XMM-XXL subsample is culled from the X-ray source catalog presented by \citet{liu16} using the X-ray reduction pipeline described by \citet{georgakakis11}, and with optical spectroscopy published by \citet{menzel16}. 

These fields are covered by a rich multiwavelength set of ancillary data spanning from the X-rays to the radio regime, which are essential to obtain robust measurements of the target properties. 
Our AGN were then selected to have spectroscopic redshift in the range $z = 2.0-2.5$, whose quality was flagged as ``secure'' in the respective catalogs as well as a coverage of AGN and galaxy properties as wide and uniform as possible. 
Overall, our sample consists of 27 AGN, namely 7 from XMM-XXL and 20 from COSMOS. 
IDs, coordinates and redshift of the sources are reported in Table~\ref{tab:sample}.

The ALMA program presented in this paper is part of SUPER \citep[SINFONI Survey for Unveiling the Physics and Effect of Radiative feedback;][PI: Mainieri]{circosta18}, which aims at studying AGN feedback at cosmic noon. SUPER started as an ESO's VLT/SINFONI Large Programme to perform the first systematic investigation of ionized outflows in a sizeable and blindly-selected sample of 39 X-ray AGN at $z\approx2$, which reaches high spatial resolutions ($\sim$2 kpc) thanks to the adaptive optics-assisted integral field spectroscopy observations. The goal is to have a reliable overview of the ionized gas component as well as AGN-driven outflows traced by [O\textsc{iii}]$\lambda$5007\AA.\, In parallel, we ran an ALMA campaign to trace the molecular gas properties of the host galaxies. The target sample was initially similar for the SINFONI and ALMA programs. However, due to scheduling constraints on the SINFONI observations the two programs evolved differently, but 43\% of the ALMA targets feature good-quality SINFONI observations. Moreover, SINFONI and ALMA targets share the same properties and selection criteria \citep[see][]{circosta18}. 
In Table~\ref{tab:sample}, the targets with good-quality SINFONI data from our Large Program are flagged (\citealt{kakkad20}; \citealt{vietri20}; Perna et al. in prep.). This distinction is made for future reference since we defer the comparison of outflow and molecular gas properties obtained from the SINFONI and ALMA datasets, respectively, to a future paper. In this work, we focus on the analysis of the CO(3-2) properties of our AGN.

\begin{table}[h!]
\footnotesize
\caption{\label{tab:sample} Summary of the target AGN sample.}
\centering
\begin{tabular}{ccccc}
\hline\hline
ID & RA[J2000] & DEC[J2000] & $z_{\textnormal{spec}}$ & SINFONI \\
$(1)$ & $(2)$ & $(3)$ & $(4)$ & $(5)$ \\
\hline
X\_N\_128\_48\tablefootmark{a} & 02:06:13.54 & $-$04:05:43.20 & 2.323 & n \\
X\_N\_81\_44 & 02:17:30.95 & $-$04:18:23.66 & 2.311 & y  \\
X\_N\_53\_3 & 02:20:29.84 & $-$02:56:23.41 & 2.434 & n  \\
X\_N\_6\_27\tablefootmark{a} & 02:23:06.32 & $-$03:39:11.07 & 2.263 & n \\
X\_N\_44\_64 & 02:27:01.46 & $-$04:05:06.73 & 2.252 & n \\
X\_N\_102\_35 & 02:29:05.94 & $-$04:02:42.99 & 2.190 & y \\
X\_N\_104\_25\tablefootmark{a} & 02:30:24.46 & $-$04:09:13.39 & 2.241 & n \\
lid\_1852\tablefootmark{a} & 09:58:26.57 & +02:42:30.22 & 2.444 & n \\
lid\_3456\tablefootmark{a} & 09:58:38.40 & +01:58:26.83 & 2.146 & n \\
cid\_166 & 09:58:58.68 & +02:01:39.22 & 2.448 & y  \\
lid\_1289 & 09:59:14.65 & +01:36:34.99 & 2.408 & n \\
cid\_1605 & 09:59:19.82 & +02:42:38.73 & 2.121 & y \\
cid\_337 & 09:59:30.39 & +02:06:56.08 & 2.226 & n \\
cid\_346 &  09:59:43.41 & +02:07:07.44 & 2.219 & y  \\
cid\_357\tablefootmark{a} & 09:59:58.02 & +02:07:55.10 & 2.136 & n \\
cid\_451 & 10:00:00.61 & +02:15:31.06 & 2.450 & y \\
cid\_1205 & 10:00:02.57 & +02:19:58.68 & 2.255 & y \\
cid\_2682 & 10:00:08.81 & +02:06:37.66 & 2.435 & y  \\
cid\_247\tablefootmark{a} & 10:00:11.23 & +01:52:00.27 & 2.412 & n \\
cid\_1215\tablefootmark{a} & 10:00:15.49 & +02:19:44.58 & 2.450 & n \\
cid\_467 & 10:00:24.48 & +02:06:19.76 & 2.288 & y  \\
cid\_852 & 10:00:44.21 & +02:02:06.76 & 2.232 & n  \\
cid\_970\tablefootmark{a} & 10:00:56.52 & +02:21:42.35 & 2.501 & n \\
cid\_971 & 10:00:59.45 & +02:19:57.44 & 2.473 & y  \\
cid\_38 & 10:01:02.83 & +02:03:16.63 & 2.192 & n \\
lid\_206 & 10:01:15.56 & +02:37:43.44 & 2.330 & n  \\
cid\_1253 & 10:01:30.57 & +02:18:42.57 & 2.147 & y \\
\hline
\end{tabular}
\tablefoot{
(1) Source identification number from the original source catalogs of \citet{menzel16} for the XMM-XXL field,  and \citet{civano16} for the COSMOS field (see also Sec. \ref{s:selection}). (2) RA and (3) DEC of optical counterparts: The XMM-XXL targets have an SDSS counterpart whose coordinates are given in \citet{menzel16}; for the COSMOS field we list the \textit{i}-band coordinates from \citet{marchesi16}. (4) Spectroscopic redshift from \citet{menzel16} and \citet{marchesi16}. (5) Good-quality SINFONI data available from SUPER. \\
\tablefootmark{a}{These targets were not presented in \citet{circosta18}.}\\

}
\end{table}

\subsection{\label{s:properties}Multiwavelength properties of the sample}

We characterized the physical properties of our sources by exploiting the multiwavelength coverage available from the X-ray to the radio regime. In particular, we followed the procedure described in \citet{circosta18} to collect the multiwavelength data and derive the properties of the targets that were not analyzed in our previous work, through X-ray spectral analysis and broad-band spectral energy distribution (SED)-fitting. We refer the reader to \citet{circosta18} for an extensive description, but we summarize in the following some key pieces of information. 

The counterparts to the X-ray sources in COSMOS are provided along with the optical-to-MIR multiwavelength photometry in the COSMOS2015 catalog \citep{laigle16}. 
We complemented this dataset with FIR data from \textit{Herschel}/PACS and SPIRE, when available, using a positional matching radius of 2$''$ (we note that we used 24 $\mu$m-priored catalogs, which in turn are IRAC-3.6 $\mu$m priored). \textit{Spitzer}/MIPS photometry at 24 $\mu$m and \textit{Herschel}/PACS at 100 and 160 $\mu$m was taken from the PEP DR1\citep{lutz11}. \textit{Herschel}/SPIRE photometry at 250, 350 and 500 $\mu$m was retrieved from the data products presented in \citet{hurley17}\footnote{\citet{hurley17} present a 24 $\mu$m prior-based catalog, obtained by using a probabilistic Bayesian method. The resulting flux probability distributions for each source in the catalog are described by the 50th, 84th and 16th percentiles. We assumed Gaussian uncertainties by taking the maximum between the 84th$-$50th percentile and the 50th$-$16th percentile.}. We also added ALMA continuum data in Band 7 and Band 3 available from the COSMOS A3 catalog \citep{liu19}, 
\citet{scholtz18} and our ALMA dataset (this work and Lamperti et al., in prep.).

As for XMM-XXL, the counterparts to the targets are known in the SDSS optical images and the corresponding associations to the X-ray sources, as well as the photometry from UV-to-mid-IR, are given in \citet{fotopoulou16}, \citet{menzel16} and \citet{georgakakis17}. \textit{Herschel}/PACS and SPIRE data are those released by the HerMES collaboration in the Data Release 4 and 3 respectively \citep{oliver12}\footnote{Both sets of data were extracted using the same \textit{Spitzer}/MIPS 24 $\mu$m prior catalog, whose fluxes are available along with the SPIRE data. We used aperture fluxes in smaller apertures, that is 4$''$ diameter. 
}. 

Mid- and far-IR photometric datapoints from 24 to 500 $\mu$m were considered as detections when the signal-to-noise ratio S/N$>$3, where the total noise was given by the sum in quadrature of both the instrumental and the confusion noise \citep{lutz11,oliver12,magnelli13}. The detections below this threshold were converted to 3$\sigma$ upper limits. All the data used in this work were corrected for Galactic extinction \citep{schlegel98}. 

The datasets described above were modeled by using the Code Investigating GALaxy Emission \citep[CIGALE\footnote{\url{https://cigale.lam.fr}};][]{boquien19,noll09}, a publicly available state-of-the-art galaxy SED-fitting technique (we used version 0.11.0). This code disentangles the AGN contribution from the emission of the host by adopting a multicomponent fitting approach and includes attenuated stellar emission, dust emission heated by star formation, AGN emission (both primary accretion disk emission and dust-heated emission), and nebular emission. Moreover, it takes into account the energy balance between the UV-optical absorption by dust and the corresponding reemission in the FIR. The parameters of interest (namely stellar masses, FIR luminosities and AGN bolometric luminosities) and their uncertainties are determined by the code through a Bayesian statistical analysis, by building up a probability distribution function (PDF) that takes into account the different models used during the fit. The fiducial results correspond to the mean value of the PDF and the associated error is the standard deviation \citep{noll09}. To reproduce the stellar component we used: \citet{bc03} stellar population models with solar metallicity, a delayed star-formation history, the modified \citet{calzetti00} attenuation law, 
and a \citet{chabrier03} initial mass function (IMF). The contributions from star-forming dust and AGN were modeled with the libraries presented by \citet{dale14} and \citet{fritz06}, respectively. Finally, the nebular emission was reproduced with the models of \citet{inoue11}. Tables 1 and 2 in \citet{circosta18} provide a list of the photometric filters and input parameters of the models used for the SED-fitting procedure.

The results of the SED-fitting analysis are reported in Table~\ref{tab:summary_results}, together with the optical spectroscopic classification in broad-line (BL) and narrow-line (NL) AGN, depending on the presence of broad ($\textnormal{FWHM} > 1000$ km s$^{-1}$ ) or narrow ($\textnormal{FWHM} < 1000$ km s$^{-1}$ ) permitted emission lines in their optical spectra, respectively. From now on, we will refer to type 1 and type 2 AGN according to the optical spectroscopic classification. We refer the reader to \citet{circosta18} for the SEDs of most of our sample. The SEDs of the targets that were not previously presented (flagged in Table~\ref{tab:sample}), are shown in Appendix~\ref{sec:SEDs}. 
When possible, SFRs were derived by assuming the \citet{kennicutt98} calibration, converted to a \citet{chabrier03} IMF by subtracting 0.23 dex, from the IR luminosity integrated in the rest-frame wavelength range 8$-$1000 $\mu$m, after removing the AGN contribution (11 targets). For non-detections at observed $\lambda>24$ $\mu$m, we provide a 3$\sigma$ upper limit on the SFR, derived as the 99.7th percentile from the PDF of the FIR luminosity (11 targets). When data were not available at observed $\lambda>24$ $\mu$m, the dust templates were not included in the fitting routine. We therefore provide the SFR as derived from the modeling of the stellar component in the UV-to-near-IR(NIR) regime with SED fitting, averaged over the last 100 Myr of the galaxy star-formation history (two targets, flagged in Table~\ref{tab:summary_results}). Three targets without data coverage at $\lambda>24$ $\mu$m are also dominated by the AGN emission at UV-to-NIR wavelengths (namely X\_N\_128\_48, X\_N\_102\_35, and X\_N\_104\_25). For these targets we could not derive an estimate of SFR and stellar mass and therefore only the AGN bolometric luminosity is reported in Table~\ref{tab:summary_results}\footnote{The limited wavelength coverage for two of these targets (X\_N\_128\_48 and X\_N\_104\_25), featuring only SDSS data, hampered an accurate estimate of the bolometric luminosity through SED fitting. We converted their X-ray luminosities (discussed in the main text) into bolometric ones by using the relation derived by \citet{duras20}. We compared these values to the estimates of bolometric luminosities derived from the continuum at 1450 \AA\, \citep{runnoe12,vietri20} and they are consistent within the scatter of the Duras et al. relation ($\sim$0.4 dex).}. These three targets are excluded from the analysis where estimates of stellar masses and SFRs are required (see Sec.~\ref{s:results}). Moreover, measurements of
stellar mass for X\_N\_53\_3, X\_N\_6\_27 and cid\_1605 could not be recovered because of the strong AGN contamination in the optical/NIR regime, but we report their SFRs. We measured stellar masses in the range $\log(M_{\ast}/M_{\odot})=9.6-11.2$, FIR luminosities $\log(L_{\textnormal{FIR}}/\textnormal{erg s}^{-1}) < 45.0-46.4$, $\textnormal{SFRs} <25-686$ $M_{\odot}$/yr, and AGN bolometric luminosities mainly in the range $\log(L_{\textnormal{bol}}/\textnormal{erg s}^{-1})=44.7-46.9$ covering two orders of magnitude. 
In Table~\ref{tab:summary_results} we also report black hole masses, collected from the literature \citep{menzel16} and measured from our SINFONI dataset \citep{vietri20} as well as spectra available from SDSS and the COSMOS survey. The results range between $10^{7.64}$ and $10^{10.04}$ $M_{\odot}$. Overall, our sample is composed of 17 type 1 and 10 type 2 AGN. 

\begin{figure}
\centering
  \includegraphics[width=9.3cm]{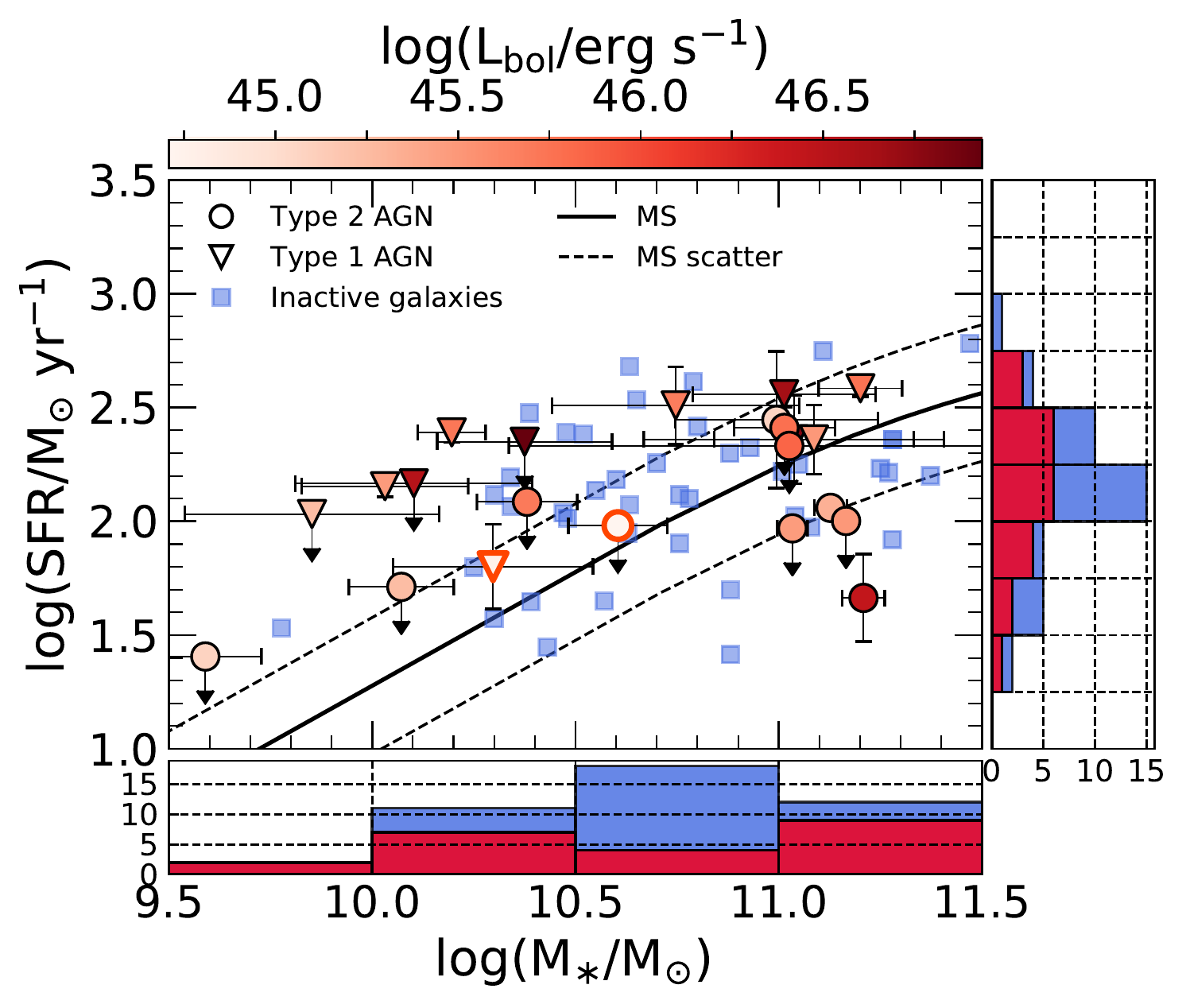}
  \caption{Distribution of host galaxy properties in the SFR-$M_{\ast}$ plane for the 21 AGN (type 1s marked by triangles and type 2s marked by circles) in our sample with an estimate of both stellar mass and SFR. The comparison sample of inactive galaxies is depicted by blue squares. The two data points with orange edges represent the targets with SFR derived through modeling of the stellar emission using SED fitting. The color coding indicates the AGN bolometric luminosity for our sample, which covers two orders of magnitude in $L_{\textnormal{bol}}$. The black solid line reproduces the main sequence of star-forming galaxies from \citet{schreiber15} at the average redshift of our target sample (i.e., $\sim$2.3). The dashed lines mark the scatter of the main sequence (equal to 0.3 dex). 
  The histograms show the projected distribution of the two quantities along each axis, in red for our sample and blue for the comparison one.}
  \label{fig:mainSeq}
\end{figure}

From the analysis of the X-ray spectra we derived obscuring column densities $N_{\textnormal{H}}$ and X-ray luminosities $L_{\textnormal{X}}$. Again, we followed the same procedure described in \citet{circosta18} to extract and fit the spectra. 
The energy ranges considered are $0.5-7.0$ keV band for \textit{Chandra} and $0.5-10$ keV band for \textit{XMM-Newton}. All the fits were performed by using XSPEC v. 12.9.1\footnote{\url{https://heasarc.gsfc.nasa.gov/xanadu/xspec/}} \citep{arnaud96} and adopting the Cash statistic \citep{cash79} and the direct background option \citep{wachter79}. The spectra were binned to 1 count per bin to exclude empty channels. 
For sources with more than 30 (50) net counts (reported in Table \ref{tab:summary_results}) for \textit{Chandra} (\textit{XMM-Newton}), we used a simple spectral model, consisting of a power law modified by both intrinsic and Galactic absorption, as well as a secondary power law to reproduce any excess in the soft band (0.5$-$2 keV), due to scattering or partial covering in obscured sources. The photon index was left free to vary (typical values within $\Gamma = 1.5-2.5$) for spectra with more than $\sim$100 net counts, otherwise we fixed it to the canonical value of 1.8 \citep[e.g.,][]{piconcelli05}. For targets with less than 30 (50) counts $N_{\textnormal{H}}$ values were derived at the source redshift from hardness ratios ($HR = \frac{H-S}{H+S}$, where $H$ and $S$ are the number of counts in the hard $2-7$ keV and soft $0.5-2$ keV bands, respectively; \citealt{lanzuisi09}). 
In both cases, we propagated the uncertainty on $N_{\textnormal{H}}$ when deriving the errors on the intrinsic luminosity. The results derived for column densities and $2-10$ keV absorption-corrected luminosities are listed in Table~\ref{tab:summary_results}. Our sample spans a range of obscuration properties including unobscured targets ($N_{\textnormal{H}}<10^{22}$ cm$^{-2}$) up to Compton-thick ones ($N_{\textnormal{H}}>10^{24}$ cm$^{-2}$) and X-ray luminosities in the range $\log(L_{\textnormal{[2-10 keV]}}/\textnormal{erg s}^{-1})=43.0-45.4$.

\section{\label{s:lit_sample}Comparison sample of inactive galaxies}

We built a comparison sample of inactive galaxies 
from the Plateau de Bure high-z Blue Sequence Survey \citep[PHIBSS;][]{tacconi18}. PHIBSS aims at investigating the gas properties of galaxies across cosmic time by using a sample of 1444 targets spanning a redshift range $z=0-4.4$ with estimates of molecular gas mass given by CO measurements, FIR SED dust measurements or 1mm dust measurements. In particular, in the redshift range of our interest ($z=2-2.5$) PHIBSS galaxies were selected to have stellar mass $\geq 2.5 \times 10^{10}$ $M_{\odot}$ and SFR$\geq 30$ $M_{\odot}$ yr$^{-1}$, in order to ensure a relatively uniform coverage of the SFR-stellar mass relation. 
At $z\approx2$, the PHIBSS sample contains CO(3-2) measurements obtained with IRAM PbBI and the updated IRAM NOEMA \citep{tacconi13, freundlich19}, as well as CO measurements of main sequence star-forming galaxies and submm galaxies (SMGs) collected from the literature \citep{tacconi06,bothwell13,saintonge13,silverman15,decarli16}, performed with both ALMA and NOEMA.
From the PHIBSS parent sample, we selected for our purposes galaxies with $z=2-2.5$ (the same redshift range of our targets), whose molecular gas mass was derived through CO(3-2) observations. 
By requiring that also the control sample is observed in CO(3-2), we can directly compare the observed CO luminosity and 
we are free from assumptions on conversion factors (e.g., the excitation correction and $\alpha_{\textnormal{CO}}$), which could bias the results. 
All stellar masses and SFRs assume a Chabrier IMF. The typical systematic uncertainties on both the stellar mass and the SFR were assumed to be 0.2 dex \citep{freundlich19}. As for the SFRs, it is worth noting that they were mainly derived from IR measurements, or a combination of IR and UV. However, a small subsample of galaxies feature SFR estimates from H$\alpha$ fluxes \citep{tacconi13}. Since we want to compare CO and FIR luminosities of these galaxies, we relied on the agreement between SFRs derived from H$\alpha$ and IR in main-sequence galaxies at cosmic noon \citep{rodighiero14,puglisi16,shivaei16} and converted all SFRs into FIR luminosities, using the \citet{kennicutt98} relation corrected for a \citet{chabrier03} IMF (i.e., by subtracting 0.23 dex). We used CO fluxes retrieved from the corresponding papers, when data were taken from the literature, or provided by the PHIBSS collaboration (L. Tacconi, private communication), since the survey paper by \citet{tacconi18} provides only the final estimates of gas masses.
We looked for the presence of AGN in the sample, by checking the corresponding papers the targets are from (see \citealt{tacconi18}). They were identified by using power-law NIR component fraction (i.e., rest-frame NIR excess in their SED), color selection, X-ray emission  and line ratios. Six targets were therefore excluded. We did not include these targets in our AGN sample because (i) the methods used to identify AGN were not the same as ours and (ii) stellar masses and SFRs were not derived by accounting for the AGN contribution hence these parameters may be overestimated. 
The inactive galaxies satisfying the selection criteria mentioned above cover the same range of stellar mass of our AGN sample within the uncertainties ($\log M_{\ast}\approx 9.5-11.5$). From this set of objects we further excluded those with extreme SFRs, that is the four targets with SFRs larger than the highest value in our sample per bin of stellar mass (width of 0.5 dex; Fig.~\ref{fig:mainSeq}), in order to have a similar distribution on the main sequence. We note that even without excluding such targets from the comparison sample, the results presented in Sec.~\ref{s:results} would not change significantly. The final sample selected from PHIBSS is made of 42 objects. The distribution on the main sequence of our AGN and the comparison sample is shown in Fig.~\ref{fig:mainSeq}. 
Although we are unavoidable limited by 50\% SFR upper limits in our AGN sample considered in Fig.~\ref{fig:mainSeq}, the comparison sample has a broadly similar distribution on the main sequence. 

\section{\label{s:alma_data}ALMA data: Observations and analysis}

The ALMA program was carried out in Cycle 4 and 5 (Project codes: 2016.1.00798.S and  2017.1.00893.S; PI: V. Mainieri) using 42$-$47 antennas and maximum baselines between 704 m and 1.1 km. Observations were taken in Band 3 between November 2016 and May 2017 (Cycle 4) and in March 2018 (Cycle 5). In each observation, one spectral window, with a bandwidth of 1.875 GHz, was centered at the expected redshifted frequency of the CO(3-2) emission line (the rest-frame frequency is 345.8 GHz) based on the spectroscopic redshift of the target (Table~\ref{tab:sample}), while the other three spectral windows were used to sample the continuum emission. Each spectral window was divided into 240 channels and had a native spectral resolution of 7.8 MHz, corresponding to 21$-$23 km/s at the line frequency. The on-source exposure time varied between $\sim$9 and $\sim$80 minutes.

ALMA visibilities were calibrated using the CASA software\footnote{\url{https://casa.nrao.edu}} version 4.7.0 for Cycle 4 data and 5.1.1 for Cycle 5, as originally used for the reduction with the pipeline. The presence of continuum emission was estimated from the continuum maps generated with \textsc{tclean} by averaging all spectral windows, and subtracted in the uv plane by using the \textsc{uvcontsub} task within CASA, by excluding the spectral range where we expected the line. Continuum emission at $\sim$100 GHz was detected for five targets, as shown in Appendix~\ref{sec:cont_maps}. For these targets, a fitting procedure using a two-dimensional Gaussian was performed on the continuum maps in the image plane to estimate the flux densities, reported in Table~\ref{tab:ALMA_results}. The final continuum-subtracted datacubes were generated with the CASA task \textsc{tclean} in velocity mode, using ``natural'' weighting, which optimizes the point-source sensitivity in the image plane, the ``Hogbom'' cleaning algorithm, a cellsize of 0.2$''$, and a velocity bin width of 24 km/s. As a threshold, we used 3 times the rms derived from the dirty cubes (reported in Table~\ref{tab:ALMA_results}). 
The emission-line cubes have angular resolutions in the range 0.7$-$1.7$''$, and 1 $\sigma$ rms in the range 0.18$-$0.72 mJy/beam per 24 km/s velocity bin, as listed in Table~\ref{tab:ALMA_results}. 

To determine the flux and size of the CO(3-2) line we proceeded as follows. For each target, we considered the continuum-subtracted cube and created moment 0 maps (i.e., velocity-integrated flux maps) by averaging, with the task \textsc{immoments}, spectral channels around the expected CO(3-2) frequency over 
increasingly larger velocity widths, from 100 km/s to 1000 km/s in steps of 100 km/s (in case of successful line detection, the procedure was repeated by averaging spectral channels around the peak of the emission line). 
The rms of the collapsed CO maps was estimated over an area approximately half that of the primary beam. We then 
assessed the significance of the emission line in the moment 0 maps. If the line was detected (i.e., S/N$\gtrsim$3), we extracted the one-dimensional spectrum from the emission-line cube using as extraction area the region above 2$\sigma$ significance in the moment 0 map ($\sim$1$-$2 arcsec, centered on the target; see contours in Appendix~\ref{sec:line_maps}). Among the different integrated maps produced, we considered the one where the line had the highest significance. 
The one-dimensional line spectrum was fitted using a single Gaussian profile (python function \textsc{curve\_fit} in scipy.optimize) to estimate parameters such as line centroid (from which $z_{\textnormal{CO}}$, Table~\ref{tab:ALMA_results}), FWHM and integrated flux density. 
The uncertainties on fluxes, FWHM, $z_{\textnormal{CO}}$, and peak fluxes were estimated by adopting a Monte Carlo approach. We created 100 mock spectra by adding to the model spectrum random noise proportional to the noise measured per channel. To estimate the noise in each channel of the spectrum we extracted 100 spectra from random regions within the cube. Such regions had an area similar to the source extraction region and were not located too close to the source. We considered the standard deviation of the flux densities, in each spectral channel, of the 100 random spectra as an indication of the rms per channel, and used these values in the Monte Carlo routine. We note that the rms resulted to be quite uniform throughout the spectrum. 
We performed the fit for each of our mock spectra and considered the standard deviation of the distribution of the resulting parameters as uncertainties. As for the integrated fluxes, the final uncertainties are given by adding in quadrature a typical ALMA flux calibration error equal to 5 per cent of the flux, as presented by \citet{bonato18}, to the uncertainties obtained from the Monte Carlo routine. We detect the CO(3-2) line in 11 targets out of 27 observed. The line is considered detected if the emission in the line-integrated map is significant at a level $\gtrsim$3$\sigma$ (the integrated rms is given in Table~\ref{tab:ALMA_results}). For the sources without CO detection, we provide a 3$\sigma$ upper limit, calculated from the rms of the velocity-integrated maps assuming a line width of 348 km/s (i.e., the mean of the FWHM values measured for our detections). The results of our analysis, namely the line flux, FWHM and redshift of the line, are reported in Table~\ref{tab:ALMA_results}. We measured FWHM of the CO line in the range 97$-$810 km/s and flux densities in the range 76$-$2427 mJy km/s. The redshift of the CO line is in good agreement with that reported in Table~\ref{tab:sample}. 
Moment 0 maps and spectra of the targets with a detected line are reported in Appendix~\ref{sec:line_maps}. 

Finally, we find that all but one target are spatially unresolved and/or the S/N is not enough to firmly constrain their CO(3-2) sizes in the current observations. cid\_346 
is spatially resolved with deconvolved CO(3-2) FWHM size $(0.61\pm0.15)''\times(0.31\pm0.28)''$ 
as derived from a two-dimensional Gaussian fit in the image plane. As a consistency check, we also analyzed the CO maps in the uv plane, and compared the results with those obtained from the analysis of the CO maps in the image plane. The analysis in the uv plane was performed with GILDAS (UVFIT). We fitted the velocity-integrated visibilities of the CO lines with either a point-source (for unresolved sources) or Gaussian functions (for resolved sources), and found that RA, DEC positions, fluxes and sizes are consistent with those obtained from the analysis in the image plane. In the maps of cid\_1253 and cid\_971 we found a companion at 2.19 and 5.61 arcsec, respectively. Instead, a tentative emission-line detection (S/N$\approx$3) was found in the field of cid\_1215 at 3.6 arcsec. As for cid\_1253 and cid\_1215, we note that the 2$\sigma$ contours of the targets include their companion and therefore the whole system was considered when extracting the CO(3-2) spectra. Presumably the large beam characterizing \textit{Herschel} observations, used to derive the FIR luminosities, cannot separate the central target from its companion. Since cid\_971 lacks FIR data and the SFR was estimated from the UV-to-NIR SED (see Sec.~\ref{s:properties}), it was sensible to exclude the companion from the extraction region.

The number of detections found (11 out of 27 targets) corresponds to a detection rate of $\sim$40\%. This means that CO luminosities for non detections are lower than what we were expecting from estimates based on samples of star-forming galaxies \citep{genzel10}. We note that most (70\%) of the non-detections are also characterized by upper limits on the FIR luminosity, which may have contributed with additional uncertainty to the predicted CO luminosity.

We finally derived CO(3-2) luminosities (in Table~\ref{tab:ALMA_results}) as given by \citet{solomon05}:

\begin{equation}
L'_{\textnormal{CO}} = 3.25 \times 10^7 I_{\textnormal{CO}} \nu_{\textnormal{obs}}^{-2} D_{\textnormal{L}}^2 (1+z)^{-3},
\end{equation}
where $I_{\textnormal{CO}}$ is the velocity-integrated flux, $D_{\textnormal{L}}$ is the luminosity distance, $\nu_{\textnormal{obs}}$ is the observed frequency of the line and $z$ is the redshift. We measured CO(3-2) luminosities in the range $\log(L'_{\textnormal{CO}}/\textnormal{K km s}^{-1} \textnormal{pc}^{2})=9.33-10.80$. For completeness, we also derived gas masses for both our AGN sample and the comparison one (in the range $\log M_{\textnormal{mol}}/M_{\odot}=10.19-11.66$) by adopting uniform assumptions, that is the excitation factor $r_{31}=L'_{\textnormal{CO(3-2)}}/L'_{\textnormal{CO(1-0)}}=0.5$ and $\alpha_{\textnormal{CO}}=3.6$ M$_{\odot}$/(K km s$^{-1}$ pc$^2$) commonly used in the literature for star-forming galaxies at similar redshifts as our AGN sample \citep{daddi10,tacconi13,kakkad17}. However, literature results show that AGN and inactive galaxies may feature different excitation factors \citep[e.g.,][but see also \citealt{riechers20}]{kirkpatrick19}. Moreover, $\alpha_{\textnormal{CO}}$ prescriptions dependent on metallicity and star-formation conditions can be adopted \citep{bolatto13}. The decision to use the same assumptions for both samples follows our goal of minimizing systematic uncertainties, which could produce artificial offsets between our samples. Molecular gas masses were derived to provide a reference to a physically motivated parameter in the plots presented in Sec.~\ref{s:results}, besides the CO(3-2) luminosity, which is the main focus of our study and is adopted as a consistent proxy for the molecular gas mass that allows us to avoid systematics involved when assuming conversion factors.

\section{\label{s:results}Results}

In this Section we perform comparisons between CO and FIR luminosities as well as stellar masses of AGN and inactive galaxies. In our analysis we focused on the CO(3-2) luminosity for both AGN and inactive galaxies to limit systematic uncertainties and avoid assumptions on conversion factors (e.g., excitation correction and $\alpha_{\textnormal{CO}}$) needed to estimate the CO(1-0) luminosity and the gas mass. Therefore, the CO(3-2) luminosity can be considered as a proxy for the molecular gas mass. Similarly, we used the FIR luminosity, from which we subtracted the AGN contribution (Sec.~\ref{s:properties}), as a proxy for the SFR. Stellar masses were derived from broad-band SED fitting for both samples. 
Bayesian methods were used in order to properly take into account the upper limits on both CO and FIR luminosities, which especially concern the AGN sample. Throughout our analysis, we relied on the lack of assumptions on conversion factors, the uniform characterization of the multiwavelength properties of our sample as well as the Bayesian methods including upper limits adopted.

\subsection{\label{s:co_fir}CO versus FIR luminosities}

\begin{figure}[h!]
\centering
  \includegraphics[width=9.3cm]{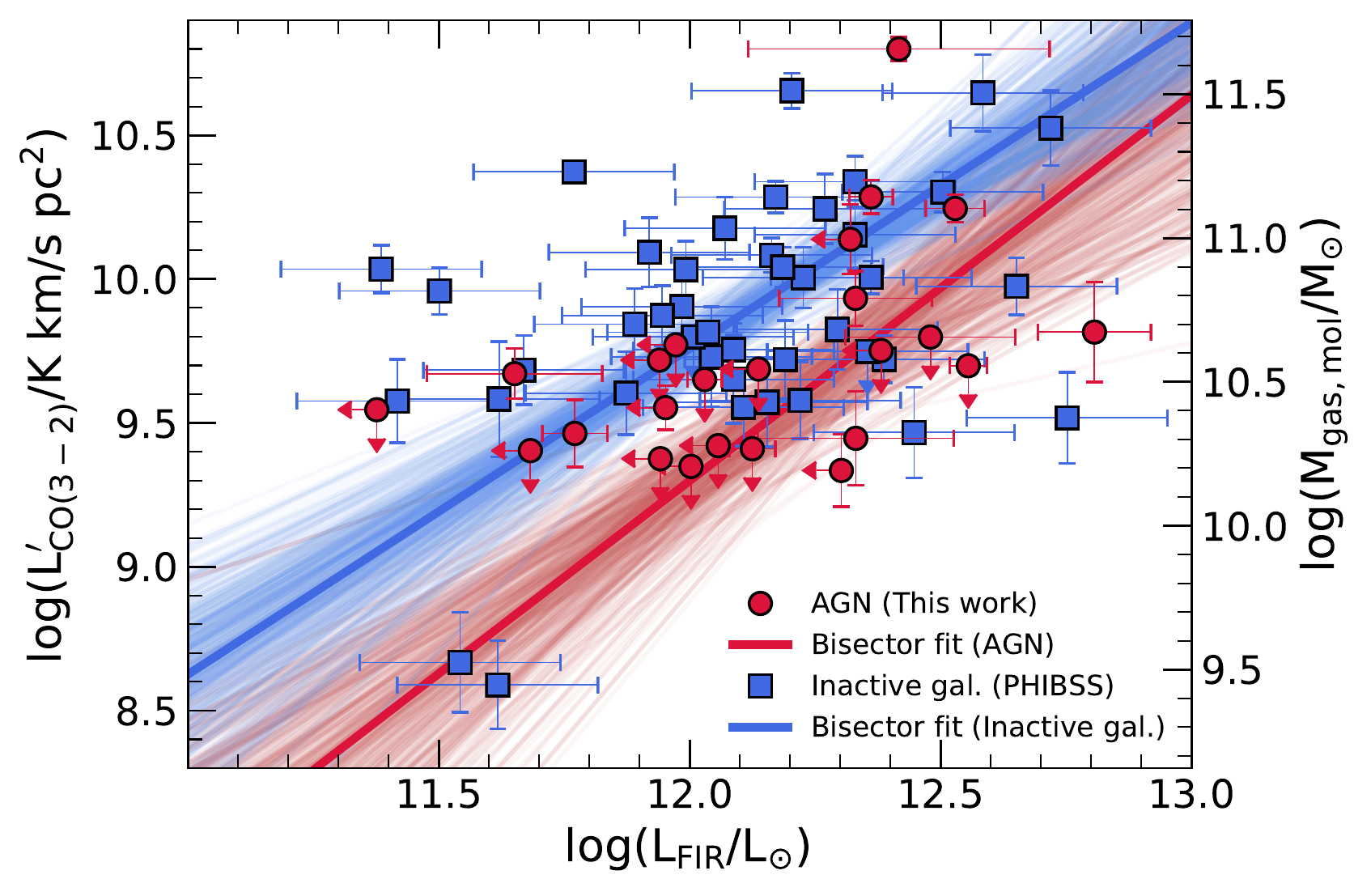}
  \caption{$L'_{\textnormal{CO}}$ and molecular gas masses as a function of $L_{\textnormal{FIR}}$ for our AGN sample (red circles) and inactive galaxies (blue squares; \citealt{tacconi18}). 
  To display the distribution of the two samples in the plot, we performed a bisector fit by adopting a Bayesian framework (see text). The dispersion of the fits is given by plotting 500 realizations of the bisector fit. Where the fits are better constrained ($\log L_{\textnormal{FIR}}/L_{\odot} = 12.2$) AGN show CO luminosities 0.43 dex lower than inactive galaxies at the $\sim$2$\sigma$ level.
  }
  \label{fig:CO_FIR}
\end{figure}

In Fig.~\ref{fig:CO_FIR} we study the correlation between CO and FIR luminosities for our sample and the comparison one. This is the integrated form of the Schmidt-Kennicutt relation, although we use the CO(3-2) transition instead of CO(1-0). In order to quantify the distribution of the two samples in this plane we fitted a linear model to the data by using the ordinary least-square (OLS) bisector fit method \citep{isobe90}, that is, by taking into account uncertainties on $L_{\textnormal{FIR}}$ and $L'_{\textnormal{CO}}$ separately and then considering the bisector of the two lines. To derive the best-fit parameters of the individual fits we adopted a Bayesian framework. When constructing the likelihood function, we assumed that uncertainties are Gaussian-distributed for detections, while upper limits were taken into account by integrating the Gaussian likelihood from minus infinity to the value of the upper limit (i.e., the 3$\sigma$ flux upper limits derived in Sec. \ref{s:alma_data}), which resulted in the error function (e.g., see \citealt{sawicki12}). 
The likelihood function was sampled using \textsc{emcee} \citep{foreman13}, a python implementation of the affine invariant MCMC (Monte Carlo Markov Chain) ensemble sampler of \citet{goodman10}. We sampled the posterior distribution in the parameter space to derive the marginalized posterior probability distribution. The initial guesses were given by the maximum likelihood estimates obtained with the python module scipy.optimize. 
We also included an intrinsic scatter to the relation as third free parameter in the fit that accounts for underestimated uncertainties, given the presence of luminosities with uncertainties spanning a large range of values, which could bias the fitting results. 
Best-fit parameters were estimated by taking the median of the sampled marginalized posterior distribution of the OLS bisector fit parameters and the 16th and 84th percentiles as uncertainties. We then used these best-fit parameters to derive slope and intercept of the bisector as given in \citet{isobe90}. 
The results of the analysis are shown in Fig.~\ref{fig:CO_FIR}, where we plot the AGN sample in red and inactive galaxies in blue, with the corresponding fits and their dispersions. The dispersion shown is obtained by taking 500 realizations of the bisector fit by considering the values of slope and intercept within one sigma of the sampled marginalized posterior distributions of the two fits (along the $x$ and $y$ axes). 
The fit ($\log L'_{\textnormal{CO}}/\textnormal{K km s}^{-1} \textnormal{pc}^{2} = m \log L_{\textnormal{FIR}}/L_{\odot} + b$) to the AGN sample has slope $m=1.35\pm0.32$ and intercept $b=-6.90^{+3.90}_{-3.96}$, while for inactive galaxies $m=1.14\pm0.19$ and $b=-3.91^{+2.16}_{-2.28}$. Because of the scatter characterizing the two samples and the small luminosity range probed, the uncertainties are quite large. Within the uncertainties, our fits indicate an almost linear logarithmic slope for both samples. 
Previous analyses in the literature show different slopes; for example, \citet{sargent14} find a value of $0.81 \pm 0.03$ over a large range of redshifts and galaxy types, while \citet{sharon16} report a linear relation for $z\approx2$ AGN and SMGs that becomes superlinear ($m \approx 1.15-1.2$) when low-redshift IR-bright galaxies are included.  
However, a detailed analysis of the slope of the relation 
and/or trends across a wide range of luminosities and redshift is beyond the scope of this paper (e.g., see \citealt{yao03,sargent14,kamenetzky16,sharon16}). Instead, we aim at quantifying any potential shift between the distributions of AGN and inactive galaxies. 
As can be seen in Fig.~\ref{fig:CO_FIR}, the dispersions of the two fits are quite large and the two distributions do not seem to be statistically different within the uncertainties. We provide an estimate of the shift between the two samples at $\log(L_{\textnormal{FIR}}/L_{\odot}) = 12.2$, where the dispersion of the fits in the $y$ direction is better constrained ($\sim$0.15 dex). AGN have CO luminosities a factor of 2.7 lower (0.43 dex), which are different at the $\sim$2$\sigma$ level. To verify how different the two distributions are, we also performed a Kolmogorov-Smirnov (KS) test for two-dimensional datasets\footnote{Throughout the paper, we consider the result of a statistical test significant if the p-value (probability $P$ given by 1$-$p-value) of the null hypothesis that the two samples are similar is $<$0.01, and moderately significant if p$\approx$0.01$-$0.05}. Following \citet{press88} we implemented a Monte Carlo approach and 100 sets of synthetic data were generated from our observed CO and FIR luminosities. As done before, a Gaussian distribution was assumed for detections. As for upper limits, we included them by using the same value (i.e., the value of the upper limit) in each synthetic dataset. We then considered the fraction of synthetic datasets showing a two-dimensional KS statistic (often indicated as $D$) larger than the value measured for the observed data. This fraction represents how significant the difference between the two samples is. 
The test was first performed by excluding upper limits, 
and the two distributions result to be different at the $\sim$28\% level, although the number of suitable AGN is just eight. When including upper limits, the significance increases to 99\%. Although $L'_{\textnormal{CO}}$ upper limits could contribute in making this result a lower limit, $L_{\textnormal{FIR}}$ upper limits may go in the opposite direction. Overall, this is an indication that taking upper limits into account is crucial to understand the difference between the two samples.

We performed the linear fit by removing from the comparison sample the two outliers with low CO luminosities ($\log L'_{\textnormal{CO}}/\textnormal{K km s}^{-1} \textnormal{pc}^{2} < 9$), which are lensed galaxies from \citet{saintonge13} and, given the known uncertainties on the magnification correction \citep[e.g.,][]{sharon16}, could be a source of bias in our analysis. Without these two targets, the slope of the fit is $0.98^{+0.18}_{-0.16}$ and the intercept $-1.86^{+1.96}_{-2.17}$, which is consistent with the previous result within the uncertainties. As for the two-dimensional KS test, we obtained a significance of $\sim$35\%(99\%) without (with) upper limits.

\subsection{\label{s:co_mstar}CO luminosities versus stellar masses}

\begin{figure}[h!]
\centering
  \includegraphics[width=9.0cm]{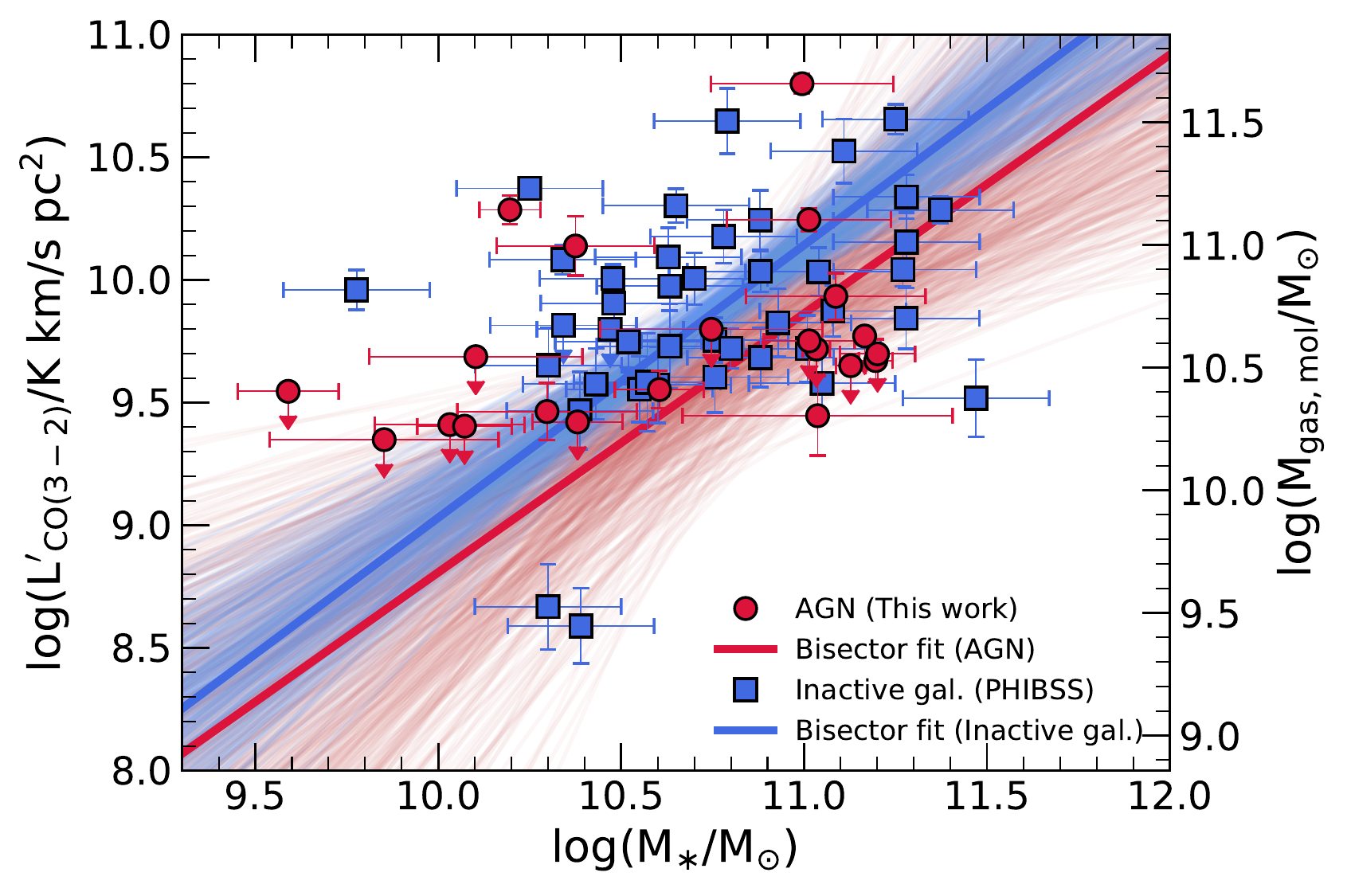}
  \vspace{2mm}
  \includegraphics[width=9.0cm]{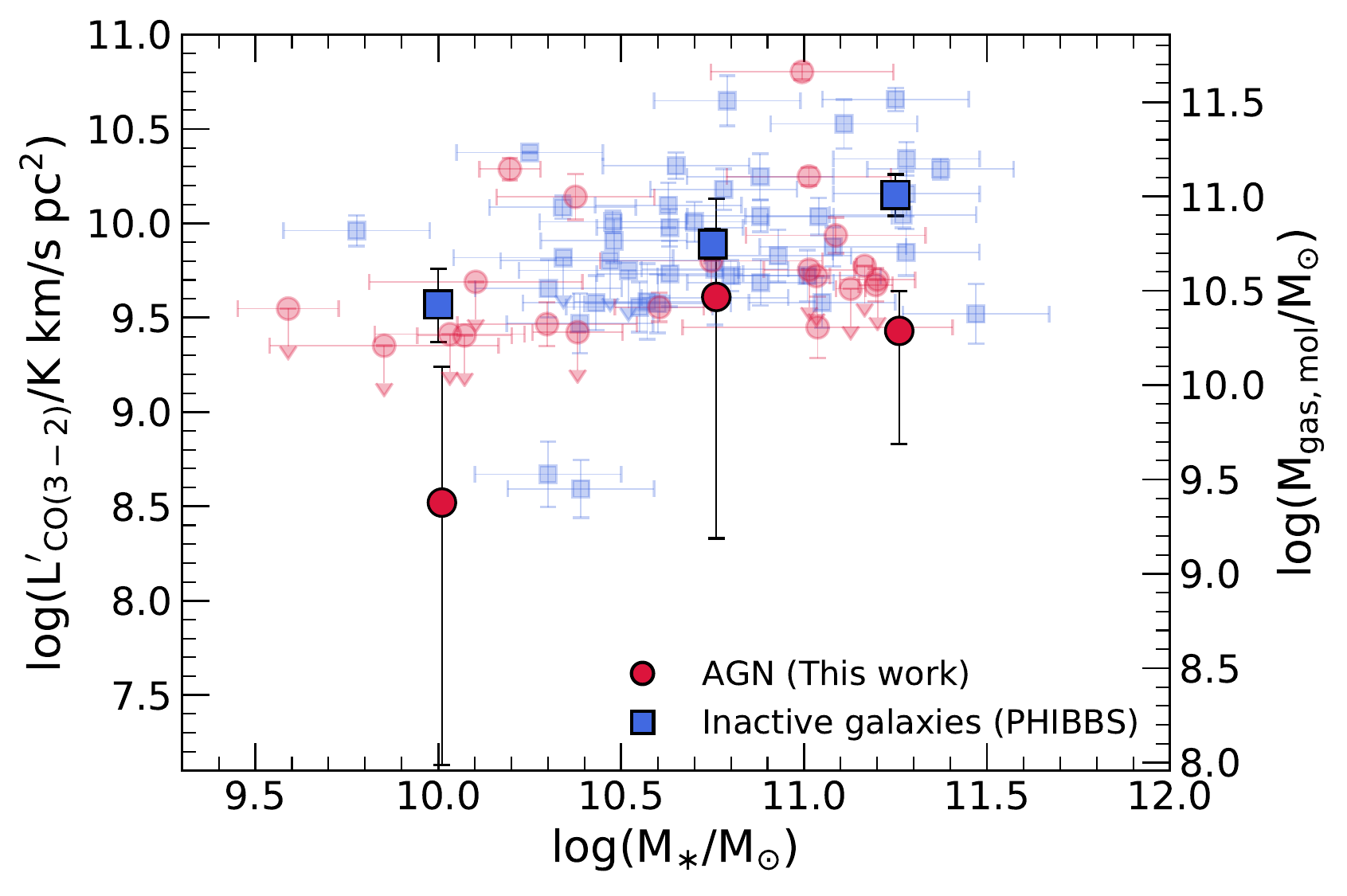}
  \caption{$L'_{\textnormal{CO}}$ and molecular gas masses as a function of $M_{\ast}$ for our AGN sample (red circles) and inactive galaxies (blue squares; \citealt{tacconi18}). 
  \textit{Top:} Bisector fit performed by adopting a Bayesian framework to display the distribution of the two samples (see text).
  The dispersion of the fits is given by plotting 500 realizations of the bisector fit. 
  We do not identify a clear difference between the two samples, since the fits are consistent within the uncertainties. \textit{Bottom:} Mean $L'_{\textnormal{CO}}$ per bin of stellar mass obtained by using a hierarchical Bayesian approach plotted in darker colors (see text). In the background, with lighter colors, we plot the distribution of the observed quantities, as in the top panel. 
  Due to the large number of upper limits in our sample and the small statistics, error bars are quite large therefore it is difficult to assess in a robust way the true distribution of AGN. 
  The difference is better constrained in the high-mass bin, where AGN show a mean CO luminosity lower than inactive galaxies by 0.72 dex at the $\sim$3$\sigma$ level. 
  }
  \label{fig:CO_mstar}
\end{figure}

We compare $L'_{\textnormal{CO}}$ with the stellar mass of AGN and inactive galaxies in Fig.~\ref{fig:CO_mstar}. 
To quantify the differences between the two samples, we performed the bisector fit as described in Section~\ref{s:co_fir}. The results of the fits ($\log L'_{\textnormal{CO}}/\textnormal{K km s}^{-1} \textnormal{pc}^{2} = m \log M_{\ast}/M_{\odot} + b$) to the AGN sample is $m=0.98^{+0.47}_{-0.44}$ and $b=-1.00^{+4.73}_{-5.10}$, while for inactive galaxies $m=1.11^{+0.19}_{-0.17}$ and $b=-2.08^{+1.82}_{-1.99}$, as displayed in the top panel of Fig.~\ref{fig:CO_mstar}. We also note that the statistics of the AGN sample is slightly reduced (21 objects) with respect to the previous fitting analysis, since three targets do not have an estimate of the stellar mass (see Sec.~\ref{s:properties} and Table~\ref{tab:summary_results}) and therefore are excluded from the fit. Our results show large uncertainties and we do not find any clear difference between the two samples. Similarly, the KS test for two-dimensional datasets implementing a Monte Carlo approach provides a significance of $\sim$32\% without upper limits. Instead, the significance including upper limits is $\sim$65\%. This result can be taken as a lower limit, since CO luminosities for the targets without a detection could be lower than the value assigned in the Monte Carlo simulation (i.e., the value of the upper limit). We performed the analysis by removing from the fit the two outliers in the comparison sample with low CO luminosities. The slope is $0.96^{+0.17}_{-0.15}$ and the intercept $-0.48^{+1.58}_{-1.79}$, which is consistent with the previous result given the large uncertainties. The two-dimensional KS test provides a significance of $\sim$30\% ($\sim$54\%) without (with) upper limits.

We performed an additional analysis in this plane, by dividing the targets in bins of stellar mass and for each we computed the mean $L'_{\textnormal{CO}}$. We joined the targets of the low-mass bins ($\log M_{\ast}/M_{\odot}=9.5-10$ and $10-10.5$) given the very low number of targets for both samples. The other mass bins have a width of 0.5 dex. We also note that this type of analysis does not take into account errors on the stellar mass, since each target is assigned to a bin based on its $M_{\ast}$ measured value. 
To derive the mean $L'_{\textnormal{CO}}$, 
we adopted a hierarchical Bayesian approach, by including the prior assumption that CO luminosities follow a common distribution, that is the prior distribution from now on. In our framework, we assumed that this prior distribution is Gaussian, described by two parameters (called hyper-parameters), mean $\mu$ and standard deviation $\sigma$. In particular, $\mu$ is the mean $L'_{\textnormal{CO}}$ per stellar mass bin that we want to estimate. The key point of this approach is that $\mu$ and $\sigma$ are not given as inputs, but we infer their distributions directly from the data, by fitting the CO luminosities of our targets simultaneously in each bin. Given the consistent number of CO upper limits in our sample, we used the hierarchical framework to provide better constraints on the distribution of upper limits, by exploiting the prior distribution followed by CO luminosities.

The validity of the assumption that CO luminosities follow a Gaussian distribution was tested on the xCOLD-GASS reference survey \citep{saintonge17}, which provides a complete sampling of the molecular gas content in galaxies across the main sequence in the nearby Universe, purely selected by mass. We considered galaxies with $M_{\ast}>10^{9.5}$ $M_{\odot}$ on the main sequence and the distribution of their CO luminosities results to be Gaussian. In order to check for redshift effects, we performed the same test on the PHIBSS survey, from which main-sequence galaxies at $2<z<2.5$ with $M_{\ast}>10^{9.5}$ $M_{\odot}$ featuring CO observations were selected. Although the statistics is much worse than xCOLD-GASS, the CO luminosities of the sample roughly follow a Gaussian distribution.
We used uniform priors (only positive values) on the hyper-parameters $\mu$ and $\sigma$ of the prior distribution.  
Similarly to the linear fit described in Section~\ref{s:co_fir}, the likelihood function was constructed by assuming that the uncertainties are Gaussian-distributed for detections. For non detections, we used the error function. 
The likelihood function was sampled with \textsc{emcee} and the mean CO luminosity per bin of stellar mass is given by the 50th percentile of the sampled marginalized posterior distribution of $\mu$ for both the AGN and inactive galaxy sample. 
The 16th and 84th percentiles of the distribution are taken as uncertainties. Our results are plotted in darker colors in Fig.~\ref{fig:CO_mstar} (\textit{bottom panel}), while the background points display the observed quantities. Overall, it seems that AGN show mean CO luminosities 0.3$-$1.0 dex lower than inactive galaxies. However, due to the large number of upper limits in our sample and the small statistics, error bars are quite large and therefore it is difficult to assess in a robust way the true distribution of AGN CO luminosities. Indeed, upper limits dominate the distribution of $\mu$ in 
the low stellar mass bins, which are characterized by a tail at low CO luminosities as proved by the low mean value (see Fig.~\ref{fig:CO_mstar}, \textit{bottom panel}). However, the difference between the two samples is better constrained at stellar masses higher than $10^{11}$ $M_{\odot}$, given the larger number of datapoints and especially detections. In this mass bin the mean is lower by 0.72 dex at the $\sim$3$\sigma$ level, with values of $9.43^{+0.21}_{-0.60}$ for AGN and $10.15\pm0.11$ for inactive galaxies. Since the two outliers of the inactive-galaxy sample reside in the low-mass bin, their removal from the analysis does not produce significant differences, given the uncertainties.

\subsection{\label{s:fgas}Gas fractions}
To gain further insight into the comparison of CO properties and stellar masses of our samples, we finally analyzed the ratio between $L'_{\textnormal{CO}}$ and stellar mass, which is a proxy for the gas fraction. A similar analysis on the ratio between $L'_{\textnormal{CO}}$ and $L_{\textnormal{FIR}}$ (which is a proxy for the depletion time) was not carried out because 40\% of the targets suitable for this investigation are characterized by an upper limit on both quantities. We applied the same hierarchical Bayesian method described in Section~\ref{s:co_mstar} with the aim of quantifying the mean gas fraction of the two samples by taking upper limits into account. 
Again, we assumed that gas fractions in the two samples follow a common Gaussian distribution and we aim at estimating the mean of the distribution ($\mu$). Through the analysis of the whole set of AGN or inactive galaxies without any binning, we want to increase the statistics and obtain better constraints on $\mu$. The other advantage of the hierarchical approach is that, by fitting the sample of gas fractions simultaneously, we derive the posterior distribution of the hyper-parameters of the prior ($\mu$ and $\sigma$) as well as the posterior distribution of the gas fraction for each target, used below to obtain the total marginalized posterior distribution.

As done before, the assumption that gas fractions follow a Gaussian distribution was verified on the xCOLD-GASS survey. We selected galaxies with $M_{\ast}>10^{9.5}$ $M_{\odot}$ on the main sequence and the distribution of their gas fractions results to be Gaussian and slightly asymmetric toward low values. This could be due to a combination of decreasing gas fractions with increasing stellar mass and the fact that above $M_{\ast}=10^{10.5}$ $M_{\odot}$ xCOLD-GASS starts to have a more important contribution from galaxies at the bottom-end of the main sequence and below, featuring lower gas fractions. The same check was performed on main-sequence galaxies with $M_{\ast}>10^{9.5}$ $M_{\odot}$ featuring CO observations from the PHIBSS survey. The distribution of gas fractions follows a Gaussian trend.
We used uniform priors (only positive values) on the standard deviation of the prior distribution. 

\begin{figure}
\centering
  \includegraphics[width=9.3cm]{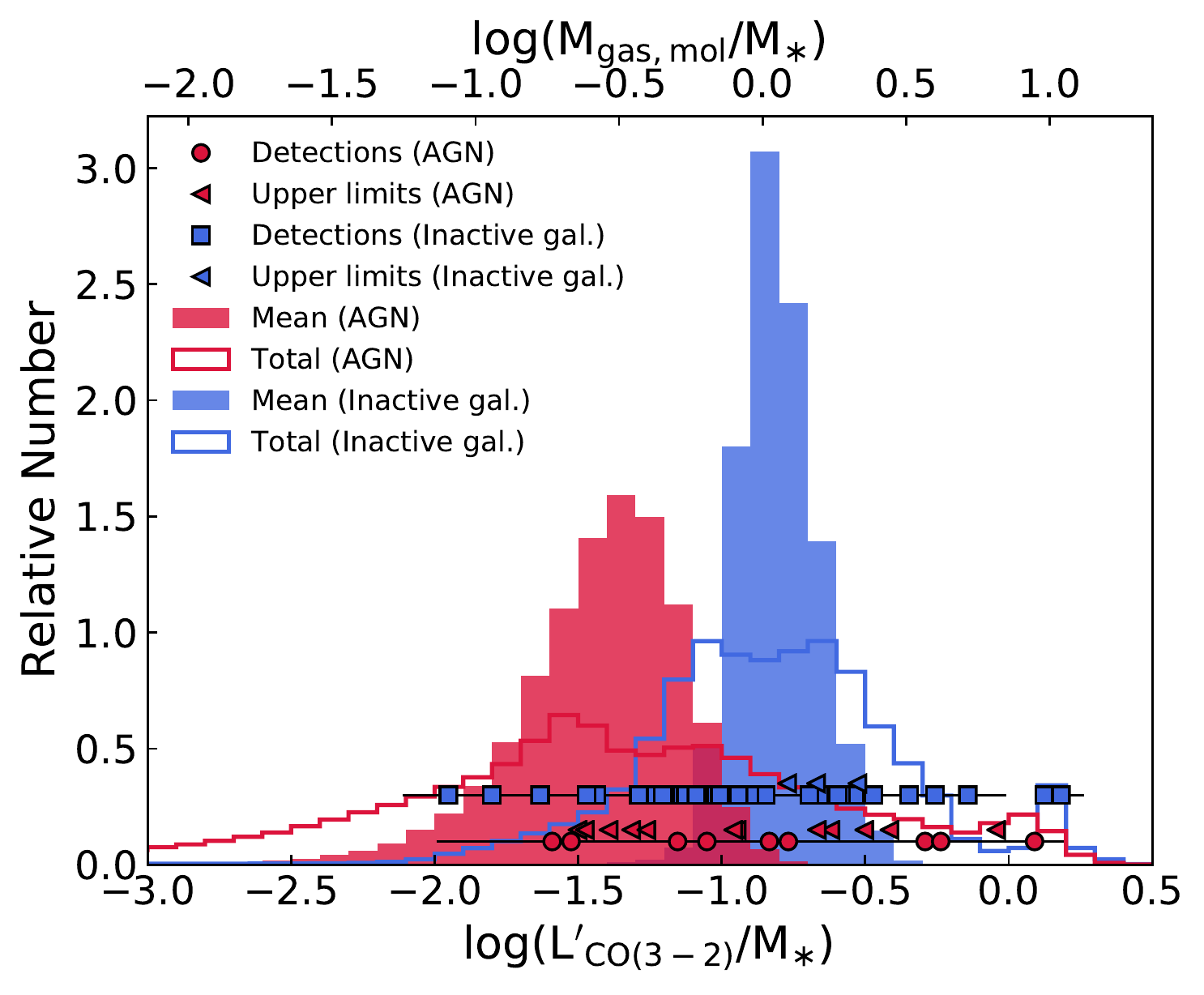}
  \caption{Distribution of the ratio between $L'_{\textnormal{CO}}$ and stellar mass (proxy of gas fraction) for AGN (red) and inactive galaxies (blue). The filled histograms show the sampled posterior distribution of the mean ($\mu$) of the hierarchical Gaussian prior adopted in our Bayesian analysis. In the bottom part of the plot, individual detections and upper limits are displayed. The unfilled histograms represent the total distributions and were obtained by joining the sampled posterior distributions of each target in the AGN or inactive galaxy sample. AGN show a mean $\log(L'_{\textnormal{CO}}/M_{\ast})$ of $-1.39^{+0.23}_{-0.30}$, while for inactive galaxies we find $-0.82^{+0.14}_{-0.11}$. The $\log(L'_{\textnormal{CO}}/M_{\ast})$ ratio of AGN is lower than inactive galaxies by a factor $\sim$3.7 (0.57 dex), at the 2.2$\sigma$ level. The two distributions are significantly different, as confirmed by statistical tests (see text).
}
  \label{fig:masses}
\end{figure}

The sampled marginalized posterior distribution of the mean $\log(L'_{\textnormal{CO}}/M_{\ast})$ for both AGN and inactive galaxies is shown by the filled histograms in Fig.~\ref{fig:masses}. As data points, we plot individual detections and upper limits in the bottom part of the Figure. The unfilled histograms represent the total sampled marginalized posterior distributions and were obtained by joining the sampled posterior distributions of each target in the AGN and inactive galaxy sample. The total posterior distributions cover the same range of values spanned by the individual datapoints, but they also incorporate the information on upper limits. The filled histograms represent the distribution of $\mu$, the mean of the prior distribution that describes the trend followed by gas fractions. 
We took the 50th, 16th and 84th percentiles from the distributions of $\mu$ as best value and uncertainties. AGN show a mean $\log(L'_{\textnormal{CO}}/M_{\ast})$ of $-1.39^{+0.23}_{-0.30}$, while for inactive galaxies we find $-0.82^{+0.14}_{-0.11}$. The AGN $\log(L'_{\textnormal{CO}}/M_{\ast})$ is lower than the inactive galaxy one by a factor $\sim$3.7 (0.57 dex), and they are different at the 2.2$\sigma$ level. By performing a two-sample KS test, the Student t-test and the Anderson-Darling test on the posterior distributions (both the mean and the total) of the two samples we can reject the null hypothesis that they are drawn from the same distribution ($P>99$\%). The logrank and Gehan's tests \citep{schmitt85} performed on the observed data provide a modestly significant result (p-value$\approx$0.02$-$0.05).

By removing the two low-CO luminosity outliers from the inactive-galaxy sample, the mean $\log(L'_{\textnormal{CO}}/M_{\ast})$ of inactive galaxies does not change substantially. Similarly, the results of the statistical tests mentioned above remain almost unchanged, although the result of the logrank test is now significant (p-value$\approx$0.004).

\section{\label{s:discussion}Discussion}

In this work we search for possible signatures of AGN feedback. We compared CO and FIR luminosities as well as stellar masses of X-ray selected AGN and inactive galaxies, and  quantified their differences by: 1. performing a linear fit in the $L'_{\textnormal{CO}}$-$L_{\textnormal{FIR}}$ plane; 2. performing a linear fit in the $L'_{\textnormal{CO}}$-$M_{\ast}$ plane; 3. dividing our samples in bins of stellar mass and computing mean CO luminosities for each bin; 4. deriving the mean distribution of $L'_{\textnormal{CO}}/M_{\ast}$ (a proxy of gas fraction) for both samples; 5. assessing how significant the differences are by performing statistical tests, such as the KS test (for both one- and two-dimensional datasets), the Student t-test, the Anderson-Darling test, and the Gehan's and logrank tests. Differently from previous work, we used the same CO transition for a fairly representative sample of AGN and inactive galaxies, controlled for assumptions and sources of bias
and treated upper limits statistically. The two samples show statistically consistent trends in the $L'_{\textnormal{CO(3-2)}}$-$L_{\textnormal{FIR}}$ and $L'_{\textnormal{CO(3-2)}}$-$M_{\ast}$ planes. However, when we focus on the subset of parameters where the results are better constrained (i.e., $L_{\textnormal{FIR}}\approx10^{12.2}$ $L_{\odot}$ and $M_{\ast}>10^{11}$ M$_{\odot}$) and on the distribution of the mean $\log(L'_{\textnormal{CO(3-2)}}/M_{\ast})$, there are indications that AGN are underluminous in CO (0.4$-$0.7 dex) with respect to inactive galaxies at the 2$-$3$\sigma$ level.

Since we presented a homogeneous characterization of the multiwavelength properties of our targets (Sec.~\ref{s:properties}), we exploited this set of physical parameters by investigating potential trends of $L'_{\textnormal{CO}}/L_{\textnormal{FIR}}$ and $L'_{\textnormal{CO}}/M_{\ast}$ with respect to AGN bolometric luminosity, obscuring column density, X-ray luminosity and Eddington ratio\footnote{$\lambda_{\textnormal{Edd}} = L_{\textnormal{bol}}/L_{\textnormal{Edd}}$, where $L_{\textnormal{Edd}} = 1.5 \times 10^{38} (M_{\textnormal{BH}}/M_{\odot})$ erg s$^{-1}$.}. We do not find any significant correlation with these properties \citep[e.g.,][]{shangguan20b}. 
As for the bolometric luminosity, there is a known limitation due to different timescales of the physical processes probed by $L_{\textnormal{FIR}}$ and $L_{\textnormal{bol}}$. Star-forming activity as traced by FIR observations has a timescale of $\sim$100 Myr while AGN activity can vary on much shorter time intervals, $<$10 Myr \citep[e.g.,][]{hickox14,stanley15}. As for the column density, evolutionary scenarios of powerful AGN predict that the unobscured phase follows the obscured one, after the AGN has removed some gas and dust from the galaxy because of feedback mechanisms such as outflows \citep[e.g.,][]{dimatteo05,hopkins06}. However, previous studies in the literature looked at possible differences between obscured and unobscured AGN in the $L'_{\textnormal{CO}}-L_{\textnormal{FIR}}$ plane but no trend with obscuration is found \citep{perna18,shangguan19}. Identifying the evolutionary phase of an AGN based on the amount of obscuration is challenging because of the nonuniform distribution of the absorbing material and rapid variations of the AGN duty cycle. Therefore, potential trends may be washed out in the scatter. Lastly, given the smaller subset (17/28) of targets with a black hole mass measurement available, the reduced statistics combined with upper limits could hamper a detailed analysis. 

AGN are normally assumed to have higher CO excitation \citep[e.g.,][]{weiss07,carilli13}, hence higher excitation correction $r_{31}$ (common values are 0.92 and 0.5 for AGN and inactive galaxies, respectively; \citealt{daddi15,tacconi18,kirkpatrick19}, but see also \citealt{riechers20}). We find on average lower CO(3-2) luminosities in AGN than in inactive galaxies (at given stellar masses and FIR luminosities), therefore even by conservatively assuming the same $r_{31}$ we would find lower molecular gas masses in AGN, and assuming higher $r_{31}$ would exacerbate this difference. 
Lastly, different assumptions on $\alpha_{\textnormal{CO}}$, based for example on metallicity and star-formation properties \citep{bolatto13}, can further increase, but also decrease, the difference. Measurements of the CO(1-0) line, for example with the JVLA, would be necessary to investigate excitation properties and obtain a reliable estimate of the total molecular gas reservoir in our AGN. 

A deficit of CO(3-2) emission in AGN has also been reported by \citet{kirkpatrick19}, who performed a study of CO emission properties of galaxies as a function of AGN contribution in the MIR, for a heterogeneous sample collected from the literature. 
Although the results are not statistically robust because of substantial uncertainties and small sample sizes, they find systematically lower CO fluxes in AGN. 
The ratio between CO(3-2) fluxes of AGN and star formation-dominated galaxies is $0.58 \pm 0.2$, for which they have a larger statistics with $\sim$30 detections in total across the two samples. 
Such a result is in agreement with what we find for our samples and would translate in lower gas masses in high-redshift AGN. 

\subsection{Comparison with previous work}
Previous work on $1<z<3$ AGN often found systematically reduced molecular gas reservoirs compared to inactive galaxies. 
For example, \citet{carilli13} and \citet{perna18} present lower CO luminosities for an heterogeneous collection of AGN from the literature (but see also \citealt{kirkpatrick19}); \citet{kakkad17} analyzed a sample of 10 AGN with homogeneous ALMA CO(2-1) observations; \citet{carniani17}, \citet{brusa18} and \citet{loiacono19} focus on a few targets with detected ionized outflows. \citet{sharon16} looked for differences between AGN and SMGs and find the two samples to be consistent in their CO and FIR properties for both CO(3-2) and CO(1-0) transitions. Conversely, \citet{spingola20} observed two lensed AGN at $z\approx2-3$ and find larger CO(1-0) luminosities with respect to their FIR emission when compared with star-forming galaxies, meaning that they may be less efficient at forming stars. 

Potential sources of bias in previous work and the way we addressed them are as follows. 

\textit{CO luminosities}. When samples are assembled from the literature by joining observations of different transitions, excitation factors have to be assumed in order to estimate the CO ground transition for all the targets. 
These correction factors depend on the SLED and, given the steeper SLED often found for AGN \citep{carilli13,kirkpatrick19}, the differences between AGN and inactive galaxies are systematically amplified when deriving the CO(1-0) luminosity. However, the CO flux at the various transitions can vary in individual targets based on the physical conditions of the ISM \citep{carilli13}.
Different assumptions on $\alpha_{\textnormal{CO}}$, for example a starburst-like value adopted for AGN in some cases \citep[e.g.,][]{perna18}, further contribute to the differences. In this work this issue was addressed by comparing directly the observed CO(3-2) luminosity for both samples. 

\textit{FIR luminosities and stellar masses}. 
A complication in deriving reliable FIR luminosities of AGN host galaxies is that the AGN can provide a significant contribution to the FIR luminosity. This issue is magnified when different methods to estimate FIR luminosities are used in the same analysis (e.g., in literature samples). For example, \citet{kirkpatrick19} and \citet{perna18} show how different methods to derive this quantity bring to different conclusions. A homogeneous SED-fitting decomposition for our AGN sample was performed in our analysis in order to derive FIR luminosities associated with star-formation activity. 
As for the comparison sample of inactive galaxies, we note that SFRs 
are taken from the literature and derived with different tracers in a few cases, although we rely on the observational evidence that main-sequence galaxies show agreement among H$\alpha$ and IR star-formation tracers \citep[][see also Sec.~\ref{s:lit_sample}]{shivaei16}. Uniformly measuring FIR luminosities for the comparison sample would allow us to quantify the entity of this bias, if any. Moreover, stellar masses are not available for many bright AGN (often the main targets of previous observations), which is a limitation toward assessing where they lie on the main sequence and performing a meaningful comparison with matched samples of inactive galaxies. 
Estimates of the stellar mass are available for the majority ($\sim$78\%) of our targets, which include more common AGN.

\textit{Sample size.} Previous conclusions on the CO properties of AGN were sometimes biased toward bright objects and therefore the more extreme ones of the AGN population \citep[e.g.,][]{carniani17,brusa18}. In analyses of larger samples, albeit more representative, upper limits were excluded in some cases \citep[e.g.,][]{kirkpatrick19}.
The study carried out in this work, focused on X-ray selected AGN, presents more common sources spanning a wide range of properties such as AGN bolometric luminosities, stellar masses and SFRs, as well as a Bayesian analysis to take upper limits into account.

We performed a careful selection of the comparison sample, with the aim of building a set of targets with similar properties to our AGN on the main sequence and looking for differences in CO properties. By taking into account the biases mentioned above, we find that the two samples do not show statistically significant differences although there are hints of AGN being deficient in their CO(3-2) luminosities at the 2$-$3$\sigma$ level. Therefore, caution must be exercised when mixing targets with observations tracing different CO transitions, since a priori assumptions on excitation corrections can artificially produce larger offsets between AGN and inactive galaxies. Finally, we note that 
our sample selection is limited by 50\% SFR upper limits of AGN on the main sequence (Fig.~\ref{fig:mainSeq}). Indeed, if the SFRs of these targets, and in turn their FIR luminosities, are actually much lower than those of the comparison sample we cannot rule out that the lower CO luminosities of AGN are consistent with their potentially lower FIR luminosities, according to the $L_{\textnormal{CO}}$-$L_{\textnormal{FIR}}$ relation \citep{sargent14}. Future improvements of our analysis rely on ALMA as the key facility to provide stronger constraints on the results presented in Sec.~\ref{s:results} and reduce the dispersion of our fits. More sensitive CO observations are needed to detect the CO(3-2) emission of our undetected AGN as well as dust-continuum observations to allow a precise characterization of the FIR emission for the targets with just upper limits from \textit{Herschel}. These observations, combined with larger and more uniformly selected samples, will allow us to assess whether there is a true difference between AGN and inactive galaxies.

\subsection{The effects of AGN activity}
Our results point toward a difference in CO luminosity between AGN and inactive galaxies, which could be ascribed to the presence of the central engine that may have a role in heating, exciting, dissociating, and/or ejecting the gas. However, understanding the physical mechanism producing such a difference requires tracing different gas phases and several CO transitions. With only the CO(3-2) data in hand, we provide possible scenarios to interpret our findings.

\textit{AGN radiation field.} 

Interesting examples of the impact of AGN activity on spatially resolved CO properties are presented by \citet{feruglio20} and \citet{rosario19}. Thanks to high-resolution observations of different gas phases (both molecular and ionized) in two nearby Seyfert galaxies, they find CO(2-1) cavities in the inner 200 pc around the AGN, which are filled with ionized and warm molecular gas (traced by $H_{2}$ NIR emission lines). They conclude that CO may be excited higher up in the rotational ladder in the CO cavities, giving rise to the apparent lack of CO emission at low J. It is also possible that CO is dissociated by the AGN radiation field and/or shocks. In particular, \citet{kawamuro20} propose that the AGN X-ray field can be responsible for weakening the CO(2-1) emission. In those Seyfert galaxies the action of the central engine is limited to the inner regions within a few hundred parsec from the AGN. It is unlikely that this cavity-effect would be detected in integrated measurements of CO transitions as it would be diluted in the beam (e.g., see \citealt{lamperti20} for a study of integrated CO(3-2)/CO(1-0) properties of $z<0.05$ AGN and star-forming galaxies). 
At high redshift, however, when the average AGN population is more active and more luminous, the AGN could have an impact out to larger spatial scales in the host galaxy, possibly out to kpc scales, and in this case it would be possible to detect an overall deficit of CO emission. Some indications of this scenario at higher redshift are provided by lensed AGN. \citet{spingola20} and \citet{paraficz18} observed a decrease in the surface brightness of the CO(1-0) and (2-1) emission (at $z\approx2$ and 0.6), respectively, at the location of the AGN emission and the center of host galaxies. This is interpreted as a potential decrement in the molecular gas at low excitation close to the AGN possibly due to AGN radiative feedback. As argued by \citet{fogasy20}, the lack of emission at a given CO transition does not necessarily mean that the source is gas poor. Indeed, they analyzed integrated ALMA observations of a $z\approx2.8$ AGN and find that the target is undetected at low-J transition lines but appears bright at high-J transitions, indicating the presence of warm and highly excited molecular gas. For our AGN, observations of more CO transitions to probe the SLED and CO observations at higher resolution would be necessary to verify the impact of AGN activity on CO excitation and smaller physical scales.

\textit{AGN-driven outflows.} Another possible effect of the AGN activity on the molecular gas is through outflows. This possibility is supported by observations of individual objects: For example, \citet{carniani17}, \citet{brusa18} and \citet{loiacono19} find low gas fractions in powerful AGN at cosmic noon hosting high-velocity molecular and ionized outflows (but see also \citealt{herreracamus19}). AGN feedback in action in these targets could be depleting the molecular gas reservoir \citep{brusa15}. \citet{forster19}, studying outflows in a large sample of $0.6<z<2.7$ galaxies through integral field spectroscopy of the H$\alpha$ emission line, find that incidence, strength, and velocity of AGN-driven winds are strongly correlated with the stellar mass. In particular, they find that high-velocity ($\sim$1000$-$2000 km s$^{-1}$) AGN-driven outflows are commonly detected at masses above $\log(M_{\ast}/M_{\odot})=10.7$, and present in up to 75\% of the population for $\log(M_{\ast}/M_{\odot}) >11.2$. Interestingly, above this stellar-mass threshold we find a significant CO luminosity deficit in our AGN sample with respect to inactive galaxies 
(Fig.~\ref{fig:CO_mstar}, \textit{bottom}). Moreover, our AGN show on average gas fractions 0.57 dex (by using uniform assumptions, Sec.~\ref{s:alma_data}) lower than inactive galaxies at the 2.2$\sigma$ level. Quantitatively, this translates into $M_{\textnormal{gas,mol}}/M_{\ast}\approx0.3$ for AGN (0.16 if we use $r_{31}=0.92$; \citealt{kirkpatrick19}) and $\approx$1 in inactive galaxies. These representative value for our AGN is in line but not as low as previous work targeting extremely powerful sources (e.g., $M_{\textnormal{gas,mol}}/M_{\ast}<0.05$ in \citealt{brusa18}). 
Our team is performing a systematic investigation of ionized gas outflows with SINFONI as part of the SUPER survey, and 11 
targets of our ALMA sample have complementary good quality SINFONI data (\citealt{kakkad20}; Perna et al., in prep.). For some of them we measured [O\textsc{iii}] line widths larger than 600 km/s, interpreted as a clear signature of the presence of an AGN-driven outflow in these objects \citep{kakkad20}. 
A detailed comparison between outflow and CO properties for these targets will be presented in a future work.
\smallskip

Distinguishing among the scenarios described above is challenging with the current dataset. AGN feedback could proceed in different ways and different mechanisms likely overlap in shaping the properties of the molecular gas reservoir. For example, AGN radiation could both heat and/or dissociate CO molecules. In this case, AGN would produce a feedback mechanism that does not require outflows but would potentially work toward inhibiting further star formation. As for AGN-driven outflows, they could impact the gas content by ejecting material out of the galaxy \citep[e.g.,][]{travascio20}, or they could produce CO heating or dissociation due to shocks. Additionally, numerical simulations predict that AGN-driven outflows may heat via shocks a significant quantity of the gas in the ISM, reaching the high temperatures required for the excitation of high-J CO transitions \citep{costa18}. To reach a deeper understanding of the impact of AGN on the molecular gas reservoir, also on longer timescales, predictions from simulations providing the spatial scales and effects of AGN activity on CO properties as a function of cosmic time are needed. 

\section{Conclusions}\label{s:Conclusions}

In this work we presented the first systematic investigation of the CO(3-2) emission of AGN at $z \approx 2$ by using ALMA observations for a sample of 27 X-ray selected AGN (Sec.~\ref{s:selection}). 
We characterized their AGN and host galaxies properties through SED-fitting and X-ray spectral analysis (Sec.~\ref{s:properties}) and measured stellar masses ($\log M_{\ast}/M_{\odot}=9.6-11.2$), FIR luminosities ($\log L_{\textnormal{FIR}}/\textnormal{erg s}^{-1} <45.0-46.4$), AGN bolometric luminosities ($\log L_{\textnormal{bol}}/\textnormal{erg s}^{-1}=44.7-46.9$), hydrogen column densities ($N_{\textnormal{H}}<2\times10^{24}$ cm$^{-2}$) and X-ray luminosities ($\log L_{\textnormal{[2-10 keV]}}/\textnormal{erg s}^{-1}=43.0-45.4$). CO emission was detected in 11 out of 27 targets and we find CO luminosities in the range $\log(L'_{\textnormal{CO}}/\textnormal{K km s}^{-1} \textnormal{pc}^{2})=9.33-10.80$, line FWHM 97$-$810 km/s and molecular gas masses $\log(M_{\textnormal{mol}}/M_{\odot})=10.19-11.66$ (Sec.~\ref{s:alma_data}). To infer whether AGN activity affects the CO emission of the host galaxy, we compared the CO properties of our sample with those of inactive galaxies with similar redshift, stellar masses and SFRs selected from the PHIBSS survey (Sec.~\ref{s:lit_sample}, Fig.~\ref{fig:mainSeq}). We further controlled for systematic differences possibly introduced by conversion factors (e.g., excitation corrections, $\alpha_{\textnormal{CO}}$) by comparing directly the CO(3-2) luminosities for both samples. In order to properly account for upper limits we adopted a Bayesian approach throughout our analysis. Our findings can be summarized as follows.

\begin{itemize}

\item We compared CO and FIR luminosities of AGN and inactive galaxies (Sec.~\ref{s:co_fir}) and quantified the distribution of the two samples in the $L'_{\textnormal{CO}}$-$L_{\textnormal{FIR}}$ plane by fitting a linear model through the ordinary least-square bisector fit method (Fig.~\ref{fig:CO_FIR}). The resulting fits are characterized by similar (almost linear) slopes and large dispersions, and do not show a significant shift within the uncertainties. However, where the results are best constrained (i.e., $\log(L_{\textnormal{FIR}}/L_{\odot})\approx12.2$), we find that AGN have CO luminosities 0.4 dex lower than inactive galaxies, different at the 2$\sigma$ level. By applying a KS test for two-dimensional datasets implementing a Monte Carlo approach, the two samples result to be different at the 99\% ($\sim$28\%) level including (excluding) upper limits.

\item A similar investigation was performed in the $L'_{\textnormal{CO}}$-$M_{\ast}$ plane (Sec.~\ref{s:co_mstar}), where the results show large uncertainties and different slopes, and we do not find any clear difference between the two samples (Fig.~\ref{fig:CO_mstar}, top). 
We further divided our samples in bins of stellar mass and computed mean CO luminosities for each bin. In order to provide constraints on the distribution of upper limits, we implemented a hierarchical Bayesian framework by adopting the prior assumption that CO luminosities follow a Gaussian distribution. Despite the error bars, the difference results to be best constrained in the high-mass bin $\log(M_{\ast}/M_{\odot})>11$ (Fig.~\ref{fig:CO_mstar}, bottom), where AGN are $\sim$0.72 dex less luminous than inactive galaxies at the 3$\sigma$ level. 

\item We finally performed our hierarchical Bayesian analysis on the total distribution of $\log(L'_{\textnormal{CO}}/M_{\ast})$, a proxy of gas fraction (Sec.~\ref{s:fgas}). AGN show a mean $\log(L'_{\textnormal{CO}}/M_{\ast})$ 0.57 dex lower than inactive galaxies at the 2.2$\sigma$ level (Fig.~\ref{fig:masses}). By applying statistical tests such as the two-sample KS, the Anderson-Darling and the Student t-test to the sampled marginalized posterior distributions, and the logrank and Gehan's tests to the observed data we assessed that the two samples are different at the 99\% level.

\end{itemize}
Summarizing, in our analysis we controlled for assumptions and sources of bias, focused on a sample of targets covering a wide range of AGN bolometric luminosity and treated upper limits statistically. When averaged over a fairly representative sample of targets including also less extreme AGN, the differences between AGN and inactive galaxies are not very large, but when data are best constrained AGN result to be underluminous in CO as a function of FIR luminosities and stellar masses at the 2$-$3$\sigma$ level. Overall, our AGN feature CO(3-2) luminosities lower than inactive galaxies (at given stellar masses and FIR luminosities), therefore even by conservatively assuming the same $r_{31}$ we would find lower molecular gas masses in AGN, and assuming higher $r_{31}$ would exacerbate this difference.
We interpreted our result as a hint toward the potential effect of AGN activity, which may be able to heat, excite, dissociate, and/or deplete the gas reservoir of the host galaxies. To help establish whether the driving mechanisms are AGN radiation and/or outflows, observations tracing the outflowing multiphase gas are needed, as well as observations of more CO transitions to probe the SLEDs and CO observations at higher resolution to verify if the impact of AGN activity on molecular gas takes place on smaller scales.

\begin{acknowledgements}
We thank the anonymous referee for carefully reading the manuscript and providing helpful comments. C. Circosta thanks Linda Tacconi for useful advice on the PHIBSS survey and providing the table with measured CO fluxes; Vinod Arumugam and Emanuele Daddi for helpful discussions about ALMA data analysis; Sotiria Fotopoulou and Antonis Georgakakis for providing the photometry of the XMM-XXL targets. C. Circosta and AS acknowledge support from the Royal Society. AP gratefully acknowledges financial support from STFC through grants ST/T000244/1 and ST/P000541/1. MP is supported by the Programa Atracci\'on de Talento de la Comunidad de Madrid via grant 2018-T2/TIC-11715.  We acknowledge support from PRIN MIUR project "Black Hole winds and the Baryon Life Cycle of Galaxies: the stone-guest at the galaxy evolution supper", contract \#2017PH3WAT. GV acknowledges financial support from Premiale 2015 MITic (PI B. Garilli). This paper makes use of the following ALMA data: ADS/JAO.ALMA\#2016.1.00798.S and ADS/JAO.ALMA\#2017.1.00893.S. ALMA is a partnership of ESO (representing its member states), NSF (USA) and NINS (Japan), together with NRC (Canada), NSC and ASIAA (Taiwan), and KASI (Republic of Korea), in cooperation with the Republic of Chile. The Joint ALMA Observatory is operated by ESO, AUI/NRAO and NAOJ. This research has made use of the following data: data based on data products from observations made with ESO Telescopes at the La Silla Paranal Observatory under ESO programme ID 179.A-2005 and on data products produced by TERAPIX and the Cambridge Astronomy Survey Unit on behalf of the UltraVISTA consortium. Data from HerMES project (http://hermes.sussex.ac.uk/). HerMES is a Herschel Key Programme utilising Guaranteed Time from the SPIRE instrument team, ESAC scientists and a mission scientist. The HerMES data was accessed through the Herschel Database in Marseille (HeDaM - http://hedam.lam.fr) operated by CeSAM and hosted by the Laboratoire d'Astrophysique de Marseille. HerMES DR3 was made possible through support of the Herschel Extragalactic Legacy Project, HELP (http://herschel.sussex.ac.uk), HELP is a European Commission Research Executive Agency funded project under the SP1-Cooperation, Collaborative project, Small or medium-scale focused research project, FP7-SPACE-2013-1 scheme.
\end{acknowledgements}


\bibliographystyle{aa}
\bibliography{biblio}

\begin{appendix}
\onecolumn
\begin{landscape}
\section{Multiwavelength properties of the sample}
\renewcommand{\arraystretch}{1.3}

\begin{longtable}{cccccccccc}
\caption{\label{tab:summary_results} Summary of AGN and host galaxy properties for the target sample.}\\
\hline\hline
ID & AGN type & $\log \frac{M_{\ast}}{M_{\odot}}$ & $\log \frac{L_{\textnormal{FIR}}}{\textnormal{erg\,s$^{-1}$}}$ & SFR [$M_{\odot}$ yr$^{-1}$] & $\log \frac{L_{\textnormal{bol}}}{\textnormal{erg\,s$^{-1}$}}$ & X-ray net counts & $\log \frac{N_{\textnormal{H}}}{\textnormal{cm$^{-2}$}}$ & $\log \frac{L_{[2-10\,\textnormal{keV}]}}{\textnormal{erg\,s$^{-1}$}}$ & $\log \frac{M_{\textnormal{BH}}}{M_{\odot}}$ \\ 
(1) & (2) & (3) & (4) & (5) & (6) & (7) & (8) & (9) & (10) \\
\hline
X\_N\_128\_48 & BL & - & - & - & $45.81 \pm 0.40$\tablefootmark{a} & $12 \pm 4$ & $<22.80$ & $44.47^{+0.28}_{-0.27}$ & $10.04 \pm 0.3$\tablefootmark{b} \\
X\_N\_81\_44 & BL & $11.04 \pm 0.37$ & $45.93 \pm 0.20$ & $229 \pm 103$ & $46.80 \pm 0.03$ & $95 \pm 10$ & $<21.86$ & $44.77^{+0.07}_{-0.09}$ & $9.05 \pm 0.30$\tablefootmark{c} \\
X\_N\_53\_3 &  BL & - & $46.41 \pm 0.11$ & $686 \pm 178$ & $46.21 \pm 0.03$ & $26 \pm 5$ & $22.77^{+0.37}_{-0.67}$ & $44.80^{+0.10}_{-0.13}$ & $8.53 \pm 0.30$\tablefootmark{c} \\
X\_N\_6\_27 & BL & - & $<45.90 $ & $<215$ & $45.85 \pm 0.05$ & $26 \pm 5$ & $<22.64$ & $44.36^{+0.22}_{-0.24}$ & $8.73 \pm 0.30$\tablefootmark{d} \\
X\_N\_44\_64 & BL & $11.09 \pm 0.25$ & $45.93 \pm 0.15$ & $229 \pm 80$ & $45.51 \pm 0.07$ & $52 \pm 7$ & $<21.97$ & $44.21^{+0.11}_{-0.17}$ & $8.76 \pm 0.31$\tablefootmark{c} \\
X\_N\_102\_35 & BL & - & - & - & $46.82 \pm 0.02$ & $79 \pm 9$ & $<22.17$ & $45.37^{+0.05}_{-0.11}$ & $8.85 \pm 0.30$\tablefootmark{c} \\
X\_N\_104\_25 & BL & - & - & - & $45.97 \pm 0.40$\tablefootmark{a} & $80 \pm 9$ & $<21.91$ & $44.60^{+0.10}_{-0.09}$ & $8.91 \pm 0.30$\tablefootmark{d} \\
lid\_1852 & NL & $10.07 \pm 0.13$ & $< 45.28$ & $< 52$  & $45.25 \pm 0.09$ & $53 \pm 7$ & $22.92^{+0.36}_{-0.74}$ & $44.46^{+0.15}_{-0.15}$ & - \\
lid\_3456 & BL & $10.75 \pm 0.30$ & $46.08 \pm 0.17$ & $323 \pm 126$ & $45.68 \pm 0.07$ & $5 \pm 2$ & $<22.00$ & $43.00^{+0.50}_{-0.50}$ & $7.64 \pm 0.33$\tablefootmark{d} \\
cid\_166 & BL & $10.38 \pm 0.22$ & $< 45.92$ & $< 224$  & $46.93 \pm 0.02$ & $718 \pm 27$ & $<21.25$ & $45.15^{+0.03}_{-0.02}$ & $9.33 \pm 0.30$\tablefootmark{c} \\
lid\_1289 & NL & $9.59 \pm 0.14$ & $< 44.98$ & $< 25$ & $45.09 \pm 0.08$ & $123 \pm 11$ & $22.50^{+0.29}_{-0.22}$ & $44.69^{+0.26}_{-0.13}$ & - \\
cid\_1605 & BL & - & $< 45.54$ & $< 94$ & $46.03 \pm 0.02$ & $328 \pm 18$ & $21.77^{+0.51}_{-0.75}$ & $44.69^{+0.06}_{-0.04}$ & $8.55 \pm 0.31$\tablefootmark{c} \\
cid\_337 & NL & $11.13 \pm 0.04$ & $45.63 \pm 0.03$ & $115 \pm 9$ & $45.34 \pm 0.09$ & $83 \pm 9$ & $<22.76$ & $44.22^{+0.11}_{-0.12}$ & - \\
cid\_346 & BL & $11.01 \pm 0.22$ & $46.13 \pm 0.06$ & $362 \pm 49$ & $46.66 \pm 0.02$ & $124 \pm 11$ & $23.05^{+0.17}_{-0.19}$ & $44.47^{+0.08}_{-0.09}$ & $9.13 \pm 0.30$\tablefootmark{c} \\
cid\_357 & BL & $9.85 \pm 0.31$ & $< 45.60$ & $< 108$  & $45.25 \pm 0.06$ & $110 \pm 11$ & $<22.87$ & $44.44^{+0.19}_{-0.15}$ & $8.46 \pm 0.30$\tablefootmark{d} \\
cid\_451 & NL & $11.21 \pm 0.05$ & $45.25 \pm 0.17$\tablefootmark{e} & $48 \pm 19$ & $46.44 \pm 0.07$ & $137 \pm 12$ &  $23.87^{+0.19}_{-0.15}$ & $45.18^{+0.23}_{-0.19}$ & - \\
cid\_1205 & BL & $11.20 \pm 0.10$ & $46.16 \pm 0.04$ & $384 \pm 33$ & $45.75 \pm 0.17$ & $34 \pm 6$ & $23.50^{+0.27}_{-0.27}$ & $44.25^{+0.21}_{-0.23}$ & $8.60 \pm 0.30$\tablefootmark{c} \\
cid\_2682 & NL & $11.03 \pm 0.04$ & $< 45.54$ & $< 93$ & $45.48 \pm 0.10$ & $36 \pm 6$ & $23.92^{+1.01}_{-0.20}$ & $44.30^{+0.96}_{-0.27}$ & - \\
cid\_247 & BL & $10.03 \pm 0.20$ & $45.73 \pm 0.05$ & $143 \pm 15$  & $45.49 \pm 0.04$ & $158 \pm 13$ & $<22.43$ & $44.43^{+0.11}_{-0.06}$ & $ 8.17 \pm 0.31$\tablefootmark{c} \\
cid\_1215 & BL & $10.20 \pm 0.08$ & $45.96 \pm 0.04$ & $246 \pm 24$  & $45.73 \pm 0.05$ & $78 \pm 9$ & $22.86^{+0.31}_{-0.50}$ & $44.34^{+0.14}_{-0.14}$ & $8.37 \pm 0.30$\tablefootmark{d} \\
cid\_467 & BL & $10.10 \pm 0.29$ & $< 45.74$ & $< 147$ & $46.53 \pm 0.04$ & $447 \pm 21$ & $22.31^{+0.23}_{-0.32}$ & $44.87^{+0.04}_{-0.05}$ & $9.28 \pm 0.31$\tablefootmark{c} \\
cid\_852 & NL & $11.17 \pm 0.02$ & $< 45.57$ & $< 100$ & $45.50 \pm 0.11$ & $25 \pm 5$ & $24.30^{+0.38}_{-0.37}$ & $45.20^{+1.14}_{-0.76}$ & - \\
cid\_970 & NL & $10.38 \pm 0.12$ & $<45.66$ & $<122$  & $45.71 \pm 0.04$ & $287 \pm 17$ & $<22.25$ & $44.69^{+0.07}_{-0.04}$ & - \\
cid\_971 & NL & $10.60 \pm 0.12$ & - & $<96$\tablefootmark{f} & $44.71 \pm 0.24$ & $33 \pm 6$ & $<23.68$ & $43.87^{+0.36}_{-0.38}$ & - \\
cid\_38 & NL & $11.01 \pm 0.12$ & $< 45.98$ & $< 258$ & $45.78 \pm 0.04$ & $159 \pm 13$ & $<22.95$ & $44.41^{+0.16}_{-0.13}$ & - \\
lid\_206 & BL & $10.30 \pm 0.25$ & - & $63 \pm 27$\tablefootmark{f} & $44.77 \pm 0.12$ & $40 \pm 6$ & $<22.55$ & $43.91^{+0.30}_{-0.29}$ & $8.37 \pm 0.30$\tablefootmark{d} \\
cid\_1253 & NL & $10.99 \pm 0.25$ & $46.02 \pm 0.30$ & $280 \pm 194$ & $45.08 \pm 0.18$ & $36 \pm 6$ & $23.22^{+0.47}_{-0.39}$ & $43.92^{+0.29}_{-0.31}$ & - \\
\hline
\end{longtable}
\tablefoot{
(1) Target ID; (2) AGN classification into broad line (BL) and narrow line (NL) according to the optical spectra; (3) Galaxy stellar mass and 1$\sigma$ error; (4) FIR luminosity (star formation only) in the $8-1000$ $\mu$m range and 1$\sigma$ error; (5) SFR from the FIR luminosity and 1$\sigma$ error; (6) AGN bolometric luminosity and 1$\sigma$ error, derived from SED fitting; (7) X-ray net counts (i.e., background subtracted) in the full band and respective error, computed assuming a Poisson statistic; (8) Absorbing hydrogen column density and 90\% confidence level error; (9) Absorption-corrected X-ray luminosity in the hard band ($2-10$ keV) and 90\% confidence level error; (10) Black hole mass and 1$\sigma$ error. \\
\tablefootmark{a}{Bolometric luminosities estimated from X-ray luminosities by using the relation of \citet{duras20}.}\\
\tablefootmark{b}{BH masses from \citet{menzel16}.}\\
\tablefootmark{c}{BH masses from \citet{vietri20}.} \\
\tablefootmark{d}{BH masses from COSMOS or SDSS spectra estimated by following the method presented in \citet{vietri20}.} \\
\tablefootmark{e}{FIR luminosity estimated by including in our SED fitting analysis ALMA Band 7 data from Lamperti et al. (in prep.). In \citet{circosta18} this value was an upper limit.} \\
\tablefootmark{f}{Average SFR over the last 100 Myr of the galaxy history as obtained from the modeling of the stellar component with SED fitting.}}
\end{landscape}
\clearpage

\onecolumn
\renewcommand{\tabcolsep}{0.15cm}
\begin{landscape}
\section{Results of the ALMA CO data analysis}
\begin{longtable}{cccccccccccc}
\caption{\label{tab:ALMA_results} Summary of ALMA CO line observations and derived quantities.}\\
\hline\hline
ID  & Beam size & rms & Integrated rms & $I_{\textnormal{CO}}$ & Peak & FWHM & $z_{\textnormal{CO}}$ & $\log(L'_{\textnormal{CO(3-2)}})$ & $\log(M_{\textnormal{mol}}/M_{\odot})$ & rms cont. & $I_{\textnormal{cont.}}$ \\ 
 & {\small[arcsec$^2$]} & {\small[mJy/beam]} & \small{[mJy/beam km/s]} & {\small[mJy km/s]} & {\small[mJy]} & {\small[km/s]} & & {\small[K km s$^{-1}$ pc$^{2}$]} & {\small[$M_{\odot}$]} & {\small[$\mu$Jy/beam]} & {\small[$\mu$Jy]} \\
$(1)$ & $(2)$ & $(3)$ & $(4)$ & $(5)$ & $(6)$ & $(7)$ & $(8)$ & $(9)$ & $(10)$ & $(11)$ & $(12)$ \\
\hline
X\_N\_128\_48 & 1.2$\times$1.1 & 0.62 & 69.2 & $<207$ & - & - & - & $<9.78$ & $<10.64$ & 27 & - \\
X\_N\_81\_44 & 1.4$\times$1.1 & 0.53 & 36.8 &  $96 \pm 36$ & $0.62\pm0.36$ & $145 \pm 76$ & $2.2950 \pm 0.0003$ & $9.45 \pm 0.16$ & $10.30 \pm 0.16$ & 24 & - \\
X\_N\_53\_3 & 1.4$\times$1.2 & 0.52 & 77.0 & $203 \pm 81$ & $0.32\pm0.19$ & $600 \pm 393$ & $2.433\pm0.003$ & $9.82 \pm 0.17$ & $10.67 \pm 0.17$ & 24 & - \\
X\_N\_6\_27 & 1.3$\times$1.1 & 0.51 & 34.9 & $76 \pm 22$ & $0.74\pm0.37$ & $97 \pm 35$ & $2.2640\pm0.0004$ & $9.33 \pm 0.13$ & $10.19 \pm 0.13$ & 22 & - \\
X\_N\_44\_64 & 1.2$\times$1.1 & 0.54 & 68.3 & $306 \pm 67$ & $0.70\pm0.32$ & $412 \pm 136$ & $2.245\pm0.001$ & $9.93 \pm 0.10$ &  $10.79 \pm 0.10$ & 22 & - \\
X\_N\_102\_35 & 1.4$\times$1.0 & 0.55 & 62.8 & $<$189 & - & - & - & $<$9.71 & $<$10.57 & 25 & $2050\pm105$ \\ 
X\_N\_104\_25 & 1.2$\times$1.1 & 0.60 & 65.2 & $<$195 & - & - & - & $<9.74$ & $<$10.60 & 25 & - \\
lid\_1852 & 1.4$\times$1.2 & 0.22 & 25.8 & $<$78 & - & - & - & $<$9.40 & $<$10.26 & 9 & - \\
lid\_3456 & 1.3$\times$1.2 & 0.72 & 80.8 & $<$243 & - & - & - & $<$9.80 & $<$10.66 & 29 & - \\
cid\_166 & 1.3$\times$0.9 & 0.37 & 60.5 & $420 \pm 116$ & $0.67\pm0.27$ & $594 \pm 186$ & $2.461 \pm 0.001$ & $10.14 \pm 0.08$ & $11.00 \pm 0.12$ & 17 & - \\
lid\_1289 & 1.4$\times$1.2 & 0.32 & 36.6 & $<$111 & - & - & - & $<$9.55 & $<$10.40 & 14 & - \\
cid\_1605 & 1.3$\times$1.2 & 0.24 & 28.7 & $<$87 & - & - & - & $<$9.38 & $<$10.23 & 9 & - \\
cid\_337 & 1.3$\times$1.1 & 0.48 & 53.9 & $<$162 & - & - & - & $<$9.65 & $<$10.51 & 20 & - \\
cid\_346 & 1.2$\times$1.1 & 0.49 & 42.9 & $632 \pm 69$ & $2.91\pm0.46$ & $204 \pm 24$ & $2.2198 \pm 0.0001$ & $10.25 \pm 0.05$ & $11.10 \pm 0.05$ & 21 & $149\pm43$ \\
cid\_357 & 1.3$\times$1.2 & 0.24 & 29.4 & $<$87 & - & - & - & $<$9.35 & $<$10.21 & 9 & - \\
cid\_451 & 1.6$\times$1.4 & 0.23 & 29.7 & $144 \pm 29$ & $0.56\pm0.21$ & $241 \pm 62$ & $2.4450\pm0.0003$ & $9.67 \pm 0.09$ & $10.53 \pm 0.09$ & 9 & $220\pm13$ \\
cid\_1205 & 1.5$\times$1.2 & 0.52 & 58.9 & $<$177 & - & - & - & $<$9.70 & $<$10.56 & 22 & - \\
cid\_2682 & 1.3$\times$1.2  & 0.47 & 54.3 & $<$162 & - & - & - & $<$9.72 & $<$10.58 & 21 & - \\
cid\_247 & 1.4$\times$1.3 & 0.22 & 27.1 & $<$81 & - & - & - & $<$9.41 & $<$10.27 & 9 & - \\
cid\_1215 & 1.7$\times$1.3 & 0.32 & 56.8 & $594 \pm 80$ & $1.00\pm0.24$ & $555 \pm 100$ & $2.446\pm0.001$ & $10.29 \pm 0.06$ & $11.14 \pm 0.06$ & 13 & $50\pm9$ \\
cid\_467 & 1.0$\times$0.7 & 0.47 & 56.1 & $<$168 & - & - & - & $<$9.69 & $<$10.54 & 22 & - \\
cid\_852 & 1.4$\times$1.1 & 0.62 & 70.8 & $<$213 & - & - & - & $<$9.77 & $<$10.63 & 26 & - \\
cid\_970 & 1.6$\times$1.3 & 0.22 & 26.0 & $<$78 & - & - & - & $<$9.42 & $<$10.28 & 9 & - \\
cid\_971 & 0.8$\times$0.8 & 0.18 & 11.4 & $108 \pm 19$ & $1.15\pm0.41$ & $88 \pm 17$ & $2.4696\pm0.0001$ & $9.55 \pm 0.08$ & $10.41 \pm 0.08$ & 8 & - \\
cid\_38 & 1.3$\times$1.2 & 0.59 & 70.4 & $<$210 & - & - & - & $<$9.75 & $<$10.61 & 24 & - \\
lid\_206 & 1.5$\times$1.2 & 0.33 & 30.0 & $97 \pm 26$ & $0.57\pm0.32$ & $159 \pm 58$ & $2.3326\pm0.0002$ & $9.46 \pm 0.12$ & $10.32 \pm 0.12$ & 15 & - \\
cid\_1253 & 1.2$\times$1.2 & 0.49 & 97.2 & $2427 \pm 227$ & $2.82\pm0.39$ & $810 \pm 93$  & $2.1508\pm0.0004$ & $10.80 \pm 0.04$ & $11.66 \pm 0.04$ & 19 & $112\pm24$\tablefootmark{a} \\
\hline
\end{longtable}
\tablefoot{(1) Source identification number; 
(2) Beam size in arcsec$^2$; (3) $1\sigma$ rms per 24 km/s velocity channel; (4) 1$\sigma$ rms of the velocity-integrated map; (5) Velocity-integrated CO flux and 1$\sigma$ error for detections and 3$\sigma$ upper limits for non detections; (6) Peak flux of the Gaussian best fit and 1$\sigma$ error; (7) FWHM of the CO line and 1$\sigma$ error; (8) Redshift of the detected CO line and 1$\sigma$ error; (9) Luminosity of the CO(3-2) line and 1$\sigma$ error or 3$\sigma$ upper limits for non-detections; (10) Molecular gas mass and 1$\sigma$ error or 3$\sigma$ upper limits for non-detections, computed adopting $r_{31}=0.5$ and $\alpha_{\textnormal{CO}}=3.6$ M$_{\odot}$/(K km s$^{-1}$ pc$^2$); (11) 1$\sigma$ rms of the continuum map; (12) Integrated flux and 1$\sigma$ error of the continuum emission.\\
\tablefootmark{a}{This value represents the flux of the main target. The flux of the companion is $158 \pm 22$ $\mu$Jy.}
}

\end{landscape}

\section{Spectral energy distributions of the sample}\label{sec:SEDs}

\begin{figure*}[h!]
\centering
  	\includegraphics[width=8.3cm]{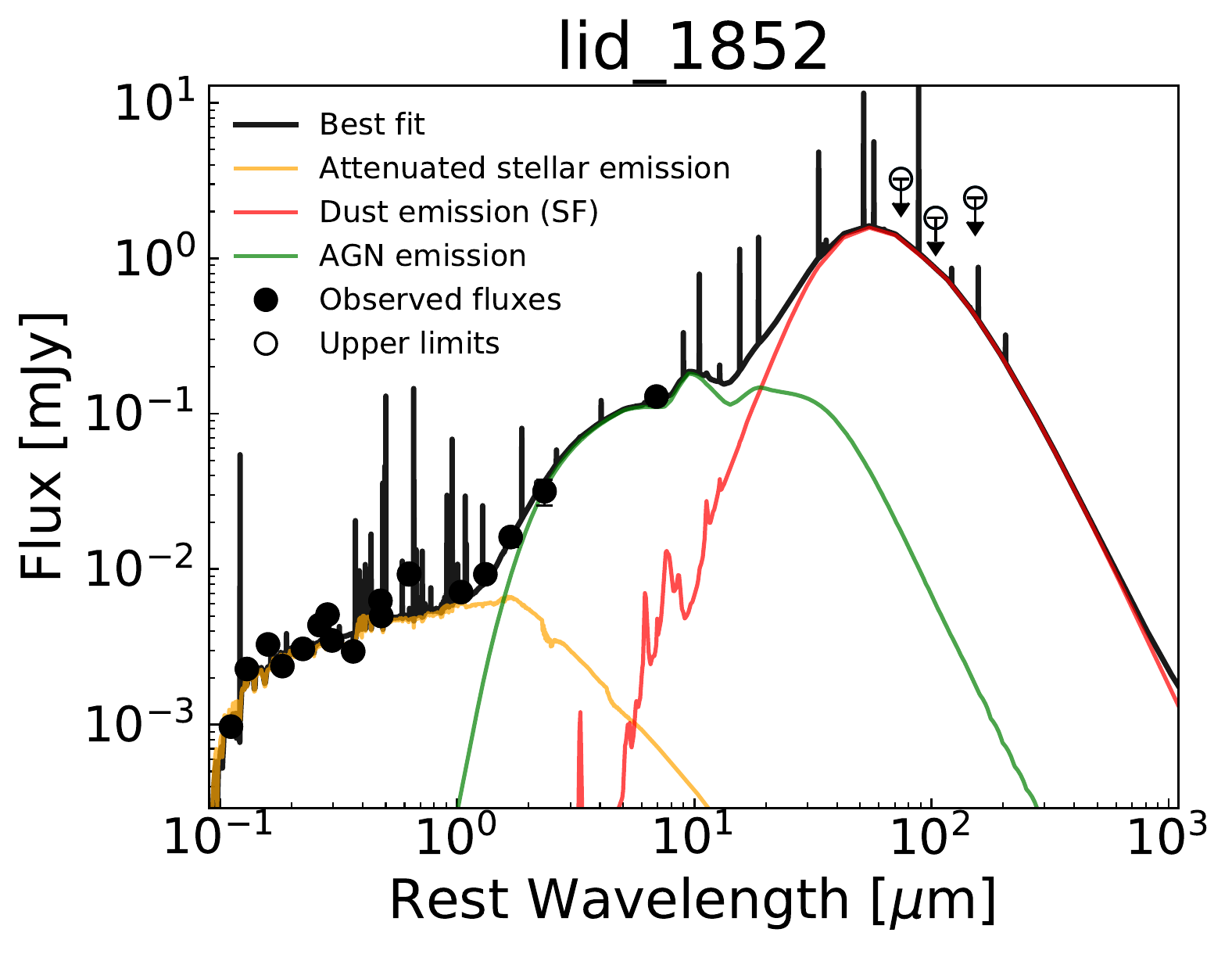} 
  	\hspace{2mm}
  	\includegraphics[width=8.3cm]{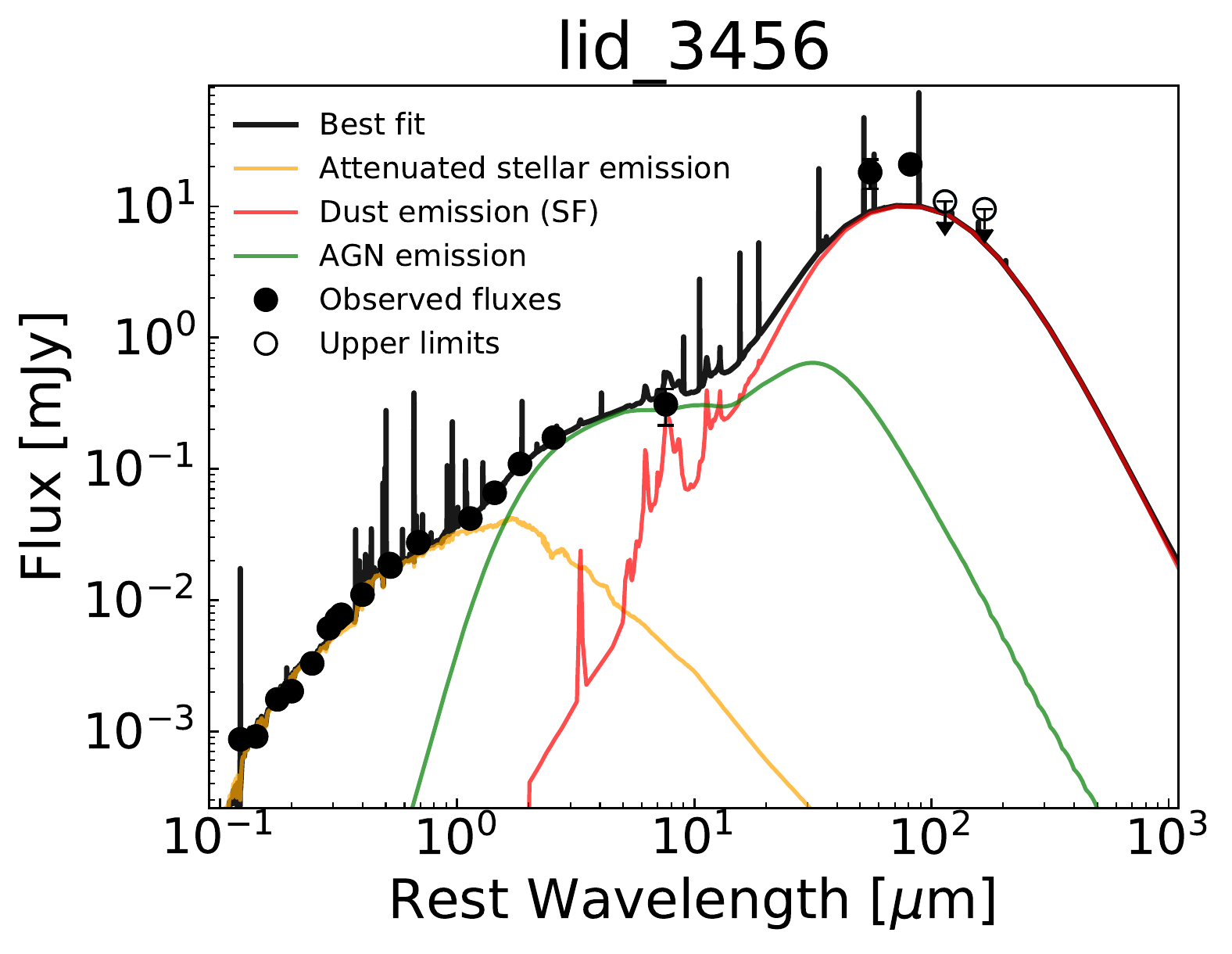} 
  	\hspace{2mm}
  	\includegraphics[width=8.3cm]{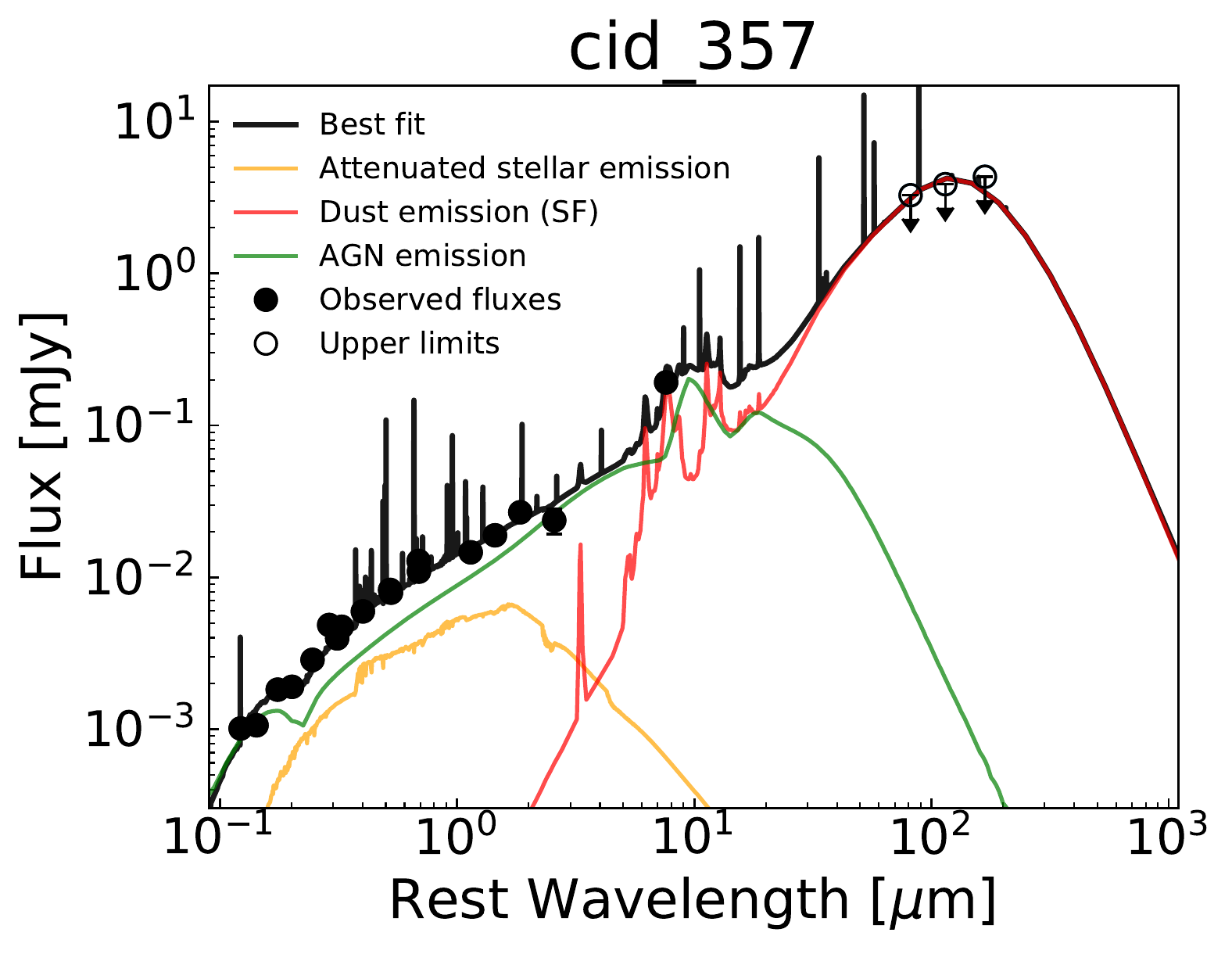} 
 	\hspace{2mm}
 	\includegraphics[width=8.3cm]{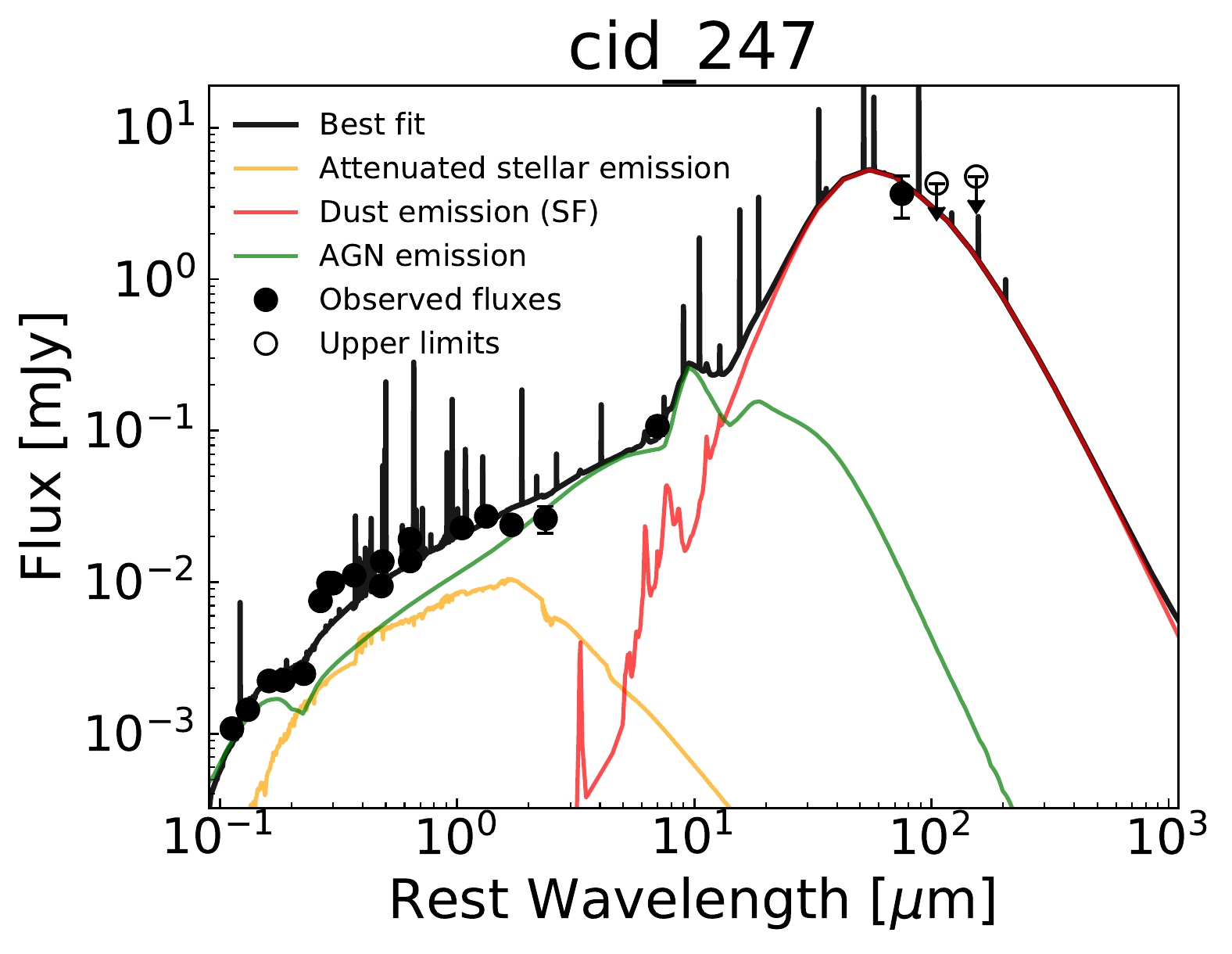} 
  	\hspace{2mm}
 	\includegraphics[width=8.3cm]{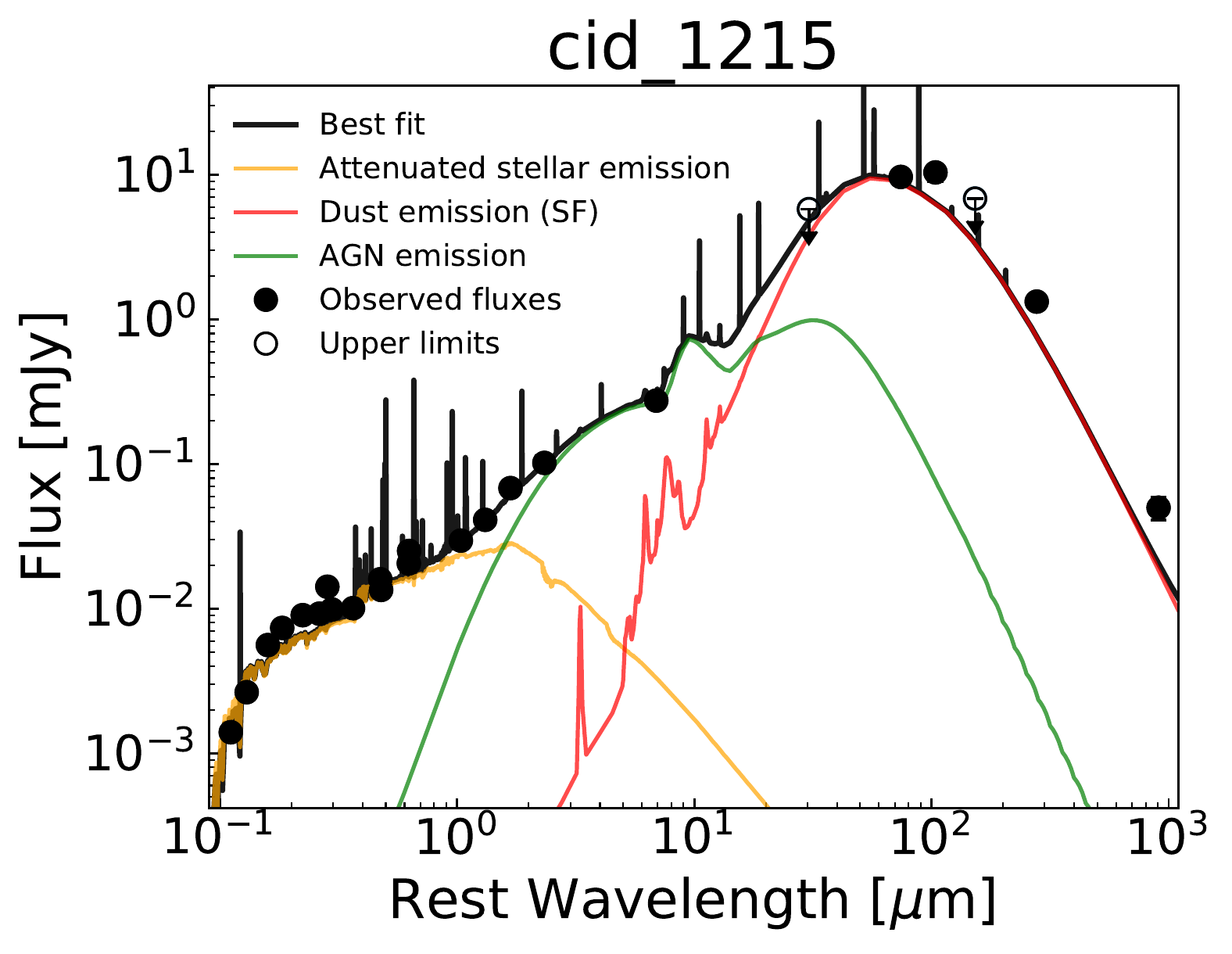} 
 	\hspace{2mm}
  	\includegraphics[width=8.3cm]{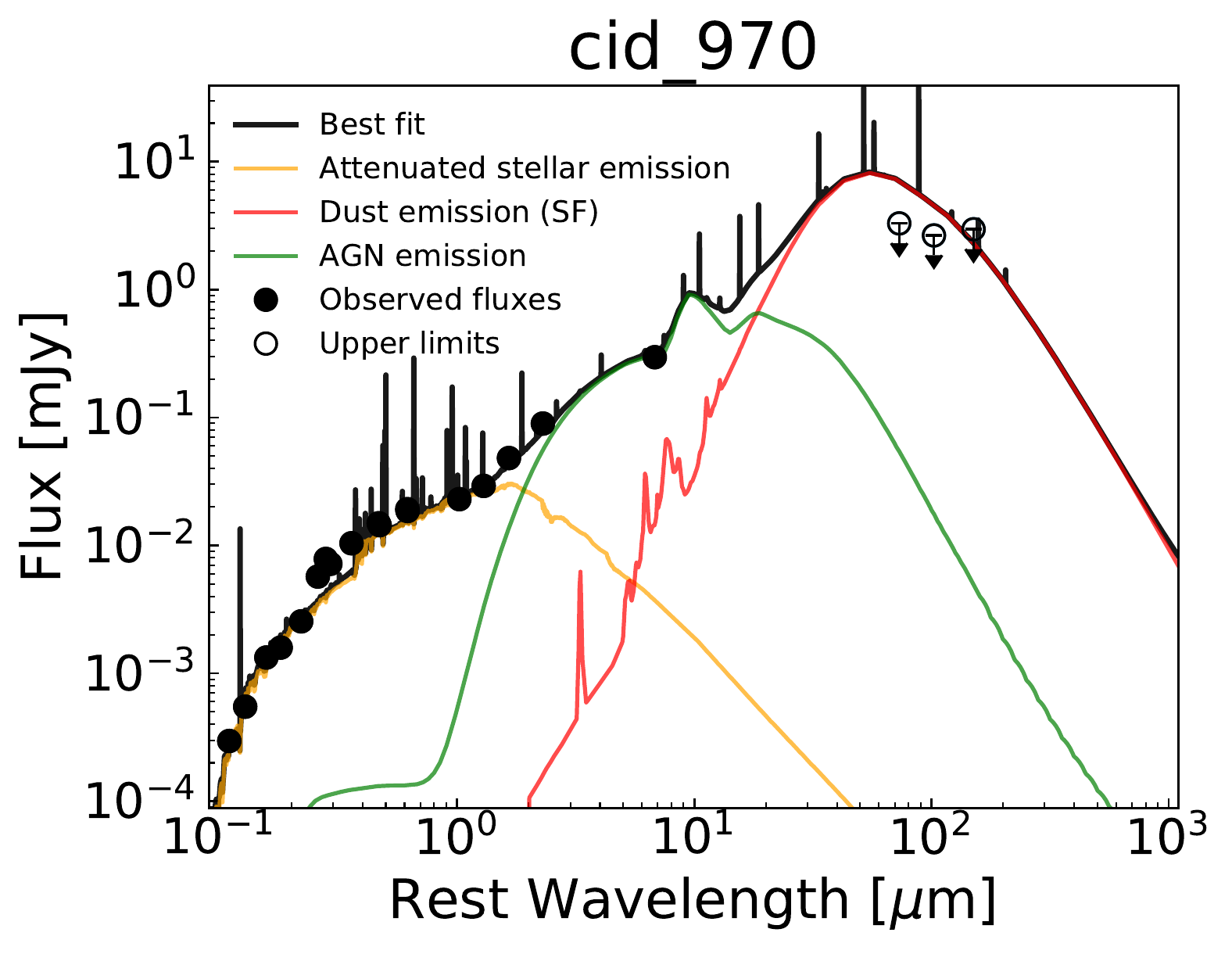}
    \caption{Rest-frame SEDs of the targets that were not presented in \citet{circosta18}. The black dots represent the observed multiwavelength photometry, while the empty dots indicate 3$\sigma$ upper limits. The black solid line is the total best-fit model, the orange curve represents the stellar emission attenuated by dust, the green template reproduces the AGN emission, the red curve accounts for dust emission heated by star formation. Emission lines in the black curves are part of the nebular emission component, included in the overall SED.}
\end{figure*}
\clearpage

\section{CO(3-2) emission}\label{sec:line_maps}
\vspace{-0.5cm}
\begin{figure*}[h!]
\centering
    \includegraphics[width=7.2cm]{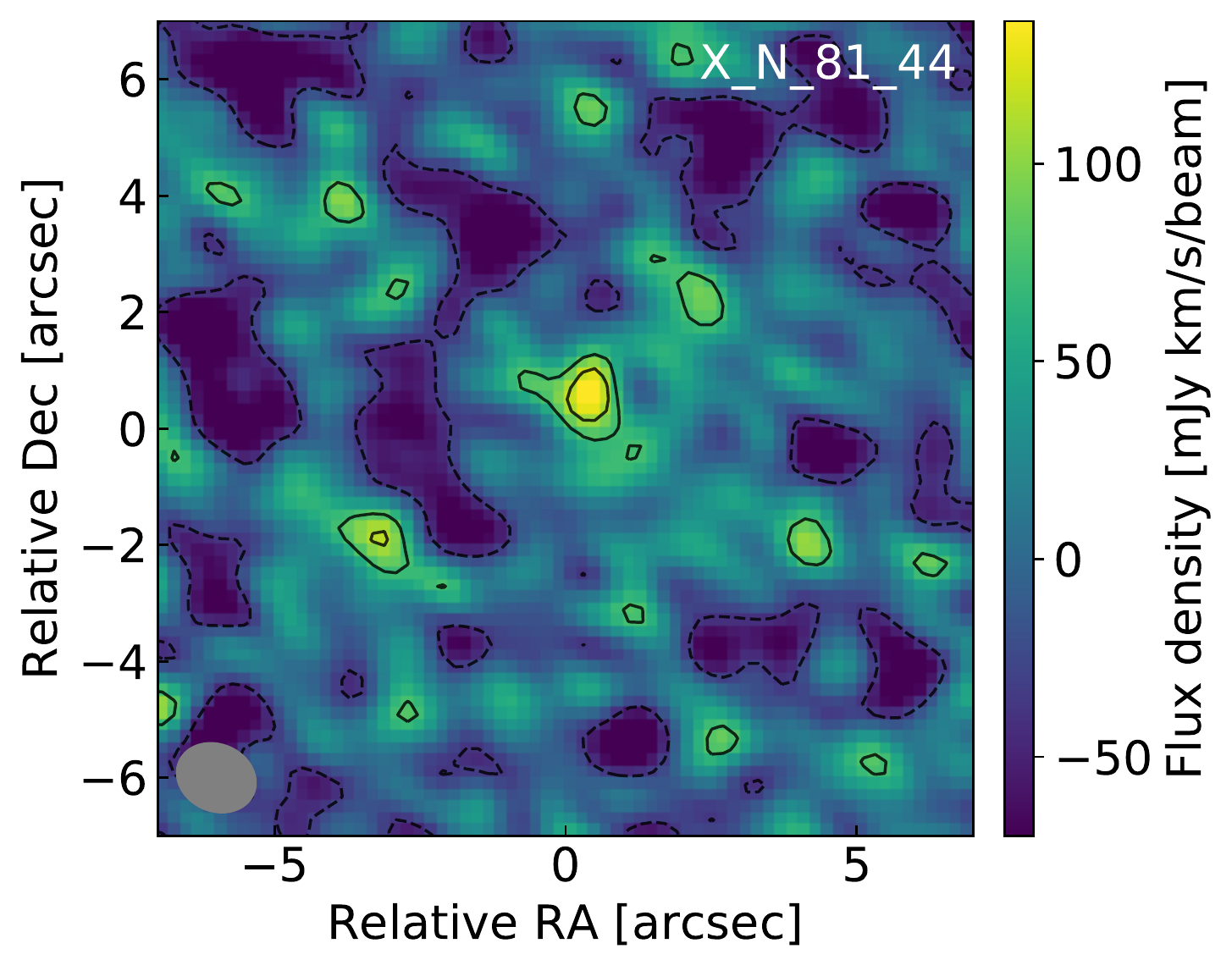} 
  	\hspace{2mm}
  	\includegraphics[width=7.8cm]{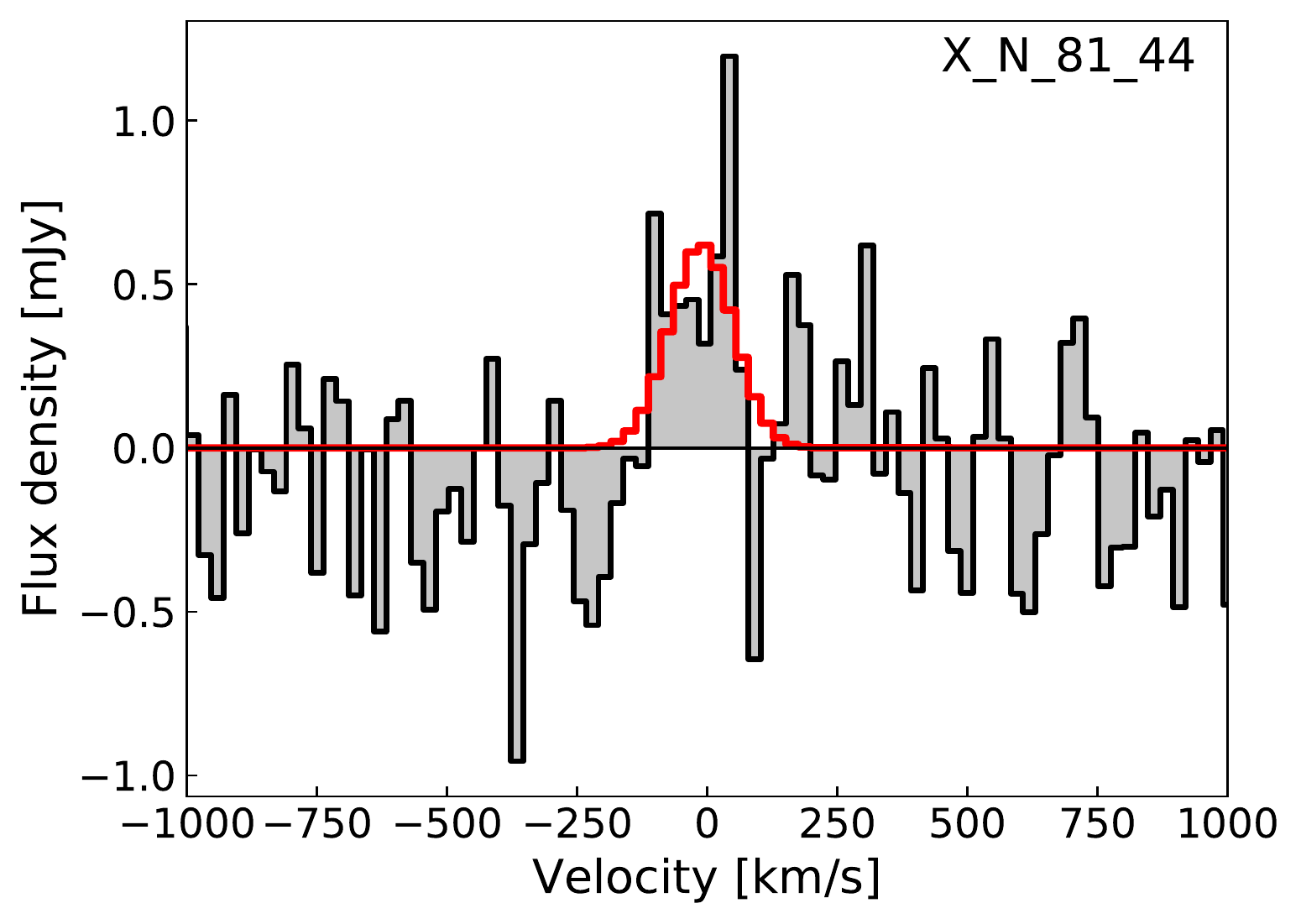} 
  	\hspace{2mm}
  	\includegraphics[width=7.2cm]{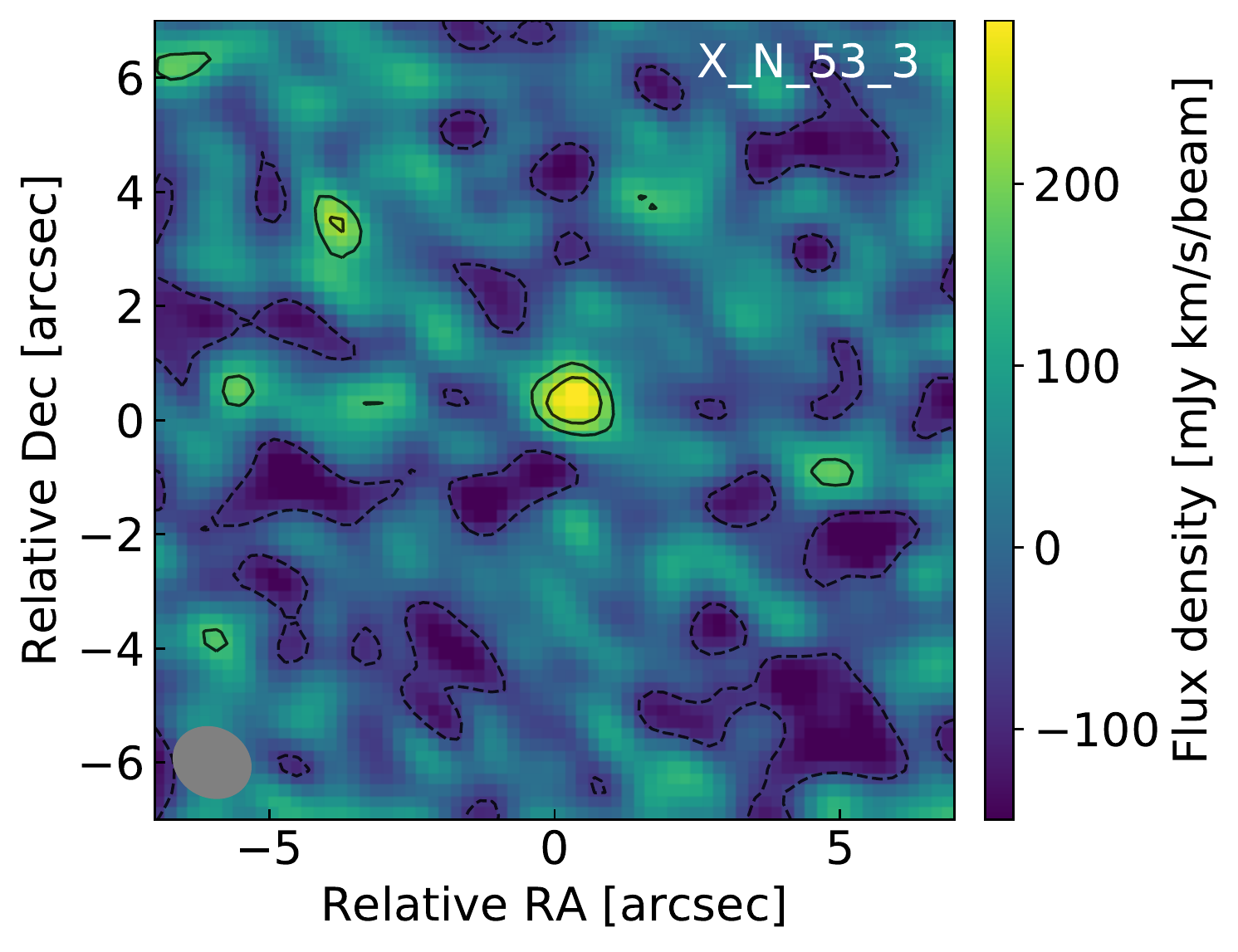} 
  	\hspace{2mm}
  	\includegraphics[width=7.8cm]{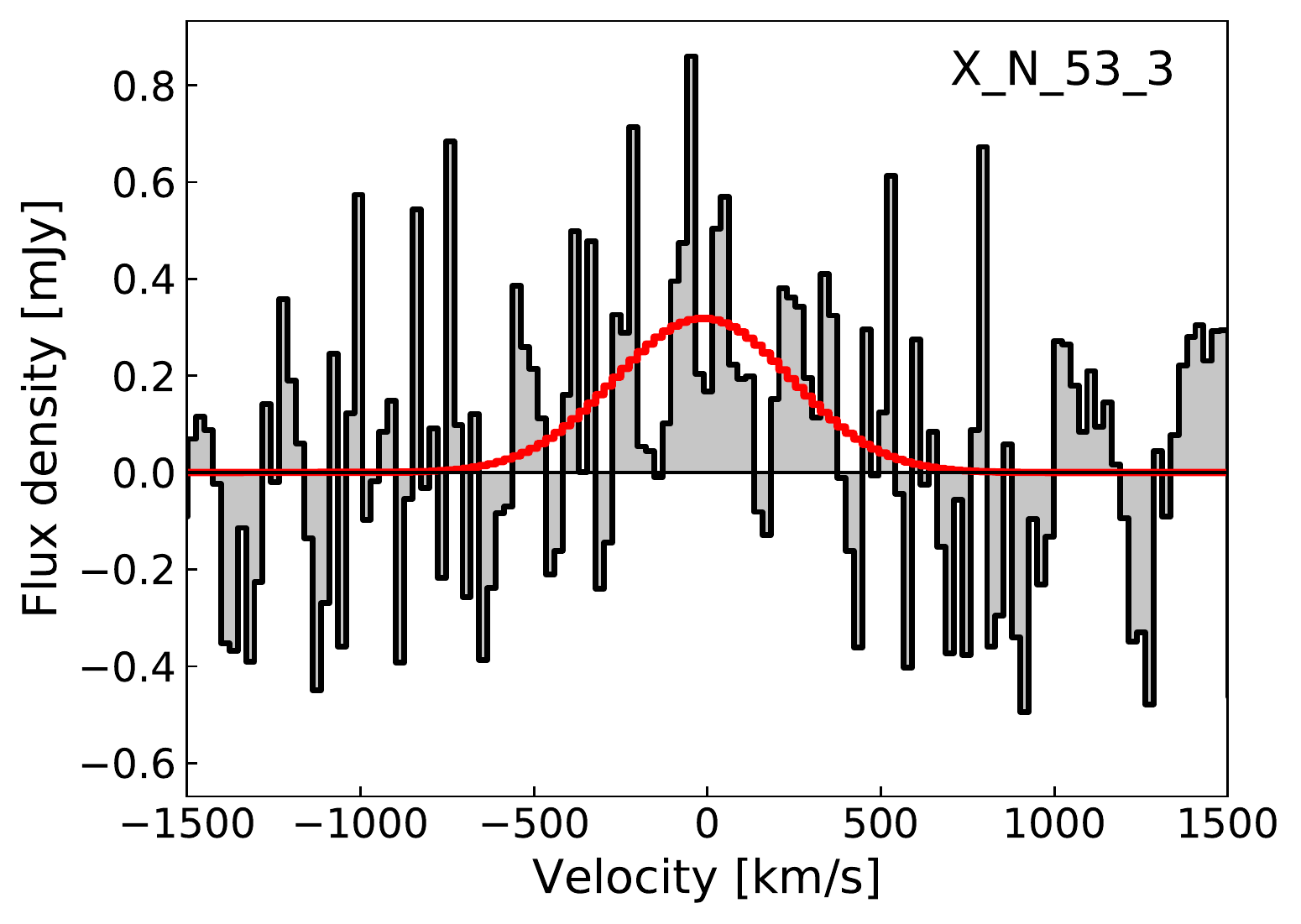} 
  	\hspace{2mm}
  	\includegraphics[width=7.2cm]{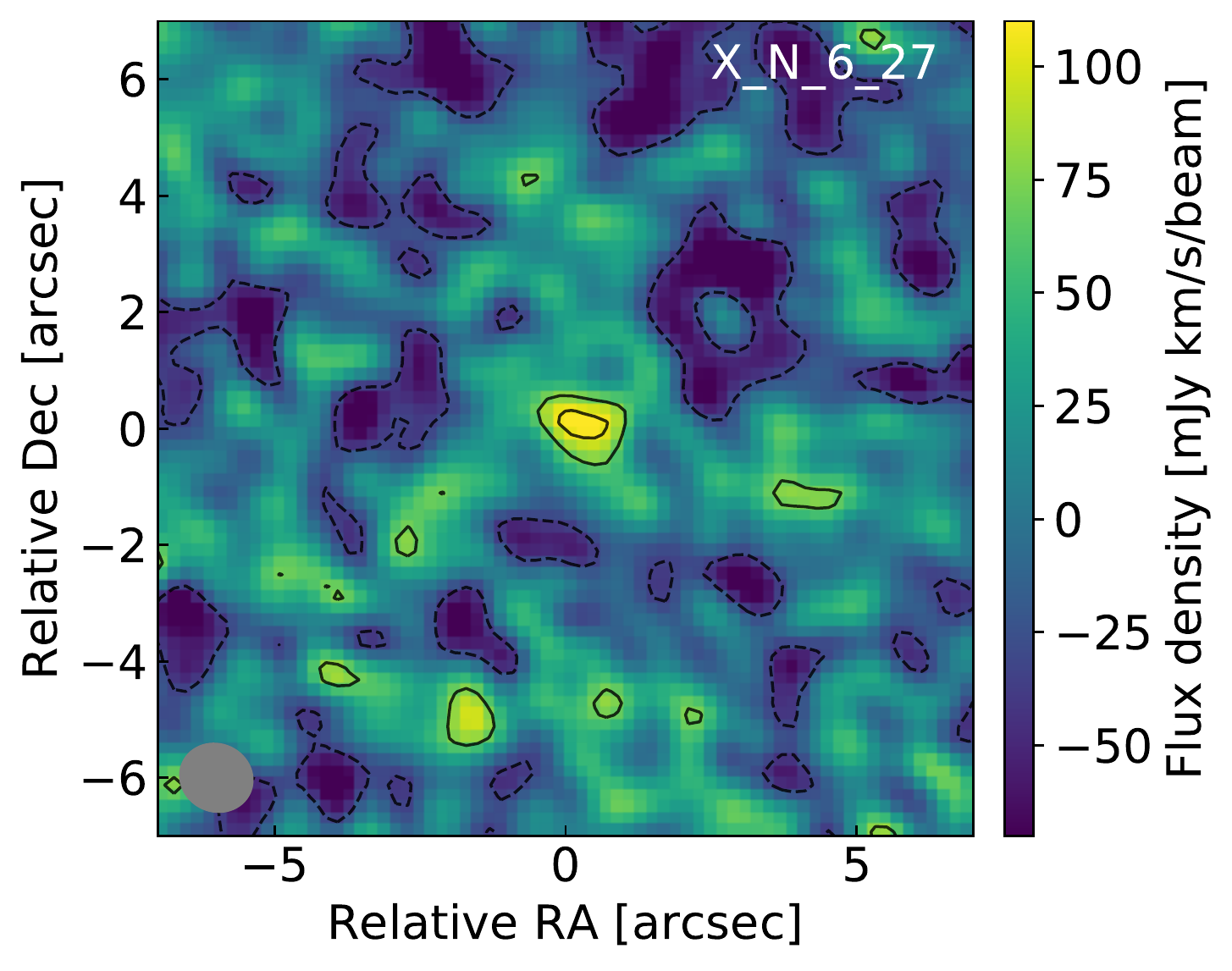} 
  	\hspace{2mm}
  	\includegraphics[width=7.8cm]{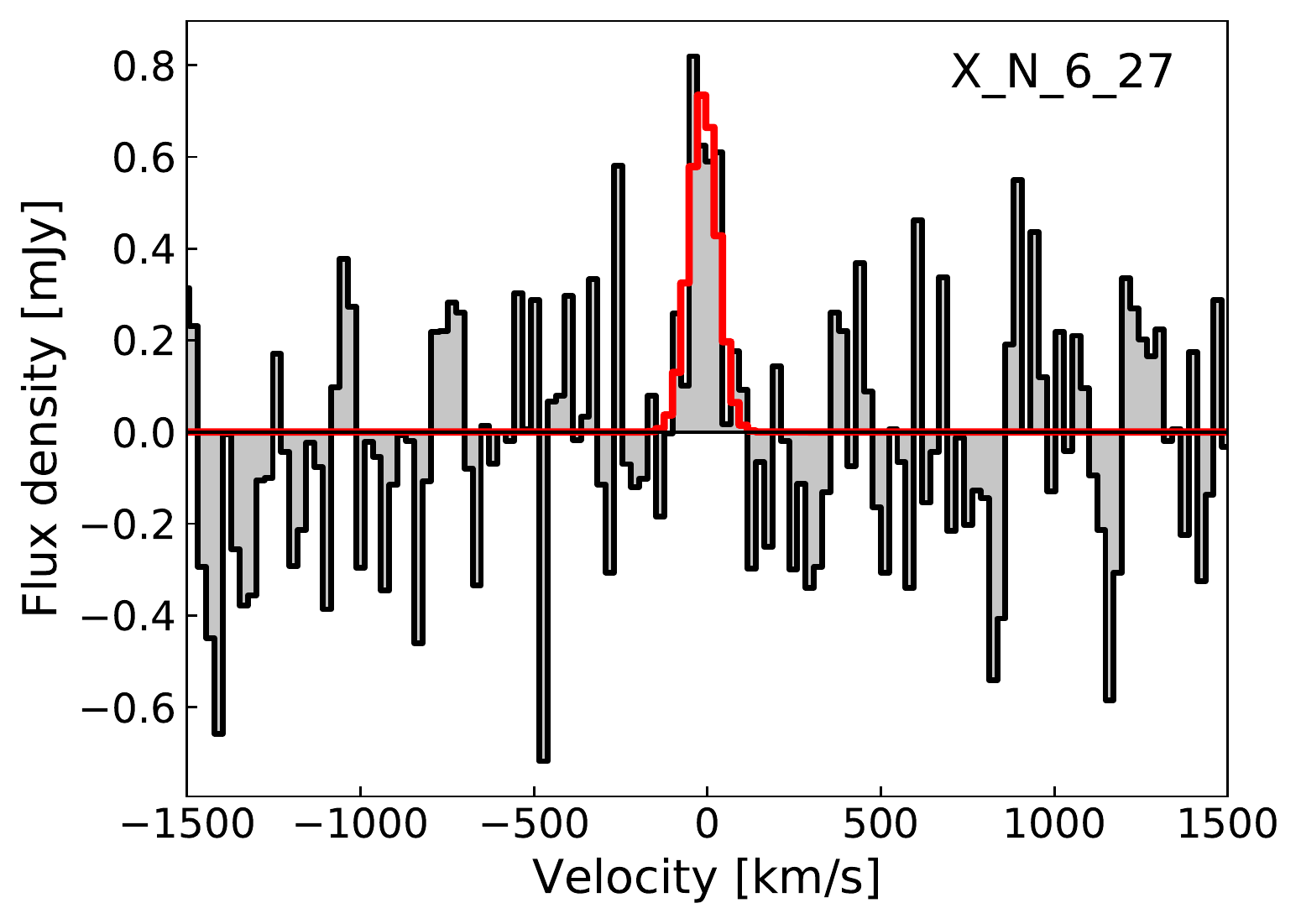} 
  	\hspace{2mm}
  	\includegraphics[width=7.2cm]{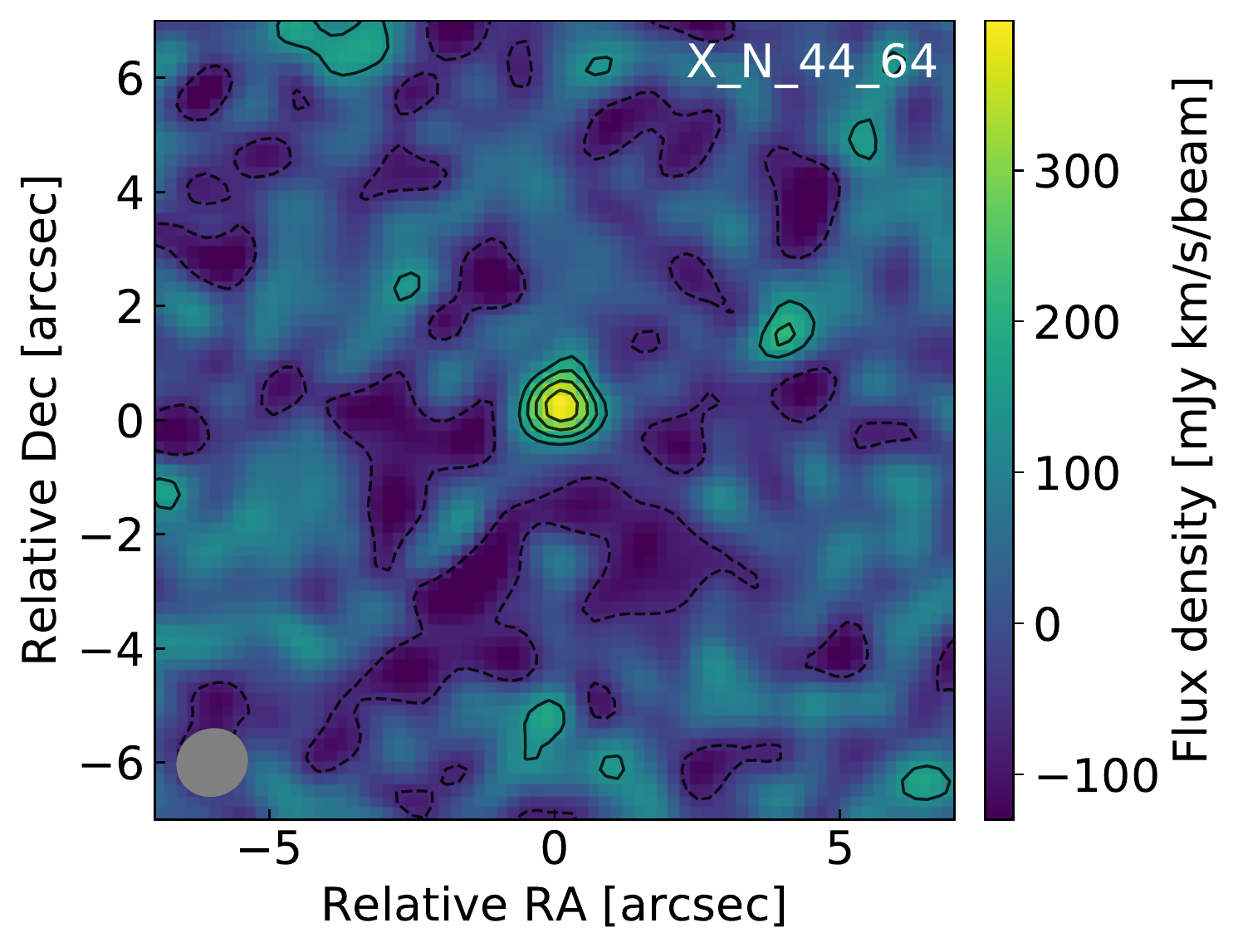} 
  	\hspace{2mm}
  	\includegraphics[width=7.8cm]{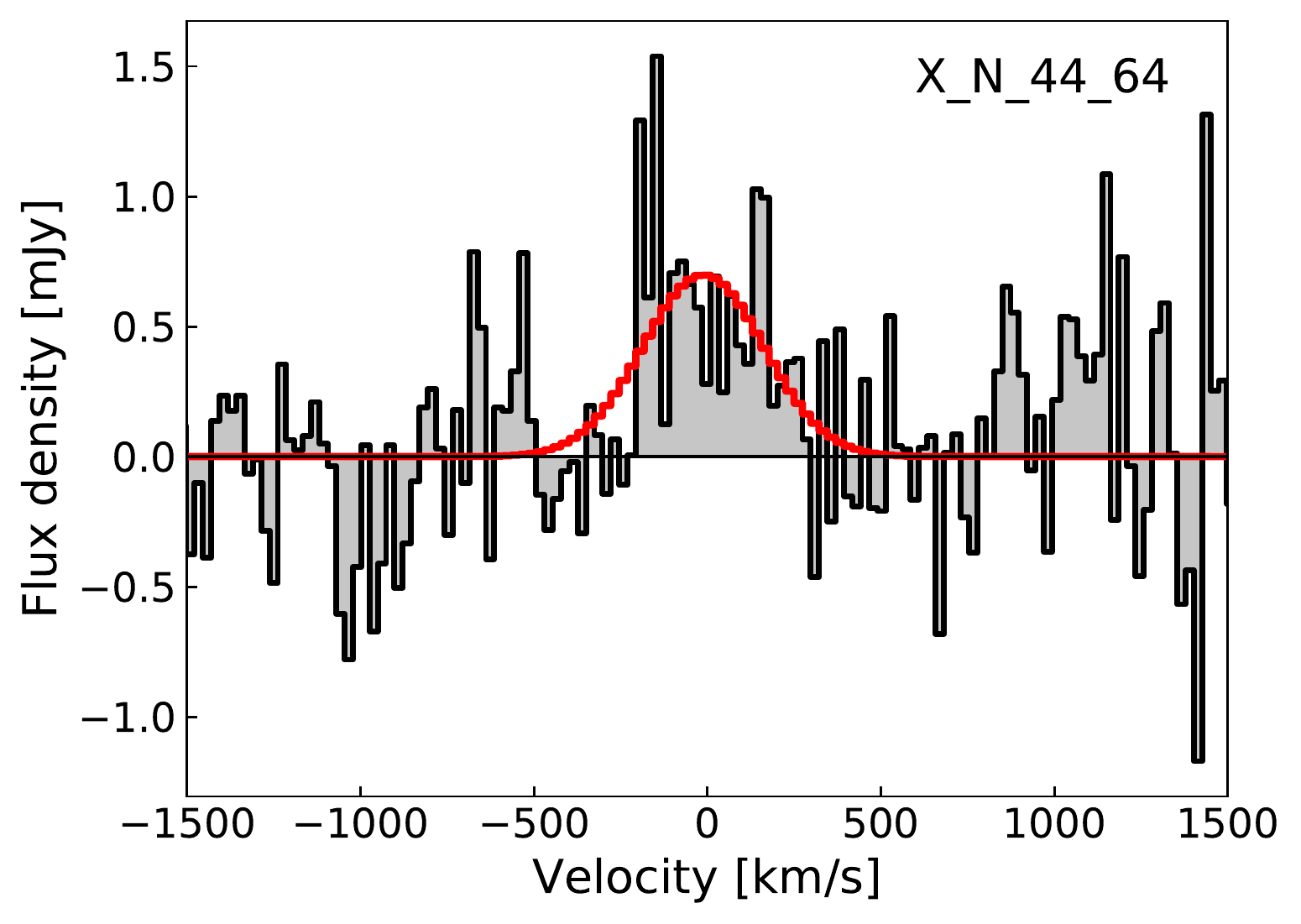} 
  	\hspace{2mm}
    \caption{CO line emission maps (\textit{left}) and spectra (\textit{right}) extracted from the region above 2$\sigma$ significance. \textit{Left}: Black contours in steps of 1$\sigma$, starting from 2$\sigma$. Dashed contours correspond to $-$1$\sigma$. The beam of each observation is shown as a gray ellipse on the bottom-left of the maps. \textit{Right:} Observed spectrum plotted in black, Gaussian best-fit model depicted in red.
    }
\end{figure*}

\begin{figure*}[h!]
\centering
    \includegraphics[width=7.2cm]{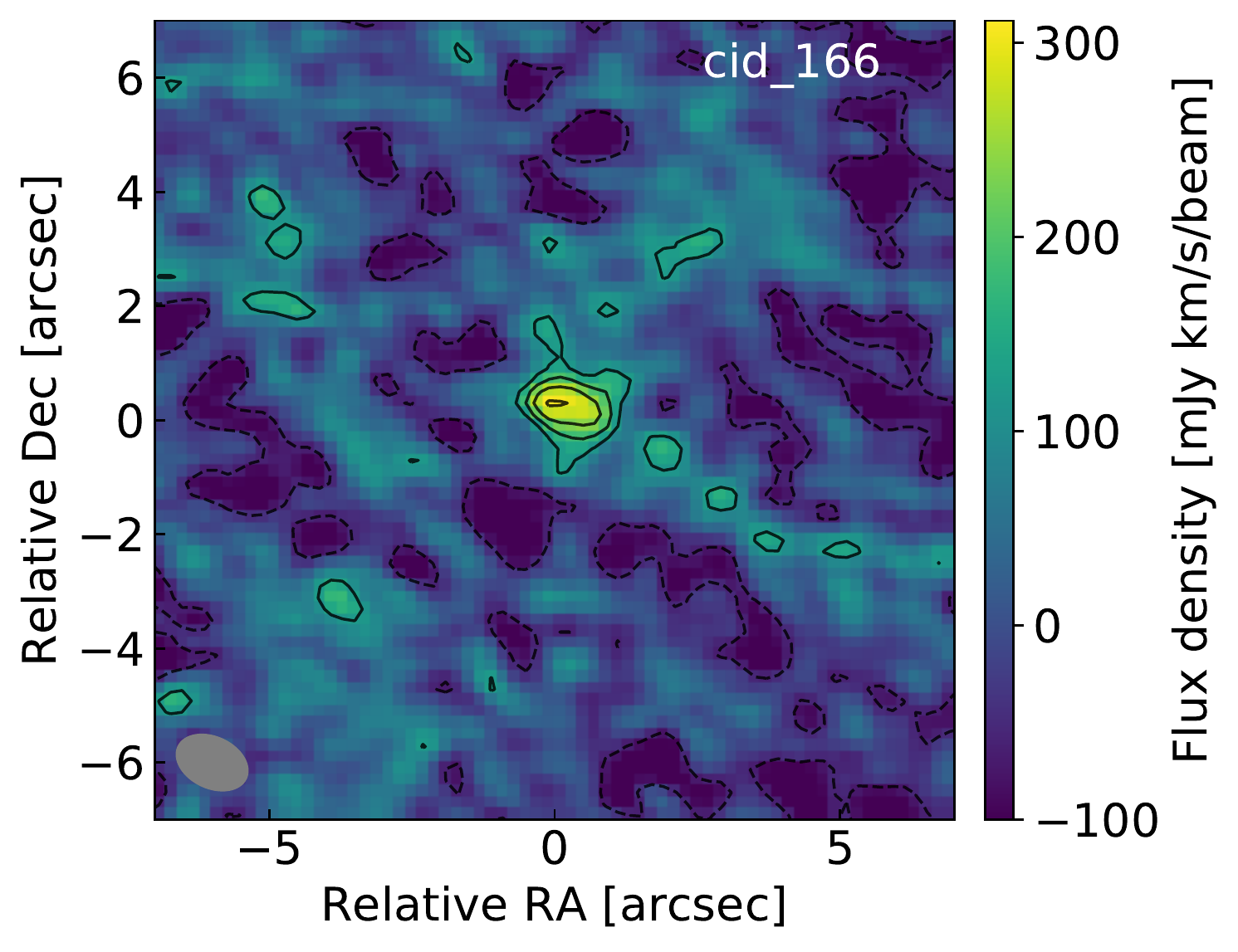} 
  	\hspace{2mm}
  	\includegraphics[width=7.8cm]{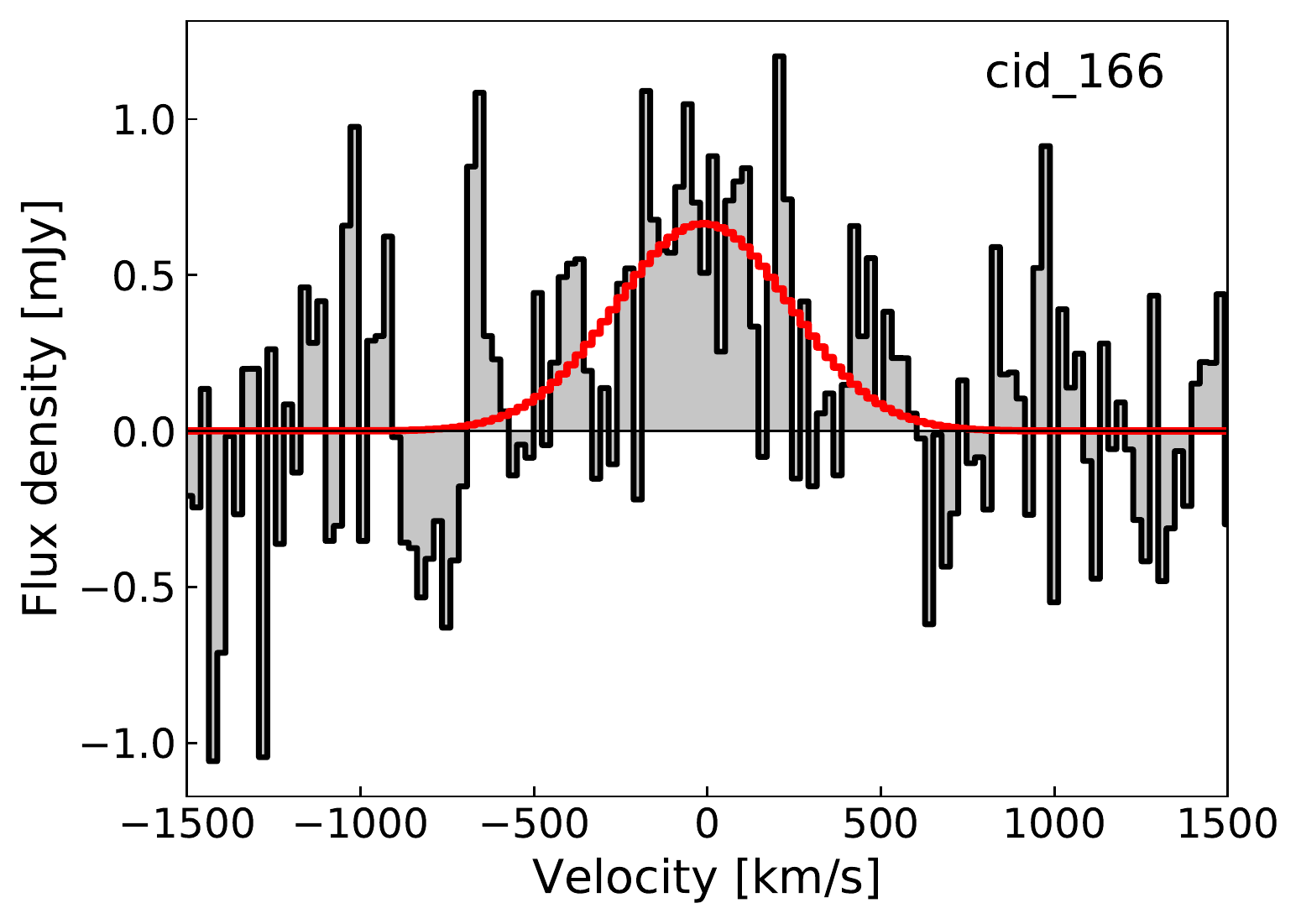} 
  	\hspace{2mm}
    \includegraphics[width=7.2cm]{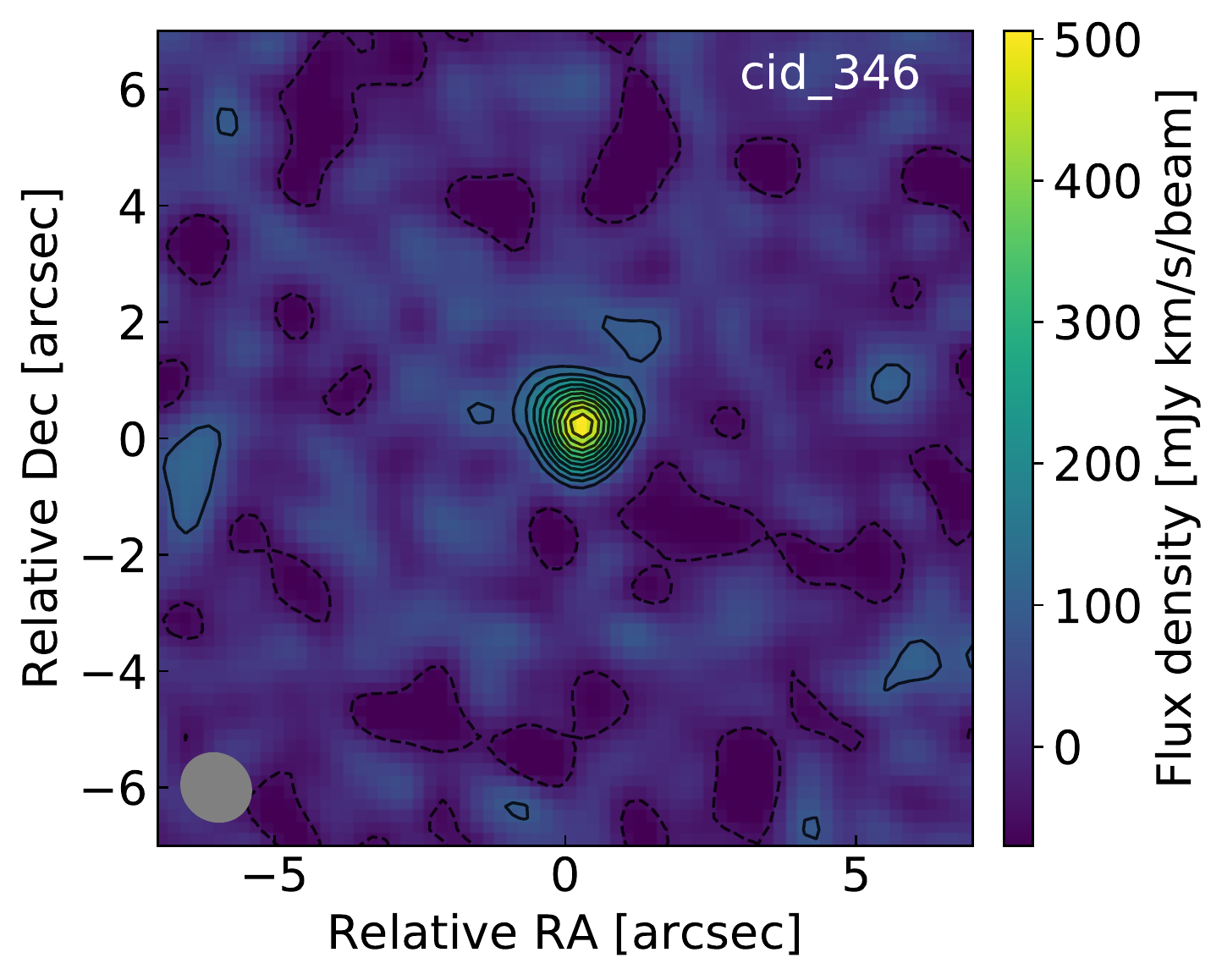} 
  	\hspace{2mm}
  	\includegraphics[width=7.8cm]{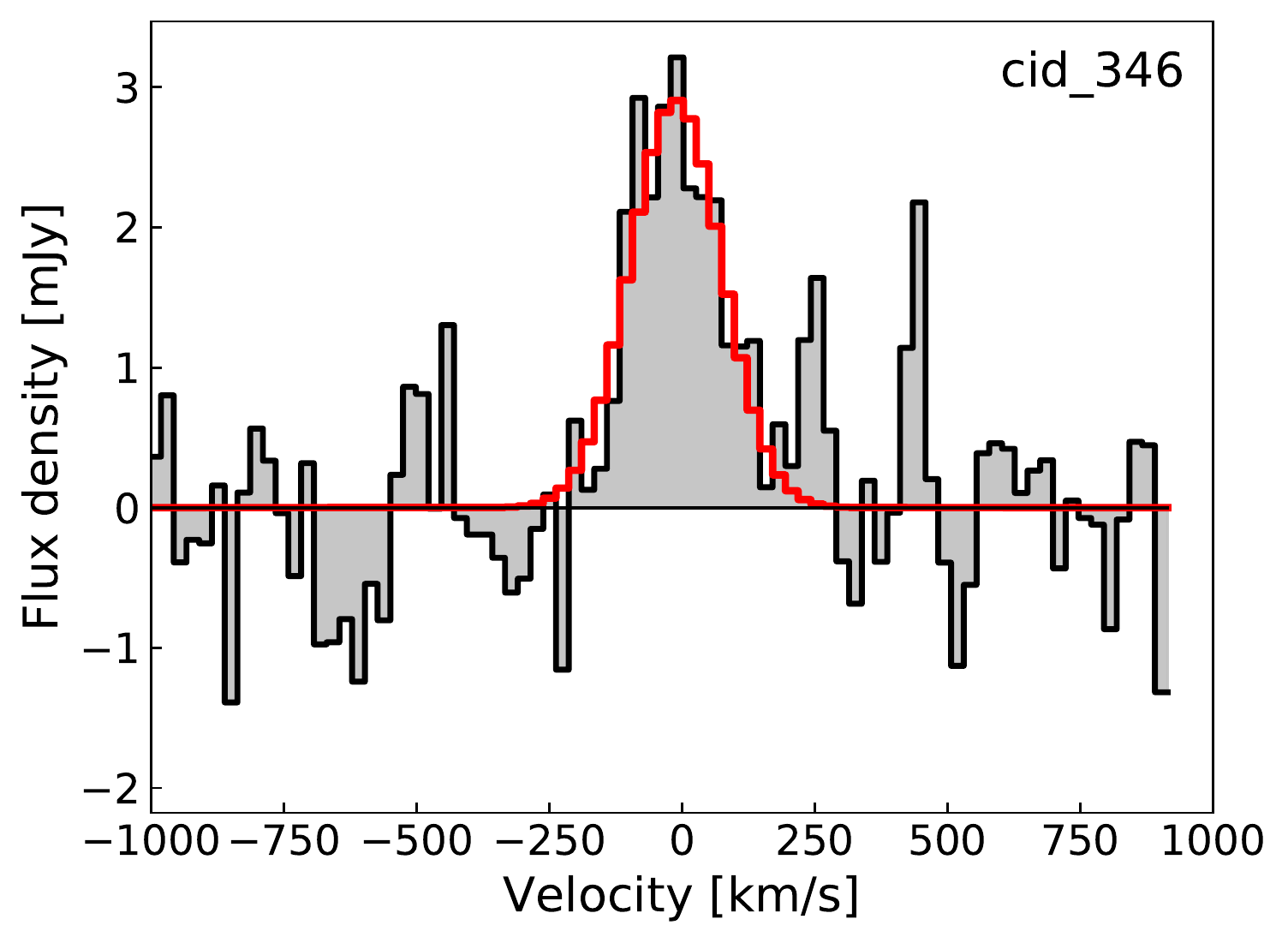} 
  	\hspace{2mm}
  	\includegraphics[width=7.2cm]{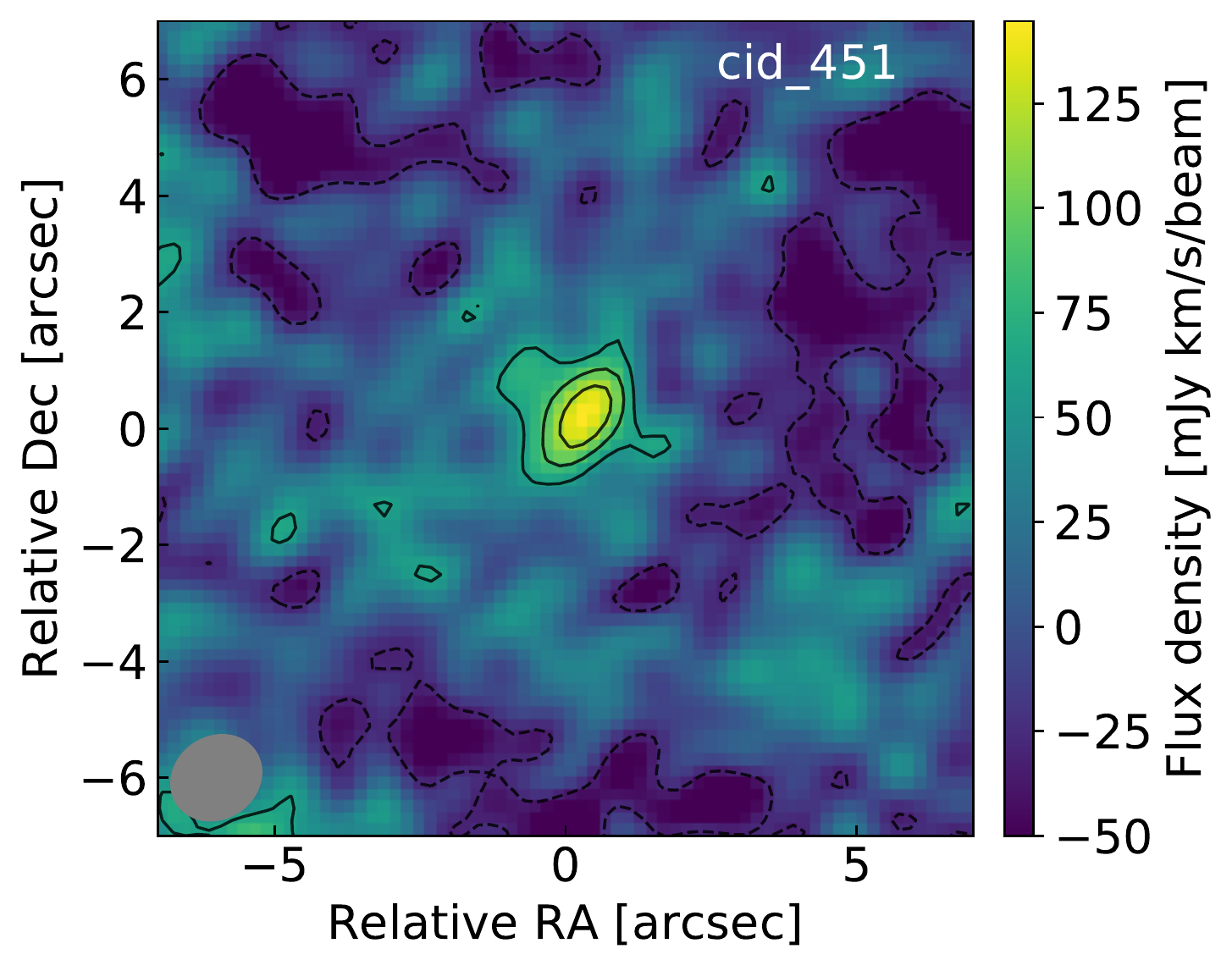} 
  	\hspace{2mm}
  	\includegraphics[width=7.8cm]{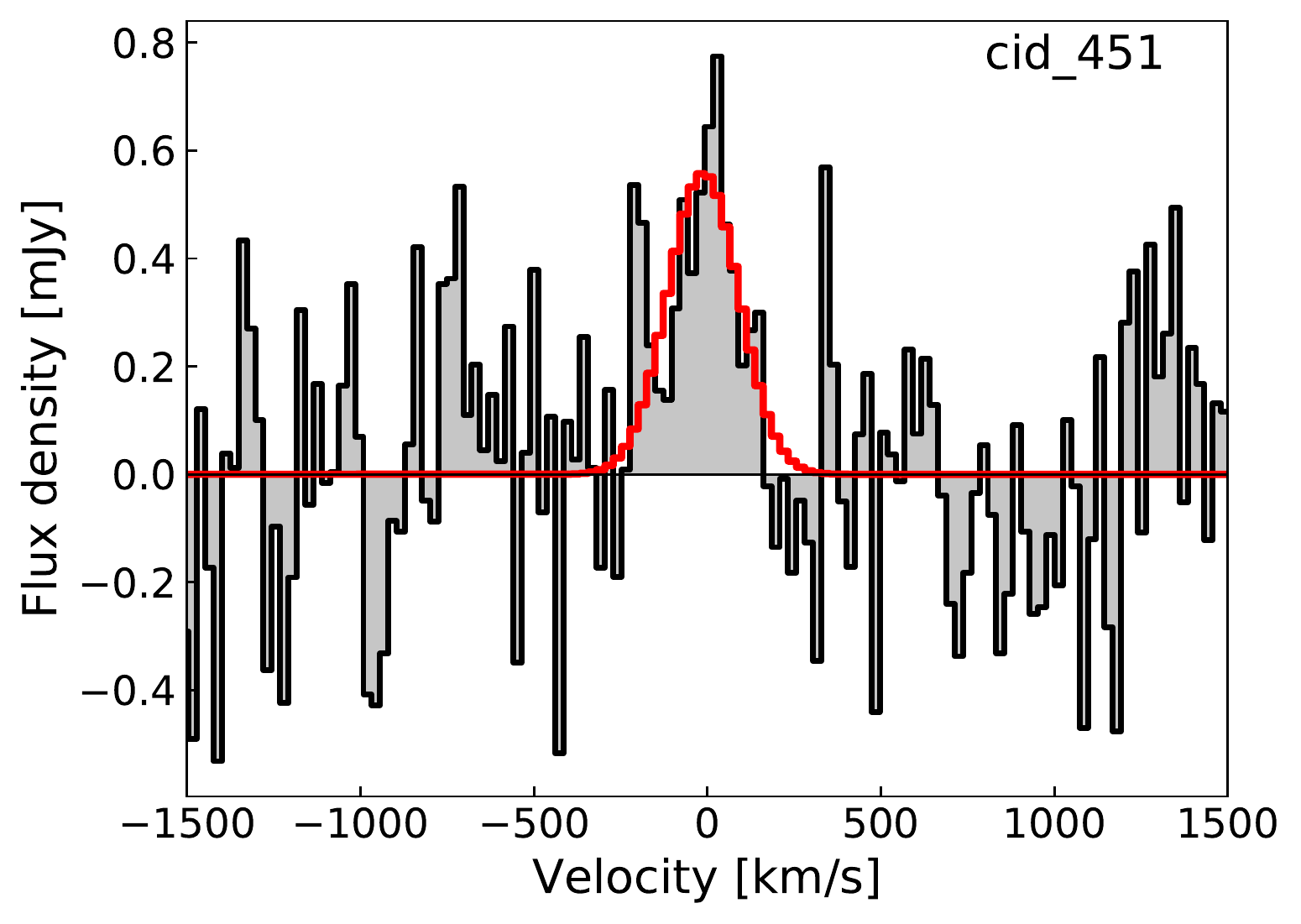} 
  	\hspace{2mm}
  	\includegraphics[width=7.2cm]{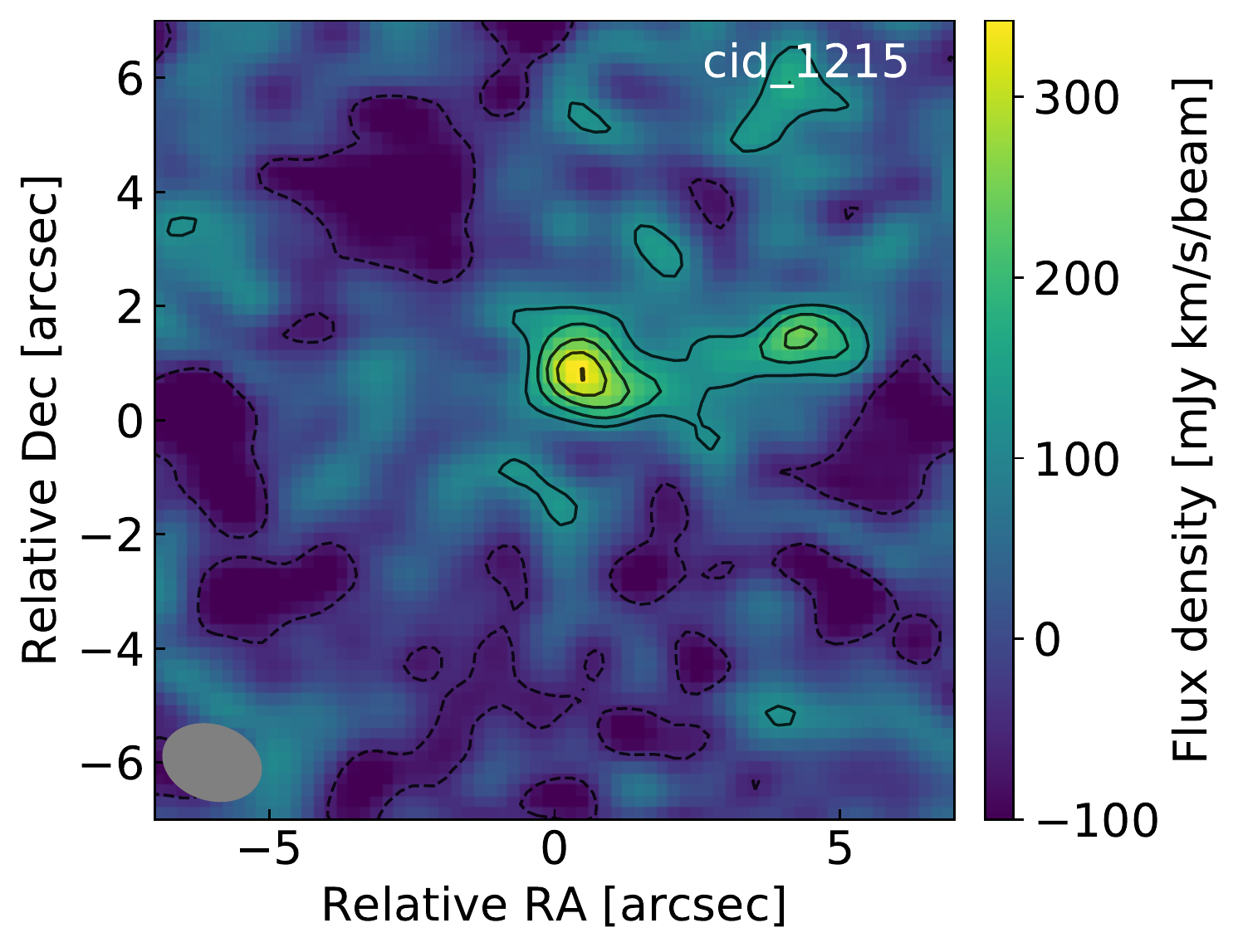} 
  	\hspace{2mm}
  	\includegraphics[width=7.8cm]{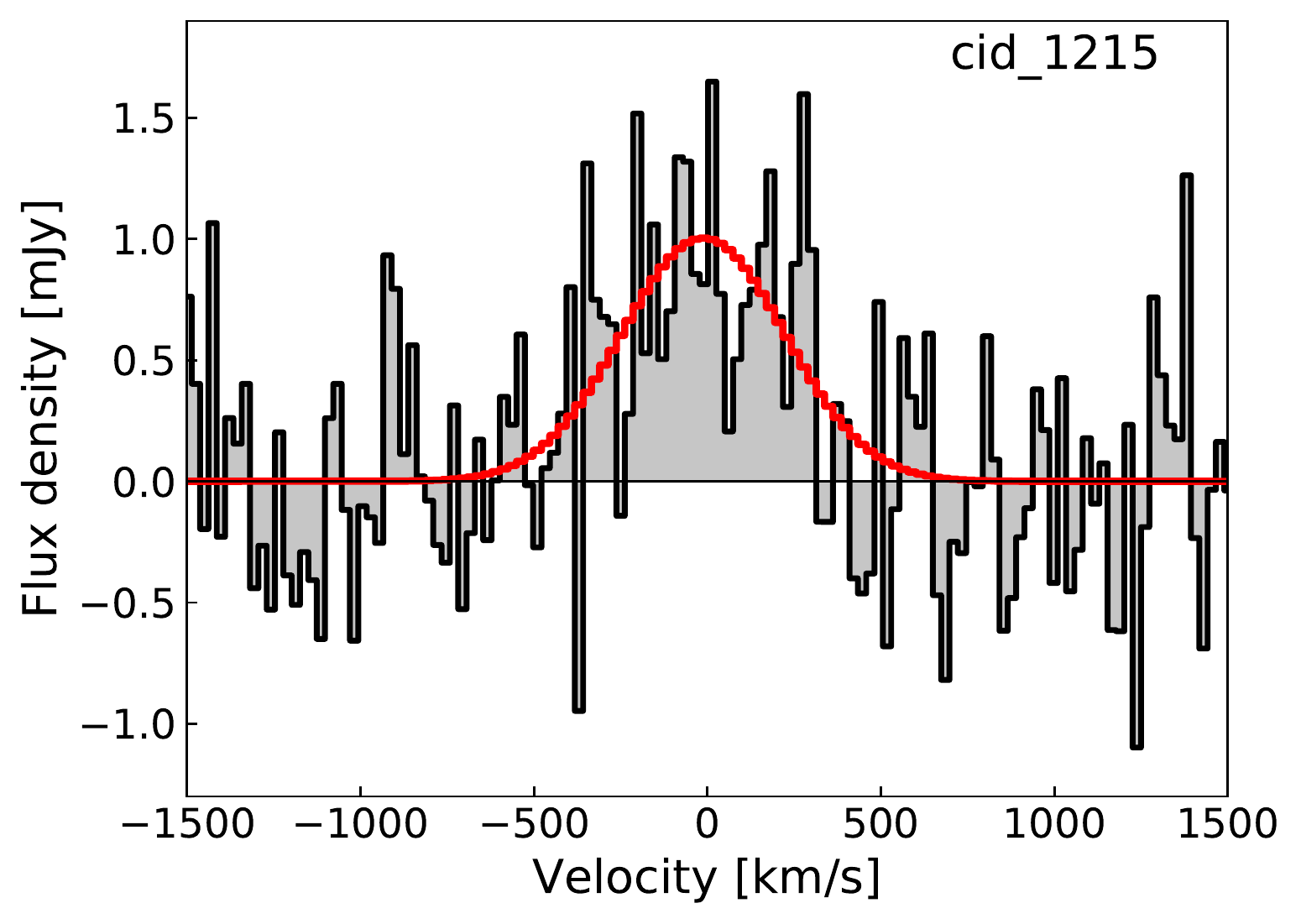} 
  	\caption{CO line emission maps (\textit{left}) and spectra (\textit{right}) extracted from the region above 2$\sigma$ significance. \textit{Left}: Black contours in steps of 1$\sigma$, starting from 2$\sigma$. Dashed contours correspond to $-$1$\sigma$. The beam of each observation is shown as a gray ellipse on the bottom-left of the maps. \textit{Right:} Observed spectrum plotted in black, Gaussian best-fit model depicted in red.
    }
\end{figure*}

\begin{figure*}[h!]
\centering
    \includegraphics[width=7.2cm]{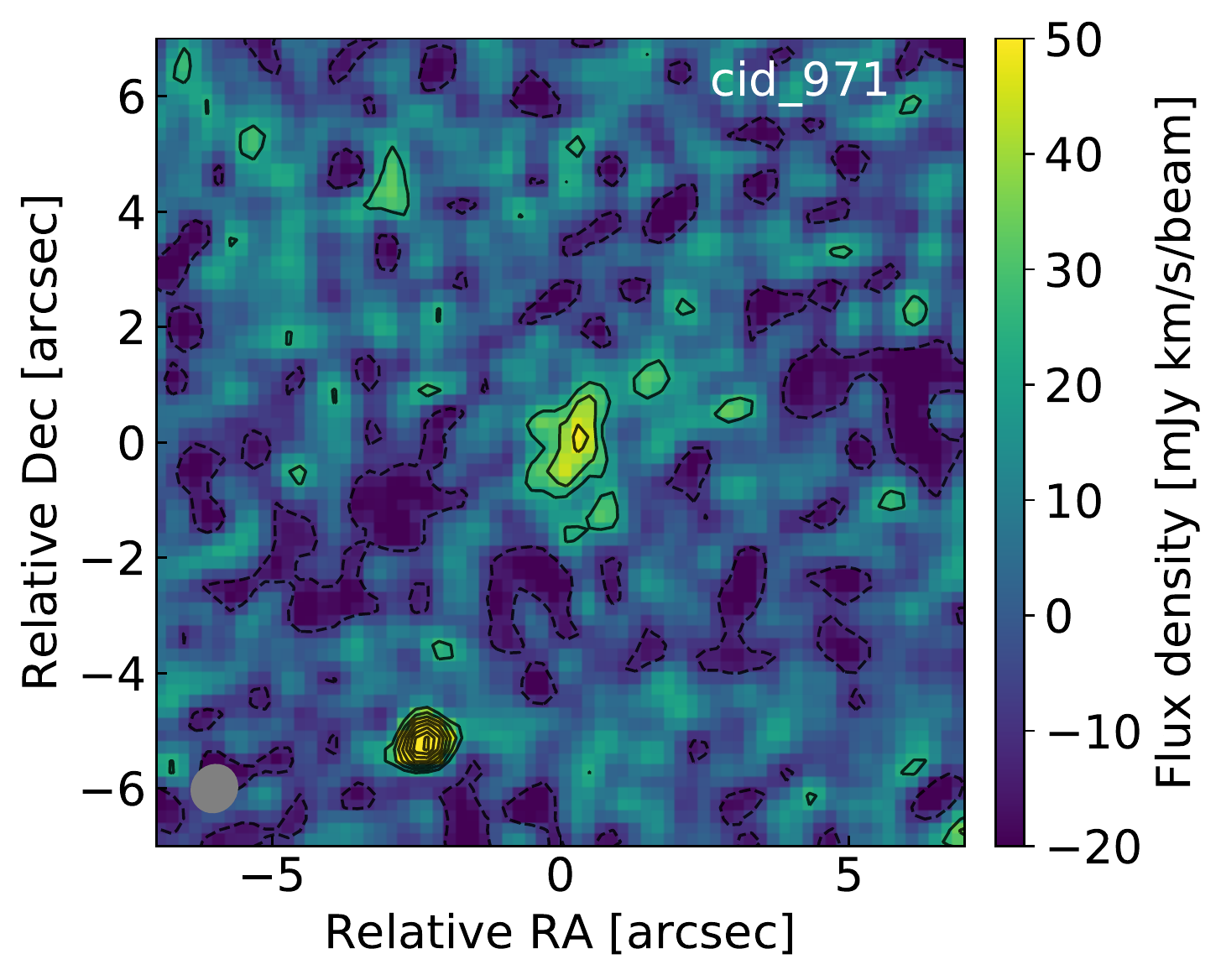} 
    \hspace{2mm}
    \includegraphics[width=7.8cm]{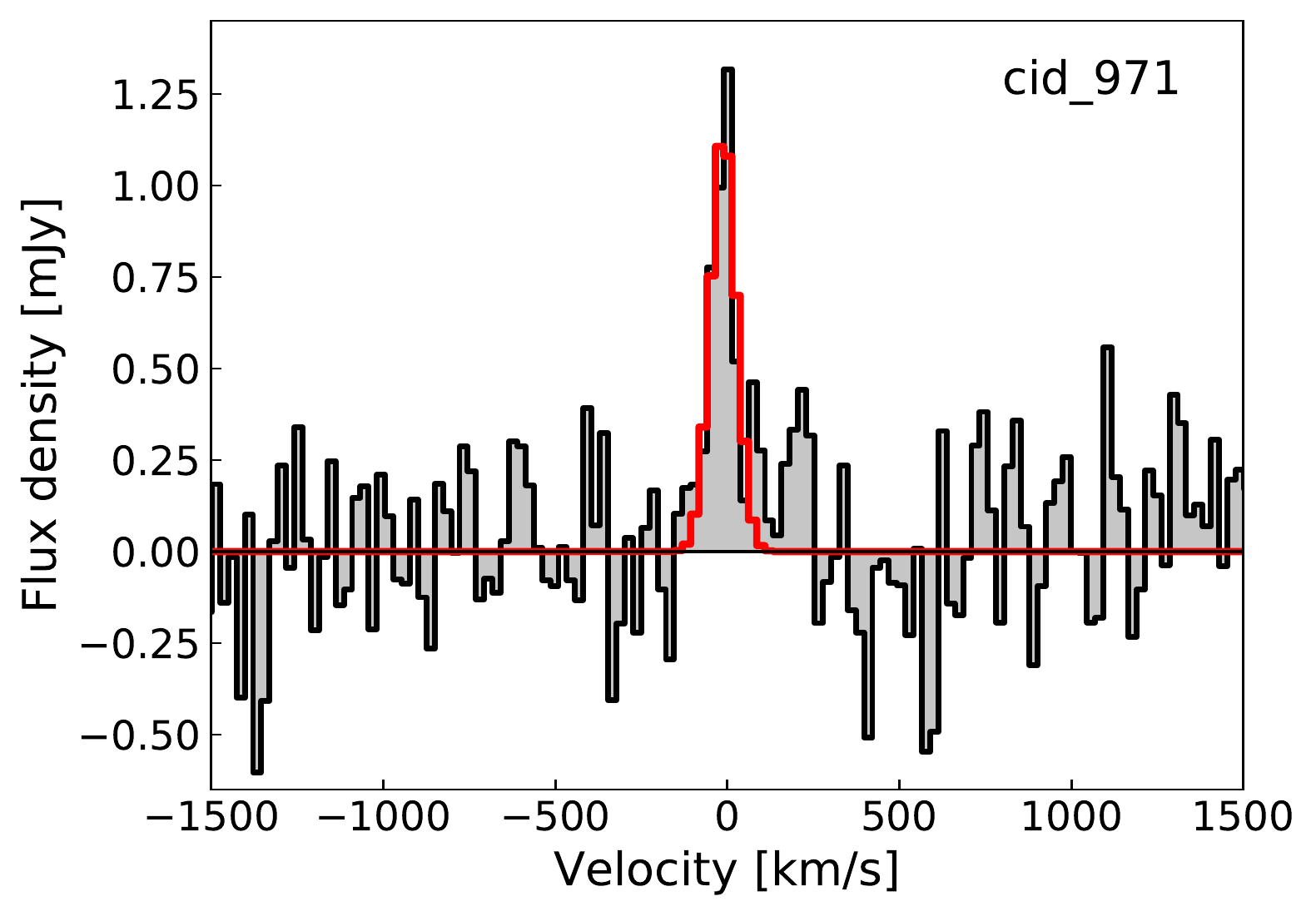}
  	\hspace{2mm}
    \includegraphics[width=7.2cm]{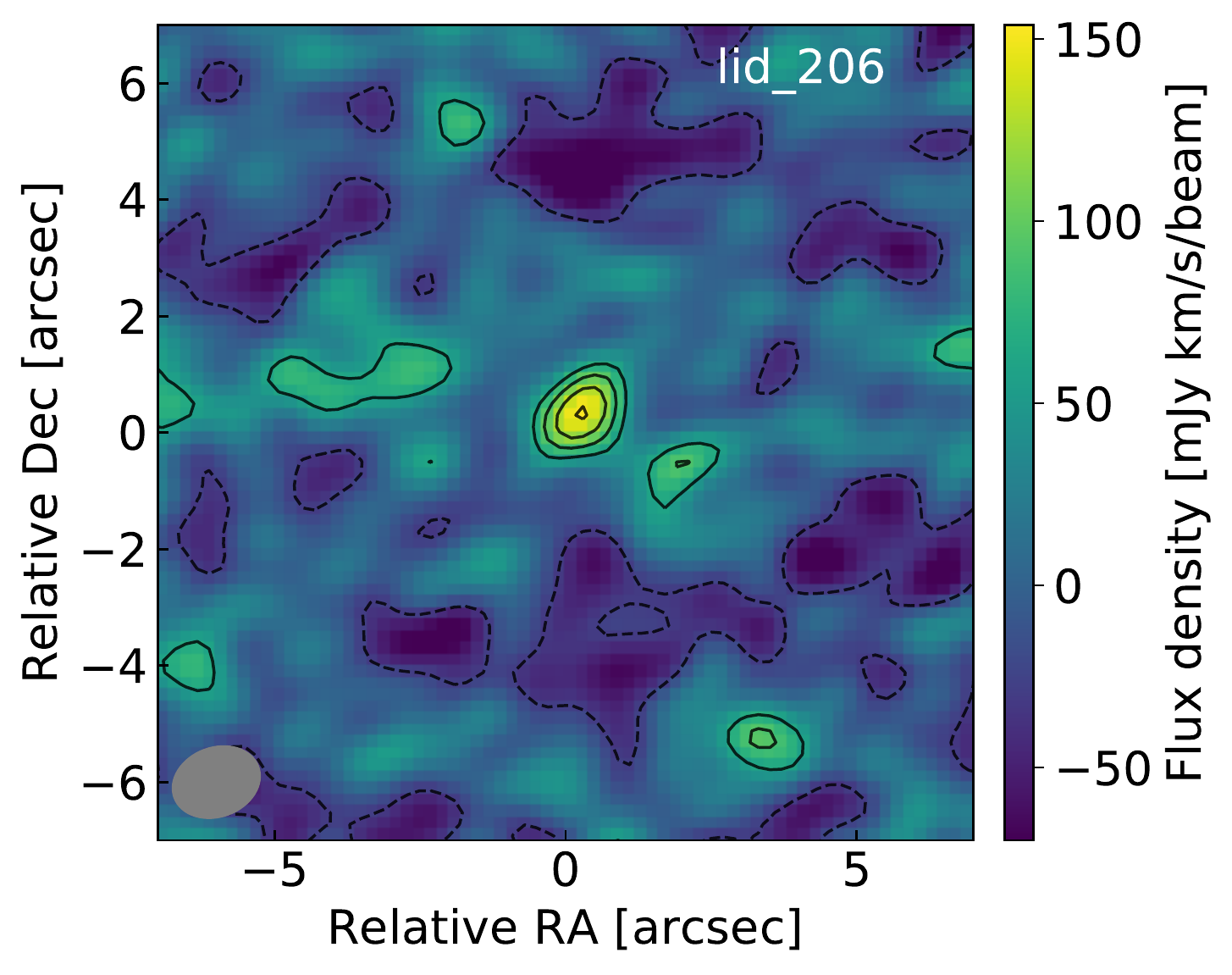} 
  	\hspace{2mm}
  	\includegraphics[width=7.8cm]{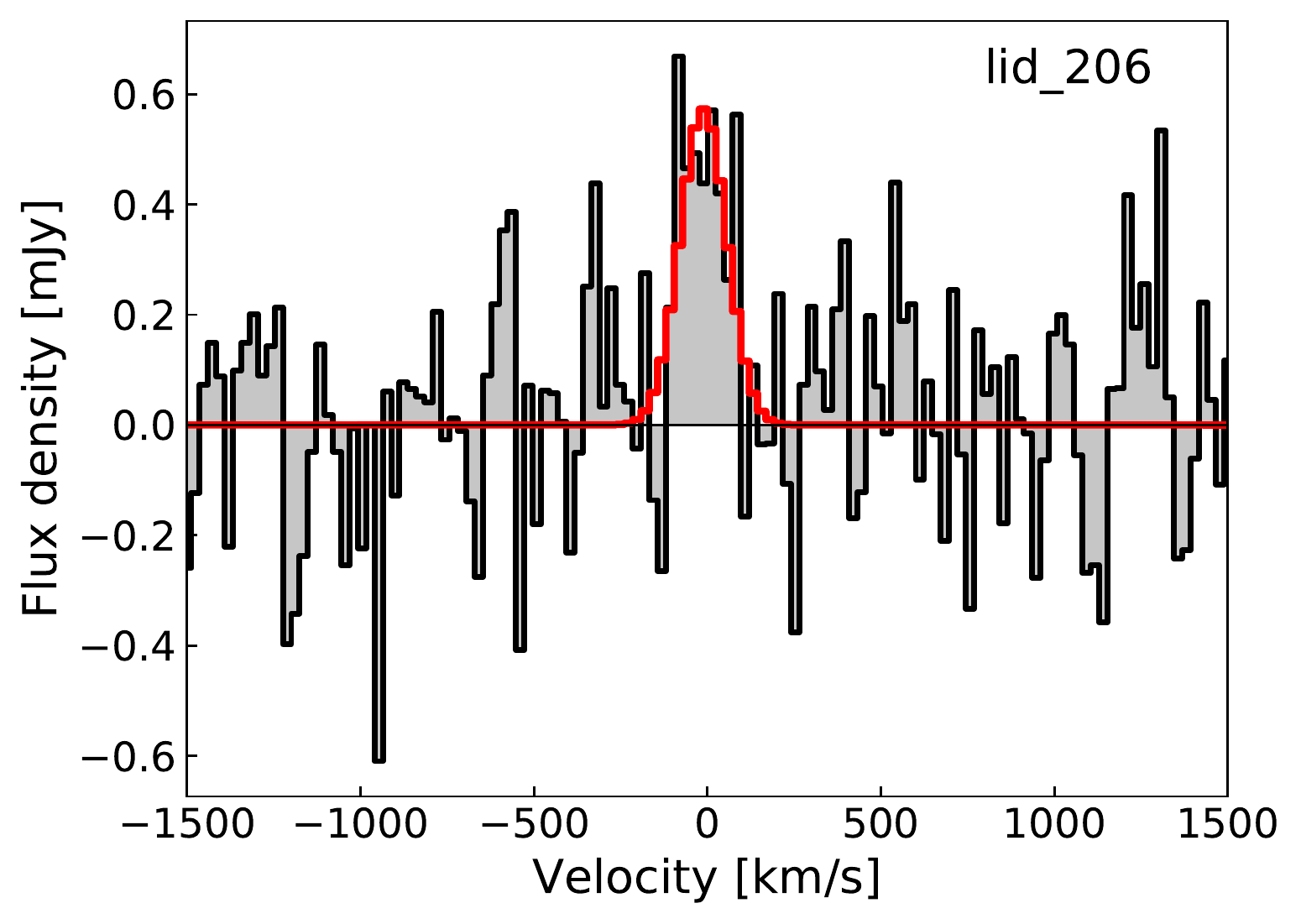}
  	\hspace{2mm}
  	\includegraphics[width=7.2cm]{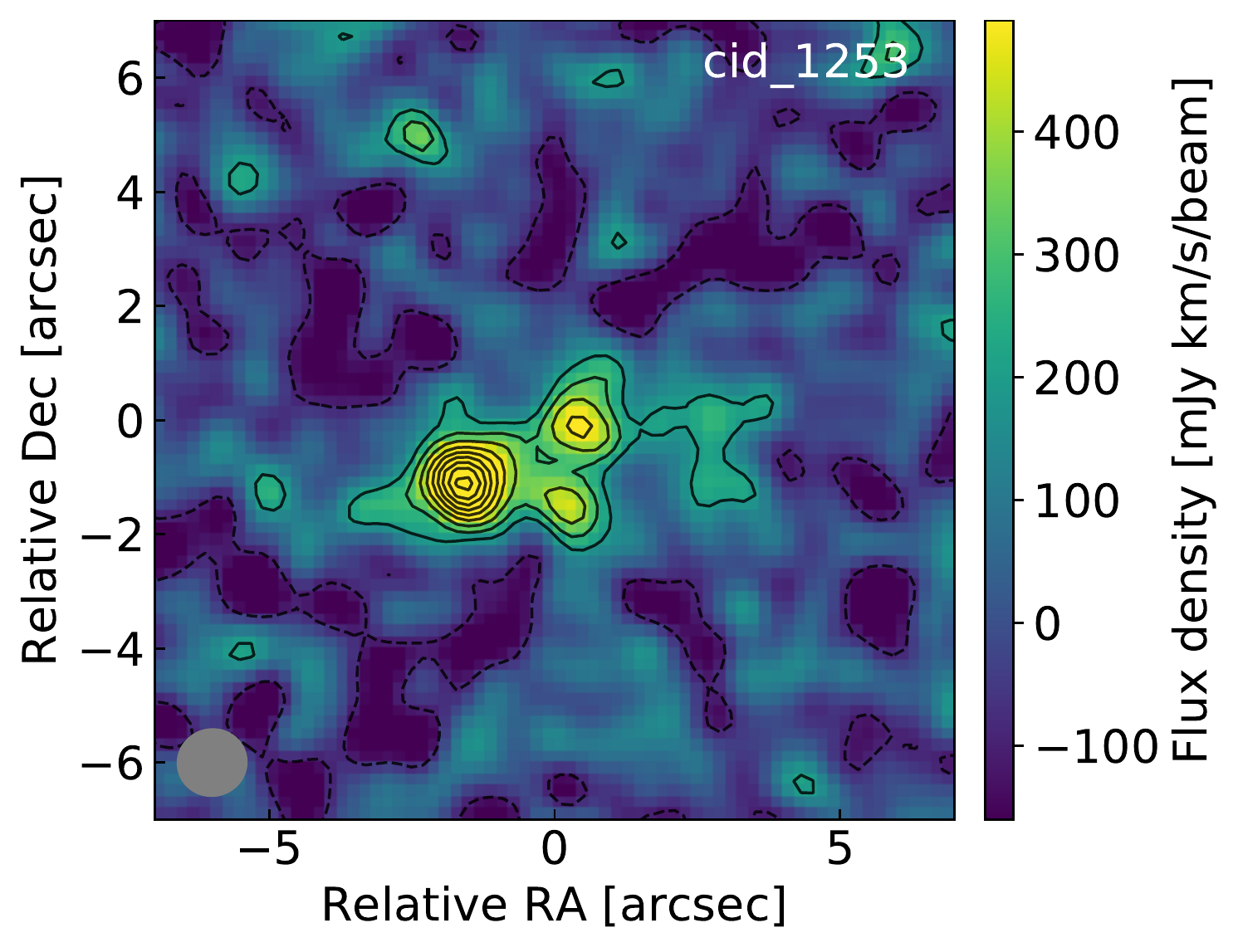} 
  	\hspace{2mm}
  	\includegraphics[width=7.8cm]{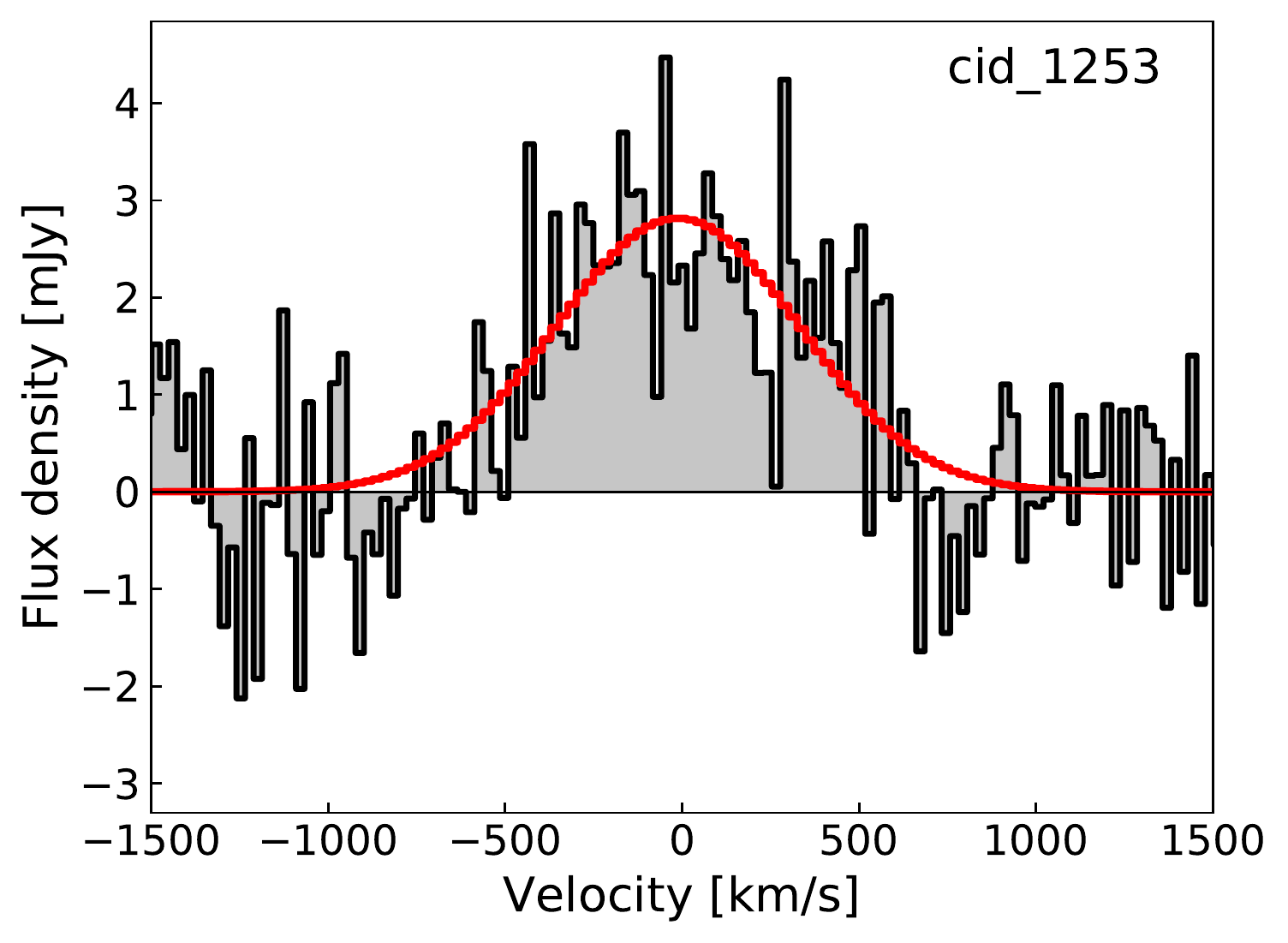}
  	\hspace{2mm}
  	\caption{CO line emission maps (\textit{left}) and spectra (\textit{right}) extracted from the region above 2$\sigma$ significance. \textit{Left}: Black contours in steps of 1$\sigma$, starting from 2$\sigma$. Dashed contours correspond to $-$1$\sigma$. The beam of each observation is shown as a gray ellipse on the bottom-left of the maps. \textit{Right:} Observed spectrum plotted in black, Gaussian best-fit model depicted in red.
    }
\end{figure*}
\clearpage

\section{Continuum emission}\label{sec:cont_maps}

\begin{figure*}[h!]
\centering
  	\includegraphics[width=7.2cm]{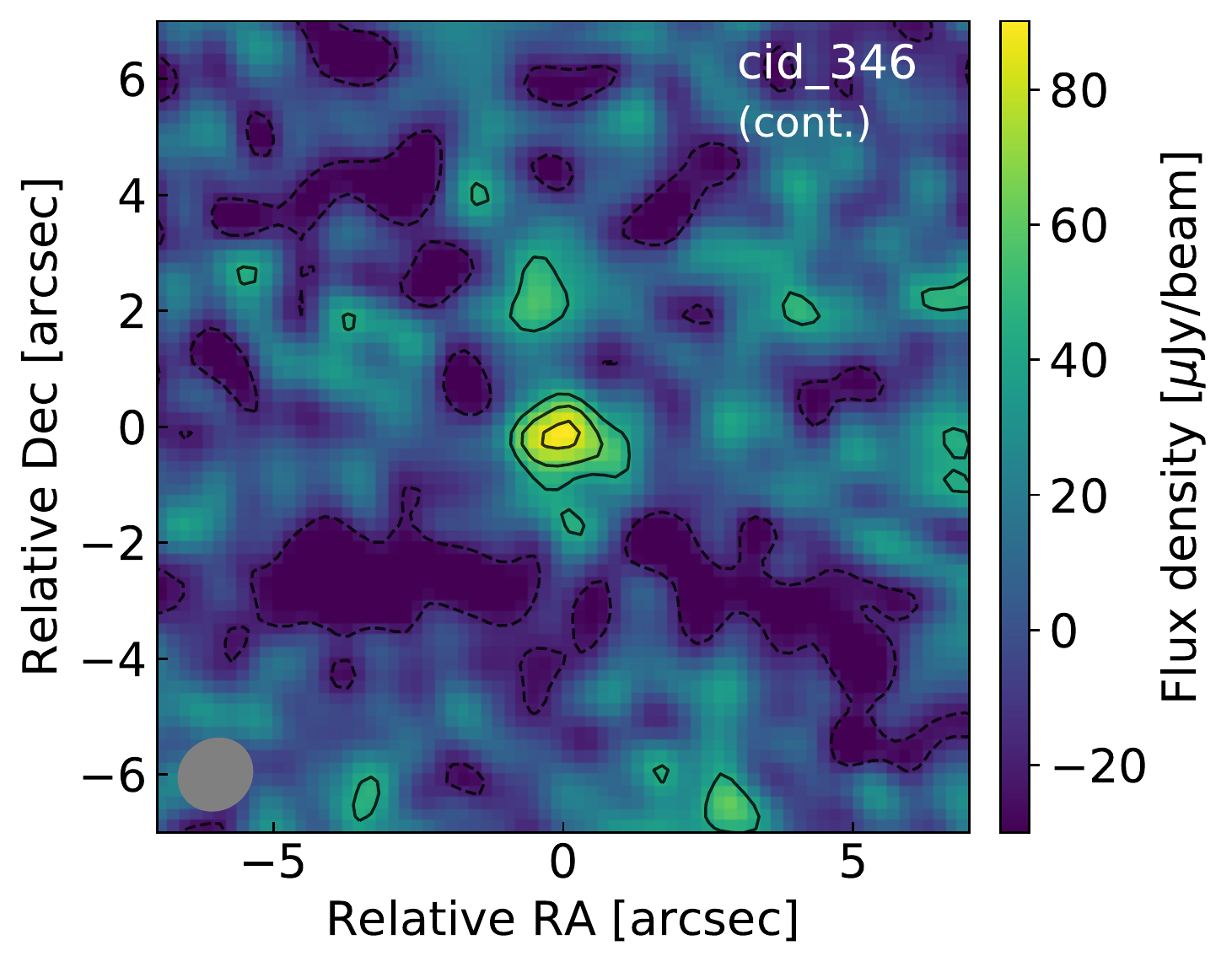} 
  	\hspace{2mm}
  	\includegraphics[width=7.2cm]{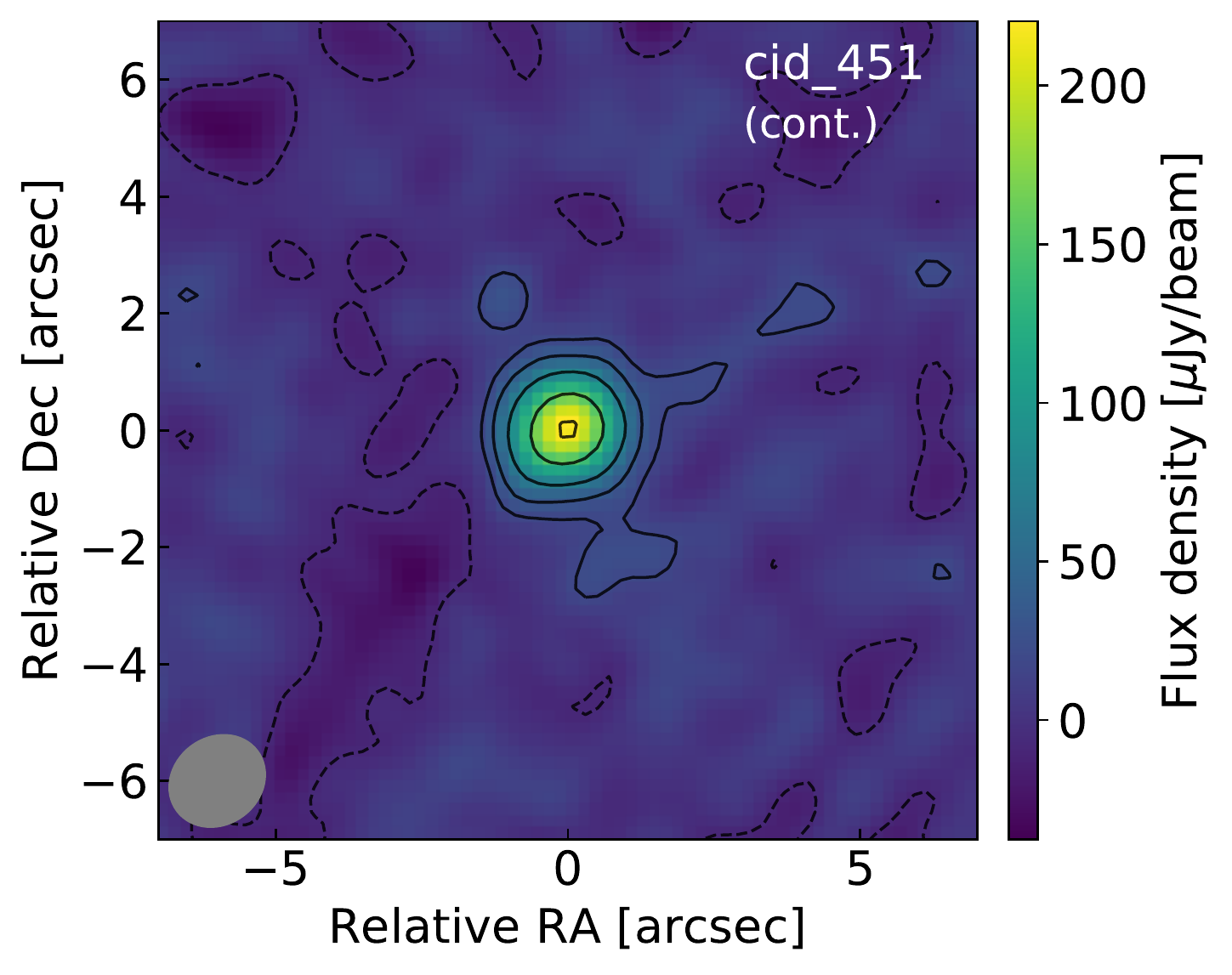} 
  	\hspace{2mm}
  	\includegraphics[width=7.2cm]{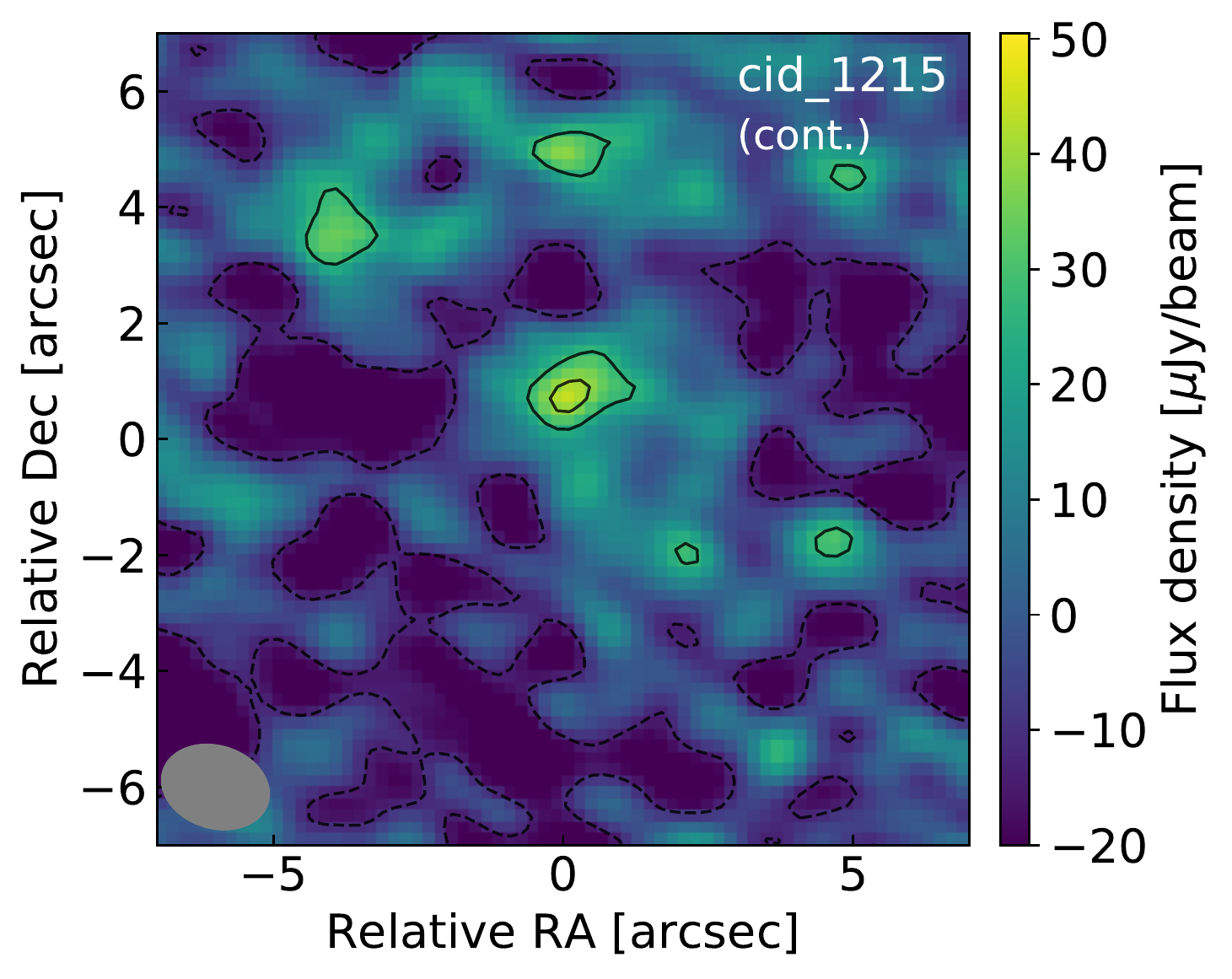} 
 	\hspace{2mm}
 	\includegraphics[width=7.2cm]{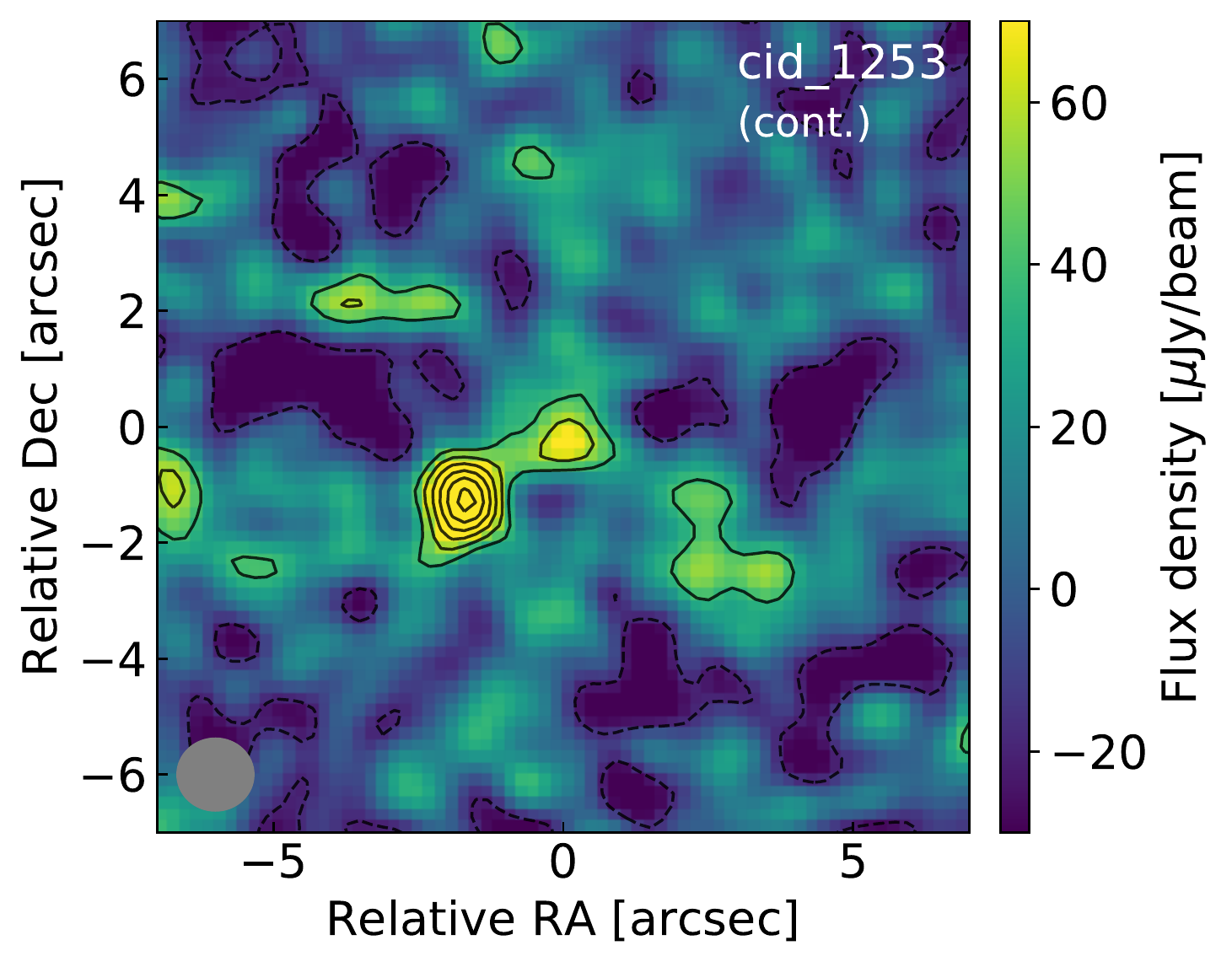} 
  	\hspace{2mm}
 	\includegraphics[width=7.2cm]{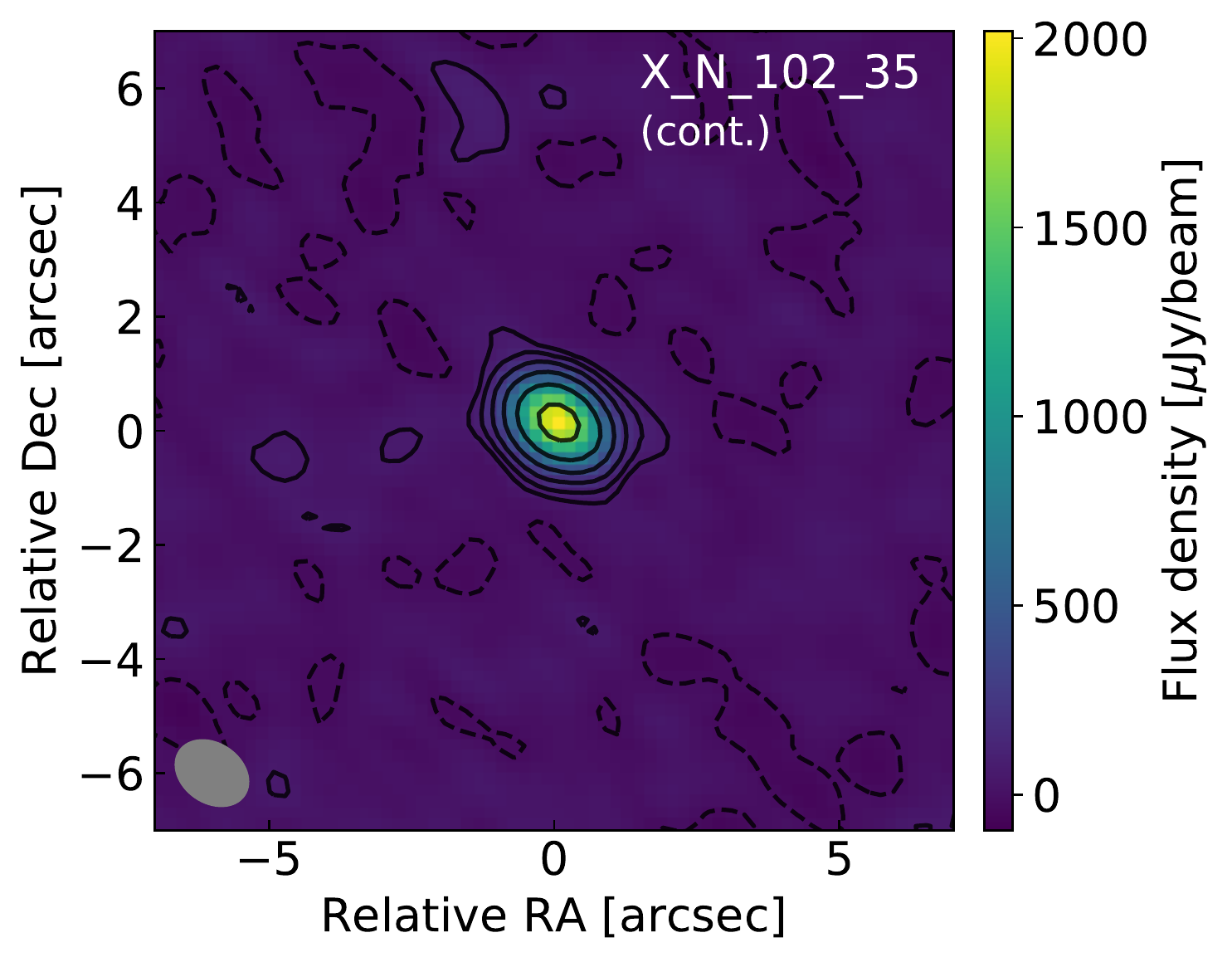} 
    \caption{Continuum emission maps at $\sim$100 GHz. Black contours are in steps of 1$\sigma$, starting from 2$\sigma$, for all targets but cid\_451 ($\sigma \times$[2, 4, 8, 16, 24]) and X\_N\_102\_35 ($\sigma \times$[2, 4, 8, 16, 32, 64]). Dashed contours correspond to $-$1$\sigma$. The beam of each observation is shown as a gray ellipse on the bottom-left of the maps. }
\end{figure*}
\end{appendix}

\end{document}